\documentclass[iicol,sn-nature,Numbered]{sn-jnl}%  
\usepackage{graphicx}%
\usepackage{multirow}%
\usepackage{amsmath,amssymb,amsfonts}%
\usepackage{amsthm}%
\usepackage{mathrsfs}%
\usepackage[title]{appendix}%
\usepackage{xcolor}%
\usepackage{textcomp}%
\usepackage{manyfoot}%
\usepackage{booktabs}%
\usepackage{algorithm}%
\usepackage{algorithmicx}%
\usepackage{algpseudocode}%
\usepackage{listings}%

\usepackage[figuresright]{rotating}
\usepackage{makecell}
\usepackage{subfig}
\usepackage{diagbox}

\begin{document}

% \title{Foundation Symbolic Regression Model on Complex Networks}
\title{Symbolic Foundation Regressor on Complex Networks}
% \title{Uncovering Interpretable Physical Representations from Observational Data via Symbolic Foundation Regressor on Complex Networks}
% \title{Mining Interpretable Compressed Representations from Spatio-Temporal Data via Foundation Symbolic Regression Model on Complex Networks}
% \title{Automatically Discovering Scientific Laws from Spatio-Temporal Data via Foundation Symbolic Regression Model on Complex Networks}
% \title{Discovering Scientific Laws from Spatio-Temporal Data via Foundation Symbolic Regression Model on Complex Networks}
% \title{
% Accelerating the Discovery of Scientific Laws from Spatio-Temporal Data via Foundation Symbolic Regression Model on Complex Networks}


\author[1,2]{\fnm{Weiting} \sur{Liu}}\email{liuwt23@mails.jlu.edu.cn}

\author*[1,2]{\fnm{Jiaxu} \sur{Cui}}\email{cjx@jlu.edu.cn}

\author[1,2]{\fnm{Jiao} \sur{Hu}}\email{hujiao22@mails.jlu.edu.cn}

% \author[1,2]{\fnm{xxx} \sur{xxx}}\email{xxx@jlu.edu.com}

\author*[1,2]{\fnm{En} \sur{Wang}}\email{wangen@jlu.edu.cn}

\author*[1,2]{\fnm{Bo} \sur{Yang}}\email{ybo@jlu.edu.cn}

% \affil*[1]{\orgdiv{Department}, \orgname{Organization}, \orgaddress{\street{Street}, \city{City}, \postcode{100190}, \state{State}, \country{Country}}}

\affil[1]{\orgdiv{College of Computer Science and Technology}, \orgname{Jilin University}, \orgaddress{ \city{Changchun}, \postcode{130012}, \country{China}}}

\affil[2]{\orgdiv{Key Laboratory of Symbolic Computation and Knowledge Engineering of Ministry of Education}, \orgname{Jilin University}, \orgaddress{\city{Changchun}, \postcode{130012}, \country{China}}}


\onecolumn

\abstract{
In science, we are interested not only in forecasting but also in understanding how predictions are made, specifically what the interpretable underlying model looks like.
Data-driven machine learning technology can significantly streamline the complex and time-consuming traditional manual process of discovering scientific laws, helping us gain insights into fundamental issues in modern science. 
In this work, we introduce a pre-trained symbolic foundation regressor that can effectively compress complex data with numerous interacting variables while producing interpretable physical representations. 
Our model has been rigorously tested on non-network symbolic regression, symbolic regression on complex networks, and the inference of network dynamics across various domains, including physics, biochemistry, ecology, and epidemiology.
The results indicate a remarkable improvement in equation inference efficiency, being three times more effective than baseline approaches while maintaining accurate predictions.
Furthermore, we apply our model to uncover more intuitive laws of interaction transmission from global epidemic outbreak data, achieving optimal data fitting. This model extends the application boundary of the pre-trained symbolic regression models to complex networks, and we believe it provides a foundational solution for revealing the hidden mechanisms behind changes in complex phenomena, enhancing interpretability, and inspiring further scientific discoveries.

% \textcolor{red}{
% In science, we are interested not only in forecasting but also in understanding how predictions are made, specifically what the interpretable potential model looks like.
% This is an effective and interpretable compressed representation of complex data.
% Data-driven machine learning has the ability to alleviate the lengthy process of manually discovering scientific laws
% help us to gain insight into fundamental problems in modern science.
% Expand application boundary
% The abstract serves both as a general introduction to the topic and as a brief, non-technical summary of the main results and their implications. Authors are advised to check the author instructions for the journal they are submitting to for word limits and if structural elements like subheadings, citations, or equations are permitted.
% }
}


\maketitle

\section*{Introduction}\label{sec1}

Mankind's pursuit of scientific laws and the eagerness to explore the unknown have never ceased.
From Galileo's principle of relativity and Hermann's law of conservation of energy to Schrödinger's wave equation in modern quantum mechanics, scientific revolutions have long been linked to the milestone discoveries of these laws.
% The process of discovering and exploring scientific laws relies heavily on careful observation and reasoning.
Traditionally, this discovery process has depended primarily on human intelligence, requiring strong empirical assumptions \cite{shan2025interference} and often taking a considerable amount of time, accompanied by elements of chance.
For instance, it took thousands of years, from ancient stargazing activities and hypotheses about celestial movement, to the discovery of the law of universal gravitation, which helps us understand the solar system.
Fortunately, with the flourishing development of artificial intelligence technology, data-driven machine learning methods show significant potential to accelerate the process of scientific discovery \cite{wang2023scientific, gao2023data, makke2024interpretable}.
For example, by using 30 years of trajectory data from our solar system’s sun, planets, and large moons, Newton’s law of gravitation has been successfully and efficiently identified through a combination of graph networks and symbolic regression \cite{lemos2023rediscovering}.
% Therefore, this work explores a new machine learning method to automatically discover scientific laws from observational or experimental data, promoting the pace of scientific discovery.

The origins of data-driven methods for mining symbolic scientific laws trace back to the 1950s, when genetic algorithms \cite{holland1992genetic} were utilized to search for target mathematical expressions fitting input-output pairs across a broad expression space.
The regression tools developed based on this principle, such as Eureqa \cite{dubvcakova2011eureqa}, GPLearn \cite{richter2022gplearn}, and PySR \cite{cranmer2023interpretable}, have been instrumental in deriving analytical formulas for the design of one-dimensional sonic crystals \cite{hruvska2025analytical}, discovering a mass scaling relationship for planar black holes in spiral galaxies \cite{davis2023discovery}, and aiding in jet background subtraction during heavy-ion collisions \cite{mengel2023interpretable}.
However, when exploring a vast expression space, search-based approaches often require extensive time and yield complex, hard-to-understand outcome expressions, limiting their practical applications.
Learning-based symbolic regression approaches introduce assumptions such as sparsity \cite{brunton2017discovering, rudy2017data, kaptanoglu2021pysindy, chen2021physics, gao2022autonomous, liu2024kan, liu2024kan2} and physical symmetry \cite{udrescu2020ai}, along with external inference systems \cite{cornelio2023combining}, to identify simple and scientifically meaningful mathematical expressions, and employ cutting-edge techniques like reinforcement learning \cite{petersen2019deep, mundhenk2021symbolic, glatt2021deep, tenachi2023deep, Beihang-NCS-2025} or Monte Carlo tree search \cite{sun2022symbolic, xu2024reinforcement, Liyong2025ICLR} to efficiently perform trial-and-error processes.
Nonetheless, they continue to struggle with defining scientifically meaningful expressions, experience long inference times, and do not fully leverage existing prior knowledge and data.

With the emergence of large language models such as ChatGPT \cite{ouyang2022training} and DeepSeek \cite{liu2024deepseek}, artificial intelligence technology has reached a new stage of development, achieving impressive results in protein design \cite{madani2023large} and natural evolution simulation \cite{hayes2025simulating}. Due to being built on extensive data and knowledge, they typically exhibit remarkable generality and can adapt to various tasks.
Inspired by this principle, a new generation of scientific discovery techniques has emerged, specifically the pre-trained symbolic regression models \cite{valipour2021symbolicgpt, biggio2021neural, vastl2024symformer, li2022transformer, d2023odeformer}, which create extensive mappings from data to equations, such as 100 million data-equation pairs \cite{biggio2021neural}, to pre-train transformer-based autoregressive models.
Through the integration of massive knowledge, these pre-trained models have the potential to emerge predictive abilities for new equations.
Meanwhile, in contrast to the hour-long search overhead associated with search-based approaches, these models do not require starting from scratch during testing. Instead, the input data is only propagated forward, allowing for the quick generation of the corresponding equation in just minutes.
However, the performance of existing pre-trained models for symbolic regression problems significantly declines when dealing with more than three variables \cite{biggio2021neural}, and they can typically handle up to a maximum of ten variables \cite{vastl2024symformer}.
This can lead to resistance in discovering scientific laws, as complex phenomena often arise from the coevolution of numerous interrelated variables.
For example, global epidemic transmission involves free variables from over 200 countries and regions \cite{cui2024learning}. If the modeling is fine enough to consider each individual as a variable, the transmission system could encompass over a billion free variables.
% If we simply test all symbolic models governing complex phenomena by increasing the equation length, it may take longer than the age of the universe to reach the target model of the system \cite{AI Feynman: A physics-inspired method for symbolic regression}.
Therefore, how to represent the numerous complex phenomena generated by so many variables and the various interactive relationships formed between variables, while effectively and efficiently uncovering the underlying simple governing laws, remains an open challenge.

In this work, we propose a symbolic foundation regressor on complex networks that can effectively handle spatio-temporal data of complex phenomena with massive correlated variables and efficiently reconstruct the corresponding governing equations to address the challenge.
We create approximately 2 billion mathematical expressions on complex networks as the training set to pre-train the foundation model, allowing it to learn a unified mapping from the data space to the expression space.
Our analysis of symbolic regression tasks involving independent (non-network) low-dimensional variables, as well as those applied to complex networks, alongside various network dynamics scenarios in fields such as physics, biochemistry, ecology, and epidemiology, shows that our model is highly applicable, effective, and efficient.
In particular, within the context of network dynamics scenarios, our model can reconstruct the network dynamics equation in only 1.2 minutes, which is more than three times faster than the search-based and learning-based approaches, and shorten the time for discovering more accurate new laws of real-world global infectious disease outbreaks to within half a minute.
To our knowledge, this is the first foundation symbolic regression model designed specifically for complex networks, extending pre-trained models to effectively address complex phenomena involving a large number of variables. 
We believe that our foundation regressor can serve as a basis for accelerating the discovery of scientific laws while enhancing the practicality of symbolic regression tools, marking an important step towards the long-term goal of developing artificial intelligence scientists.

\begin{figure*}[t]
\centering
\includegraphics[width=1.0\textwidth]{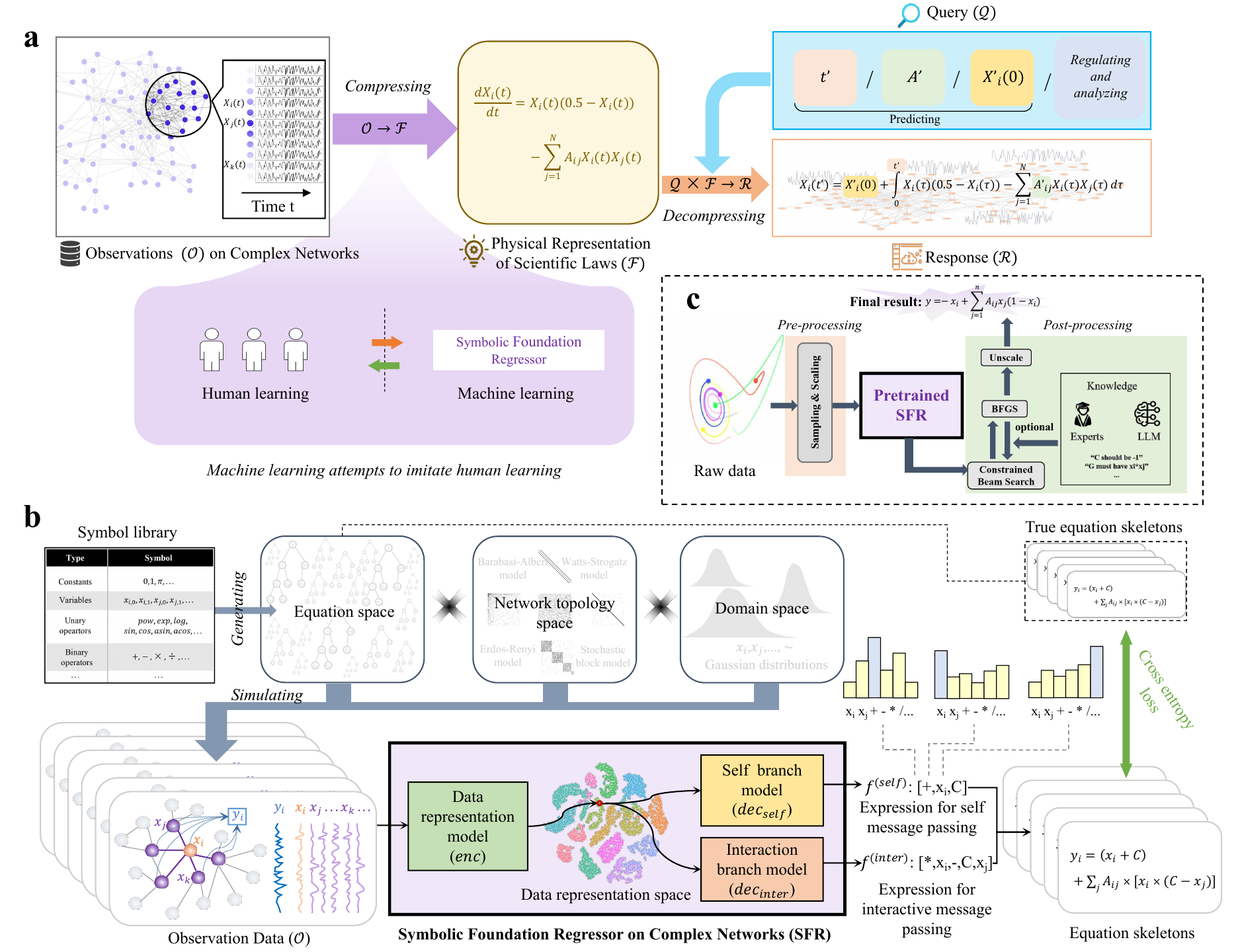}
\caption{
{\textbf{a.} Human beings can condense observations into scientific laws, and then use these compressed laws to analyze and regulate various systems.
We attempt to imitate this process of compressing interpretable physical representations in human learning through machine learning. 
\textbf{b.} The overall process of our Symbolic Foundation Regressor (SFR), including the generation of massive high-quality synthetic data-equation pairs, model architecture with dual branches, and the pre-training process. \textbf{c.} 
After pre-training the SFR, we can effectively derive the target equation for the unseen downstream task through a single forward propagation. Additionally, data pre-processing and equation post-processing are included to enhance the accuracy of the recovery equation.}}
\label{intro}
\label{fig:1}
\end{figure*}

\section*{Results}\label{sec2}

The discovery process of scientific laws can be broadly modeled as $\mathcal{O} \rightarrow \mathcal{F}$, where $\mathcal{O}$ represents the observed data, $\mathcal{F}$ represents the scientific laws.
Human scientists can induce spatio-temporal observations of complex phenomena into an interpretable physical representation \cite{iten2020discovering}, i.e., \textit{compressing} ($\mathcal{O} \rightarrow \mathcal{F}$).
In fact, recent neuroscience research reveals that the human brain has the ability to compress and encode complex behaviors and phenomena into basis functions to facilitate social decisions \cite{wittmann2025basis}.
When querying questions ($\mathcal{Q}$) about physical settings later, such as predicting, regulating, and analyzing, scientists should be able to provide correct responses ($\mathcal{R}$) using only representations rather than raw data, i.e., \textit{decompressing} ($\mathcal{Q}\times \mathcal{F} \rightarrow \mathcal{R}$).
In this work, we mainly focus on using machine learning to attempt to mimic the human learning process, as shown in Fig.~\ref{fig:1}(a), especially in the compressing part, which is more challenging than directly solving with numerical methods for decompressing.
Specifically, we introduce a pre-trained symbolic foundation regressor to accomplish this challenging task by integrating massive amounts of data and knowledge of equations, revealing scientific laws from observations on complex networks.

% \subsection*{Universality in Symbolic Regression on Complex Networks}\label{subsec21}
\subsection*{Universality in symbolic regression on complex networks}\label{subsec21}

Uncovering the underlying scientific laws from complex phenomena can be modeled as a symbolic regression task on complex networks, formulated as $y_i = \mathcal{F}(x_i, \{x_j\}_{j\in \mathcal{N}_i})$, where
$x_i$ represents the observation state at node $i$, $\mathcal{N}_i$ is the set of neighbors of $i$, and $y_i$ denotes the output at that node. 
$\mathcal{F}$ represents the mathematical expression to be regressed using input-output data pairs.
Note that the input $(x_i, \{x_j\}_{j\in \mathcal{N}_i})$ consists not only of the state of the node itself but also the states of its neighboring nodes.
This implies that classical symbolic regression can be considered a special case of symbolic regression on complex networks in which there is only one independent node and no neighboring nodes involved.
Typically, the number of neighbors for nodes within a network varies, denoted as $k$, and this distribution is closely tied to the network's topology.
Furthermore, the states of the nodes can be multidimensional, i.e., $d$. 
To effectively address this complexity, symbolic regression on complex networks is necessary to reconstruct mathematical equations that are both varying and high-dimensional from the data, i.e., $k\times d$.
Consequently, the variety of network topologies and the flexible combination equations arising from high-dimensional variables create significant challenges for pre-training a model that can accurately represent the vast spaces of network topologies and equations.

\textbf{Decoupling network topology through local sampling}.
% Network topology serves as a common abstract representation tool for illustrating the interconnected relationships present in the complex real world and encompasses various types, including grid \cite{???}, power law \cite{???}, small world \cite{???}, random \cite{???}, and community \cite{???} structures.
Representing a large topological structure space, produced by various scales, types, degree distributions, and graph generation parameters is seemingly impossible.
To bypass this, we introduce a sampling strategy to collect the local observation states that are independent of global topological features.
Due to the fact that nodes and their neighbors are the fundamental building blocks of topological structures,
we thus model an observation sample by concatenating the central node ($x_i$) with its directly connected neighbors ($\{x_j\}_{j\in\mathcal{N}_i}$) and its corresponding output signal ($y_i$), as a triplet, i.e., 
$
o_i=(x_i, \{x_j\}_{j\in\mathcal{N}_i}, y_i).
$
By sampling various central nodes, we can utilize this strategy to produce observation data, i.e., $\mathcal{O}=\{o_1,o_2,...,o_i,...\}$.
It enables us to decouple the observations from factors like topological scale and type, 
allowing us to concentrate on local regions, which is particularly important in real-world situations where incomplete observations are often encountered.
Furthermore, it enhances modeling flexibility, which simplifies the representation of diverse scenarios, such as heterogeneous propagation equations on complex networks.

\textbf{Simplifying mathematical equations through physical priors}.
When confronted with complex, high-dimensional mathematical equations, simply testing each equation by increasing its length could take longer than the age of the universe to reach the desired outcome \cite{udrescu2020ai}. To alleviate the curse of dimensionality, we propose using a physical prior, suggesting that network states can be influenced both by their own states and by the states of their neighbors \cite{barzel2013universality, gao2022autonomous, liu2023we, cui2024learning}.
Specifically, we can decompose the mathematical equation ($\mathcal{F}$) into two coupled components: the self part ($f^{(self)}$) and the interaction part ($f^{(inter)}$), i.e., 
$
y_i=f^{(self)}(x_i)+\sum_{j=1}^{N}{A_{ij}f^{(inter)}(x_i,x_j)},
$
where the subscripts $i$ and $j$ represent the corresponding nodes, $N$ is the network size, and $A_{ij}$ is the adjacency matrix.
Here, $f^{(self)}$ captures the evolution of an individual node's state, while $f^{(inter)}$ describes the dynamics governing the interactions between neighboring nodes.
Thus, \( \mathcal{F} \) consists of two functions, i.e., \( \mathcal{F} := \{f^{(self)}, f^{(inter)}\} \).
Actually, the feasibility and universality of this formulation have been confirmed \cite{barzel2013universality,barzel2015constructing}, which can accommodate a broad spectrum of real-world information propagation mechanisms on complex networks with appropriate selections of $f^{(self)}$ and $f^{(inter)}$.
Significantly, such formulation can achieve the desired dimensionality reduction for high-dimensional mathematical equations, by learning the fixed $d$-variate $f^{(self)}$ and $2d$-variate $f^{(inter)}$, rather than directly inferring the varying $(k\times d)$-variate $\mathcal{F}$.
The universality provided by local sampling and physical priors lays the basis of our model construction.

\subsection*{The overall flow of the symbolic foundation regressor (SFR)}

Our implementation of the symbolic foundation regressor (SFR), which maps observed data to interpretable physical representations, consists of three main steps: generating training samples, building and training the model, and applying it to various downstream scenarios.

\textbf{Creating a corpus of equations and synthesizing observational data.} 
The quality of data plays a crucial role in determining model performance. Unlike natural language processing models that can utilize extensive real-world data, obtaining large volumes of complete data from actual network observation scenarios poses a challenge. This difficulty stems not only from the complexities of data collection but also from the scarcity of labeled data—specifically, real equations required for supervised training models. Therefore, we create a sufficiently rich corpus of equations and sample data from these equations to meet training needs.
To achieve this, we randomly generate an expression tree using a symbol library that includes constants, variables, unary operators, binary operators, and more. In this tree, the leaf nodes represent constants or variables, while the non-leaf nodes represent operators.
To ensure that the generated equations, characterized by the expression tree, are meaningful, we implement specific rules, including avoiding nested trigonometric and exponential functions and ensuring that the interaction function \( f^{(inter)} \) contains at least one \( x_{j} \) term. 
As a result, we can generate a large number of \( f^{(self)} \) and \( f^{(inter)} \), ultimately synthesizing approximately 2 billion effective equations on complex networks. 
This process involves randomly sampling topological structures ($A_{ij}$) from the network topology space formed by networks with various sizes and types, including grid, random, power law, small world, and community.
Built on these equations, we sample their inputs ($x_{i}$) from a domain space, e.g., a standard normal distribution, to simulate the outputs ($y_{i}$), thereby forming the training samples, i.e., $(\mathcal{O}, \mathcal{F})$.
Additional details regarding the equation generation process and the associated validity rules can be found in the Method section.

\textbf{Constructing a set-to-sequence model with dual branches.}
Since the input $\mathcal{O}$ of the model is a set, we encounter a translation problem from a set to a sequence.
We thus propose a set-to-sequence model with dual branches for symbolic regression on complex networks, consisting of a data representation model ($enc$), a self branch model ($dec_{self}$), and an interaction branch model ($dec_{inter}$).
The data representation model is primarily implemented based on Set Transformer \cite{lee2019set}, which maps the sampled set of observed data to the representation space, i.e., $h=enc(\mathcal{O})$, where $h$ is the data representation.
As the data is simulated using equations, the similarities in the data representation space can also reflect the similarities among the underlying equations, providing a solid initial representation for the subsequent branch decoders.
To enable the generation of functions \( f^{(self)} \) and \( f^{(inter)} \), we design two transformer-based branch decoders to synchronously produce prefix expression sequences for \( f^{(self)} \) and \( f^{(inter)} \) in an autoregressive manner, i.e., 
$p(e^{(self)}_{k+1}|h,e^{(self)}_{1},...,e^{(self)}_{k})=dec_{self}(h,e^{(self)}_{1},...,e^{(self)}_{k})$
and
$p(e^{(inter)}_{k+1}|h,e^{(inter)}_{1},...,e^{(inter)}_{k})=dec_{inter}(h,e^{(inter)}_{1},...,e^{(inter)}_{k})$,
where $e_k$ represents the expression symbol generated at step $k$.
Using the extensive data-equation pairs generated earlier, we employ cross-entropy loss to pre-train the model thoroughly, as demonstrated in Fig.~\ref{fig:1}(b). The detailed model architecture can be found in the Method section.

\textbf{Applying to various downstream scenarios.}
After pre-training the SFR, we can efficiently derive the target equation for unseen downstream tasks using a single forward propagation, as illustrated in Fig.~\ref{fig:1}(c). 
It is important to highlight that the training data is constructed by sampling \(x_i\) from a known distribution. However, in real-world scenarios, this distribution is often unknown and may differ from the one used during pre-training. To tackle this issue, we apply a scaling transformation operation before inputting the data into the model, converting real data to match the model’s desired distribution.
Once the model generates the output equation, we optimize the constants within the equation using optimization algorithms such as BFGS \cite{nocedal2006numerical}, which enhances the precision of the equation. The final equation is then obtained through an unscaling operation. 
Additionally, during the process of the constrained beam search, we have the option to incorporate domain knowledge from experts or large language models, facilitating the equation generation. This not only creates an interface for integrating our method with existing large language models but also opens avenues for increased flexibility in our model.

\subsection*{Validation on classical non-network symbolic regression}\label{subsec22}

% \subsection*{Symbolic Regression on Independent Variables.}\label{subsec22}
\begin{figure*}[t]
\centering
\includegraphics[width=0.9\textwidth]{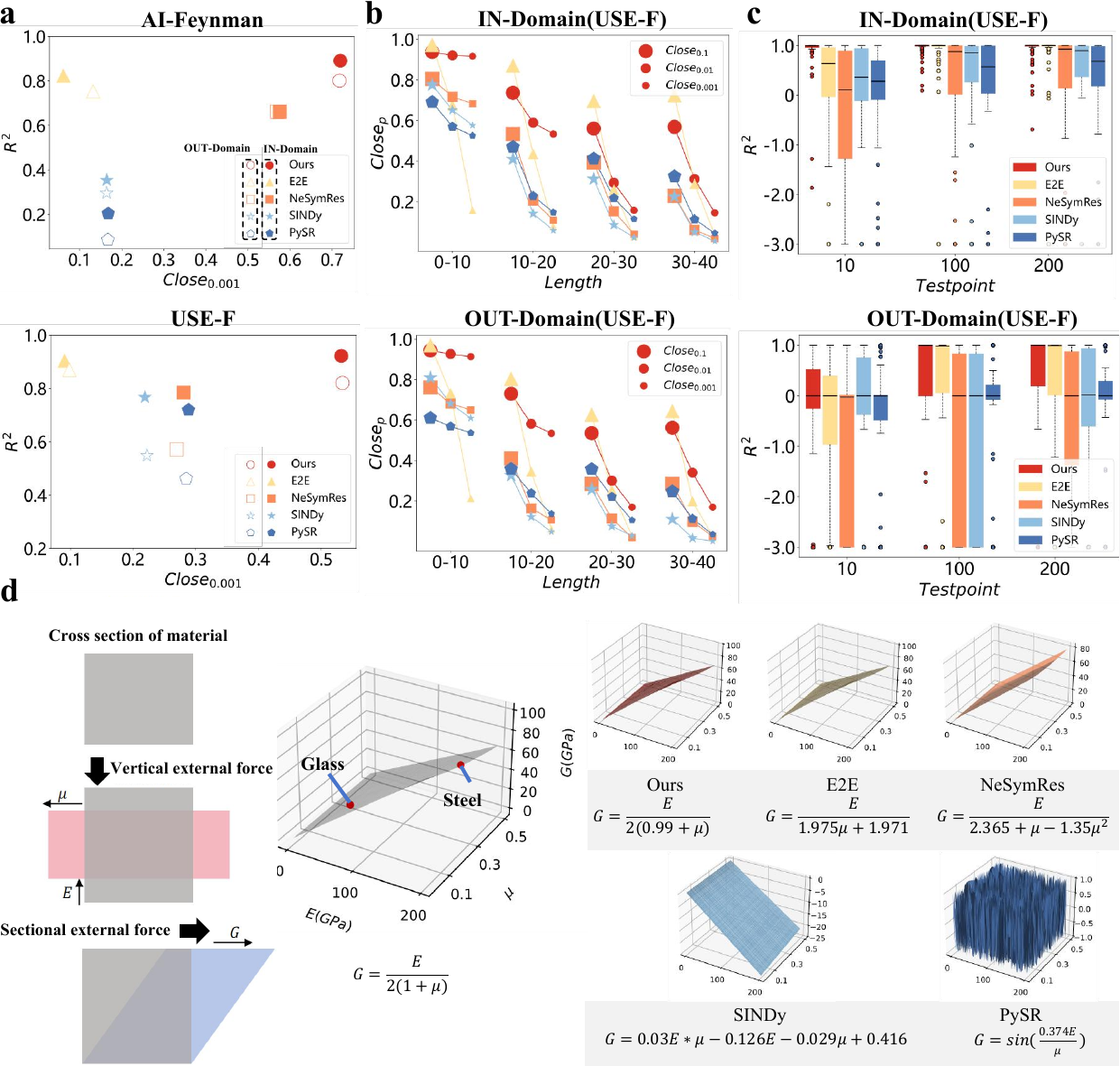}
\caption{
{Results on classical non-network symbolic regression tasks. 
\textbf{a.} Comparison of the execution accuracy ($R^2$, $Close_{0.001}$) from various methods (PySR, SINDy, NeSymRes, E2E, and Ours) on $2$ datasets (AI-Feynman and USE-F).
\textbf{b.} Comparison of the execution accuracy from various methods for equations with different lengths in USE-F.
\textbf{c.} The influence of the number of test data points on the results.
\textbf{d.} A physical equation from the AI-Feynman dataset that describes the relationship between the modulus of rigidity $G$, modulus of elasticity $E$, and Poisson's ratio $\mu$ in material science for regression analysis. our SFR can reconstruct the equation closest to ground truth with the same amount of data, demonstrating the applicability and potential of our model in classical symbolic regression tasks. 
}}
\label{exp1}
\end{figure*}

To evaluate the effectiveness of the SFR in handling classical non-network symbolic regression, we test it on two datasets, i.e., the AI-Feynman \cite{udrescu2020ai} and USE-F (Unseen Synthetic Equations with only Self parts).
The former is a commonly used standard dataset for symbolic regression tasks, which collects 100 physical equations describing natural phenomena from the Feynman Lectures \cite{feynman2018feynman}.
The latter refers to a testing set we created randomly, which has not appeared in the training set before, to enhance testing completeness, comprising approximately 30,000 equations with varying dimensions and complexities.
When applying our SFR to classical symbolic regression, we mask the neighboring states of the input data and only retain the output equation of the self branch decoder.
Evaluation tasks include reconstruction of input data (IN-Domain) and prediction of unknown data (OUT-Domain) using regression equations.
By comparing our model, SFR, to cutting-edge search-based approaches such as PySR \cite{pysr_github,cranmer2020pysr}, learning-based methods like SINDy \cite{de2020pysindy}, and pre-trained models including NeSymRes \cite{biggio2021neural} and E2E \cite{vastl2024symformer}, we can see from Fig.~\ref{exp1}(a) that SFR significantly outperforms these baselines in terms of $R^2$ and $Close_p$. The $R^2$ value measures the fit between the predictions produced by the reconstructed equation and the actual data, while $Close_p$ provides a more rigorous assessment by quantifying the percentage of data points for which the relative error between the predictions and the real data is less than $p$.
It should be noted that we also compared our results with Kolmogorov–Arnold Networks (KAN) \cite{liu2024kan, liu2024kan2}, but adjusting the network structure and parameters for each equation proved challenging. Even with a generalized structure, we could not achieve satisfactory results, as indicated by an average $R^2$ of less than -10. We thus have chosen not to include its results in our analysis.
Through multiple experiments on the USE-F, we found that both the length of the equation and the number of test data points affect the results of symbolic regression methods. 
As the length of the equation increases, the performance of these methods tends to decline (see Fig.~\ref{exp1}(b)). 
Conversely, an increase in the number of test data points enhances performance (see Fig.~\ref{exp1}(c)). 
Notably, our SFR consistently outperforms baselines, regardless of changes in these factors, especially in more stringent measurement. 
We pick up a physical equation from the AI-Feynman dataset that describes the relationship between the modulus of rigidity $G$, modulus of elasticity $E$, and Poisson's ratio $\mu$ in material science for regression analysis. As shown in Fig.~\ref{exp1}(d), our SFR can reconstruct the equation closest to ground truth with the same amount of data, demonstrating the applicability and potential of our model in classical symbolic regression tasks. 
For further experimental results concerning other influencing factors, such as the number of operator and dimensions, as well as additional regression analyses and visualizations, please refer to Appendix \ref{secA2}.

\begin{figure*}[htbp]
\centering
\includegraphics[width=0.9\textwidth]{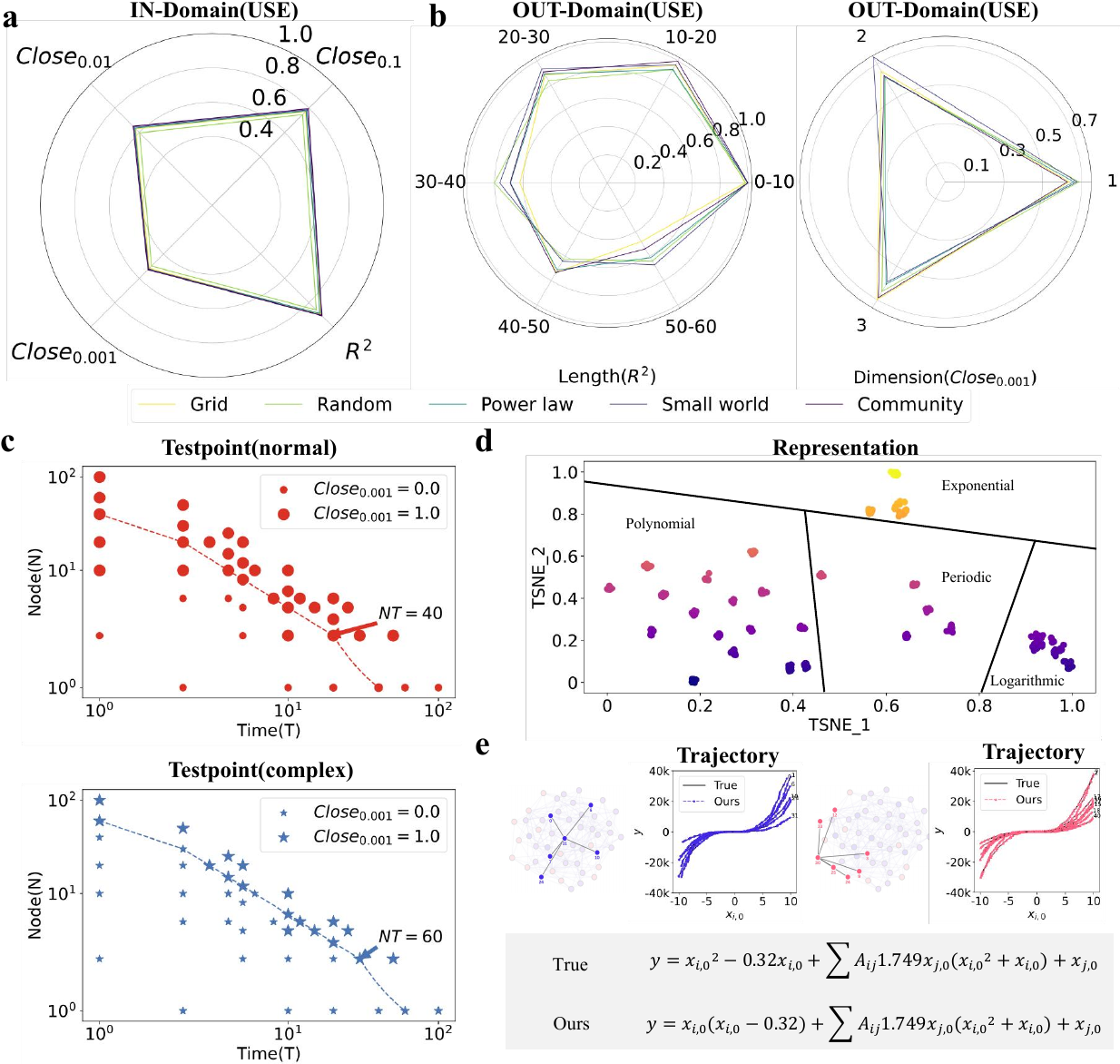}
\caption{
{Results of symbolic regression on complex networks. 
\textbf{a.} Comparison of the performance ($R^2$ and $Close_p$) of our model on USE with different topologies (Grid, Random, Power Law, Small World and Community). 
\textbf{b.} Comparison of the performance of our model on equations with different lengths and dimensions.
\textbf{c.} The impact of the number of test data points on equations with different complexity. 
\textbf{d.} Data representations (denoted as $h$) generated by equations with various characteristics are visualized through projection using t-SNE.
\textbf{e.} A specific example of symbolic regression from USE, demonstrating the ability of our model to regress high-precision equation on complex networks through local observations.}}
\label{exp2}
\end{figure*}

\subsection*{Validation on symbolic regression on complex networks}\label{subsec23}

To assess the effectiveness of our SFR in performing symbolic regression on complex networks, we have expanded the USE-F dataset to USE (Unseen Synthetic Equations), which now includes not only the self and interaction components with various lengths, dimensions, and operators, but also five types of topological structures: grid, random, power law, small world, and community. 
This results in approximately {5,000} symbolic regression testing tasks on these networks, each containing around {10 to 200} coupled variables (nodes).
Through extensive evaluations, we have observed that our SFR performs well across multiple metrics, including $R^2$ and $Close_p$, when regressing governing equations on complex networks (see Fig.~\ref{exp2}(a)). More importantly, its performance appears to be topology independent, achieved by integrating the local sampling strategy and decoupling interaction terms based on prior knowledge.
Although the length of the equation and the dimension of the node states are factors that affect performance, it still shows consistent results across different topological types (see Fig.~\ref{exp2}(b), more evaluations can refer to Fig.~\ref{a_exp2_2} in Appendix \ref{secA3}).
We also examine how the number of test data points impacts this task. The testing equations are categorized into two types based on their length: normal ($\leq$30) and complex ($>$30). 
From Fig.~\ref{exp2}(c), we can see that for normal equations, satisfactory regression results can be achieved with a product of observed nodes ($N$) and time steps ($T$) that exceeds 40. In contrast, complex equations require 60 or more test data points. This demonstrates that our SFR is advantageous in terms of the number of test data points needed, making it suitable for real-world scenarios where data is often sparse and difficult to obtain.
In addition, we feed the data generated by equations with various characteristics into the data representation model $enc$, obtain the data representation $h$, and perform a dimensionality reduction projection on $h$ using t-SNE.
We observed that equations with the same structure but different constants tend to cluster together, and those that share similar characteristics are also more likely to be grouped in the same region (see Fig.~\ref{exp2}(d)), illustrating the effective and distinctive representation power of our model, offering a clear semantic representation as a foundation for future decoders.
We also visualize a scenario of symbolic regression on complex networks in Fig. \ref{exp2}(e), showing the ability of our model to regress high-precision equations.
For large-scale networks (up to 5000 nodes), our model still has acceptable performance (see Fig.~\ref{a_exp2_5}).
Additional visualizations of data representations and scenario illustrations can be found in Appendix \ref{secA3}.

\subsection*{Application on inferring interpretable network dynamics}\label{subsec24}

\begin{figure*}[t]
\centering
\includegraphics[width=0.9\textwidth]{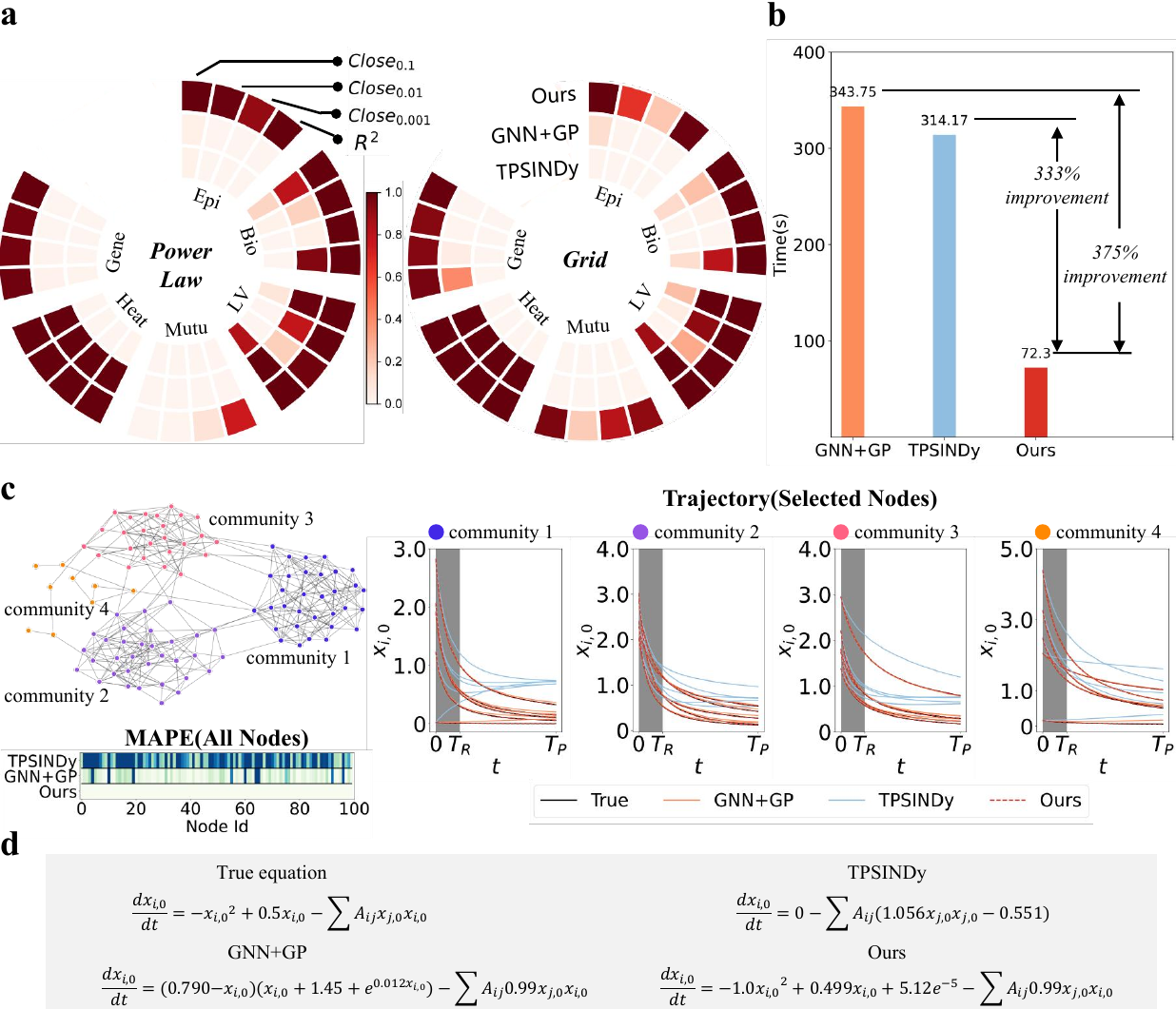}
\caption{
{Results on inferring interpretable network dynamics. 
\textbf{a} Comparison of the performance ($Close_p$, $R^2$) for reconstructing dynamics from six scenarios, including Epidemic (Epi), Biochemical (Bio), Lotka-Volterra (LV), Mutualistic interaction (Mutu), Heat diffusion (Heat), and Gene regulatory (Gene) dynamics.
\textbf{b.} Comparison of the average execution time across all dynamics for various methods.
\textbf{c.} The MAPE (Mean Absolute Percentage Error) between the predictive results produced by the discovered governing equations and ground truth in the LV scenario with four communities, and comparison of state prediction curves on selected nodes, where $T_R$ and $T_P$ are the termination times of IN-Domain and OUT-Domain respectively.
\textbf{d.} Comparison of governing equations inferred by various methods.}
}
\label{exp3}
\end{figure*}

As a typical application of symbolic regression on complex networks, inferring interpretable network dynamics mainly aims to learn the equations that govern network dynamics from observed data \cite{cranmer2020discovering,gao2022autonomous,lemos2023rediscovering}.
We apply our SFR to identify six representative network dynamics from physics, biochemistry, and ecology, i.e., gene regulatory (Gene) \cite{mazur2009reconstructing}, heat diffusion (Heat) \cite{zang2020neural}, epidemics (Epi) \cite{pastor2015epidemic}, biochemical dynamics (Bio) \cite{voit2000computational}, Lotka-Volterra model (LV) \cite{macarthur1970species}, and mutualistic interaction (Mutu) \cite{gao2016universal}.
It is important to perform difference calculations on the observed data ($x_i$), e.g., a five point finite difference method, to obtain $y_i := dx_i/dt$ before conducting the inference task.
And we use a mixture of Gaussian clustering sampling and distribution scaling to alleviate the offset between the training data distribution and that in the application scenario.
By comparing against state-of-the-art baselines in this task, i.e., a search-based GNN+GP \cite{cranmer2020discovering} and a learning-based TPSINDy \cite{gao2022autonomous}, our SFR exhibits the best performance in terms of all metrics, regardless of topology types, whether in linear or nonlinear dynamics (see Fig.~\ref{exp3}(a) and more evaluations can be found in Fig.~\ref{a_exp3_1} in Appendix \ref{secA4}).
More importantly, our SFR significantly enhances efficiency by restoring the most accurate control equations in just 1.2 minutes (see Fig.~\ref{exp3}(b)).
In comparison, this represents a 375\% improvement in efficiency over the GNN+GP, which takes approximately 5.7 minutes, and a 333\% improvement over the TPSINDy, which takes about 5.2 minutes. 
This showcases our SFR's ability to accurately predict new equations, leveraging the benefits from pre-training on a large number of data-equation pairs.
We also present the comparative results of different methods for the LV dynaimcs on a community structure, as illustrated in Fig.~\ref{exp3}(c). In terms of data fitting, both GNN+GP and ours significantly outperform the TPSINDy. However, our SFR model generates the smallest prediction error and produces the outcome equation that is more accurate and closer to the actual equation (Fig.~\ref{exp3}(d)).
Other network dynamics scenarios also lead to consistent conclusions (see Appendix \ref{secA4}).

\subsection*{Application on inferring the transmission laws of epidemics}
\label{subsec25}

\begin{figure*}[t]
\centering
\includegraphics[width=0.9\textwidth]{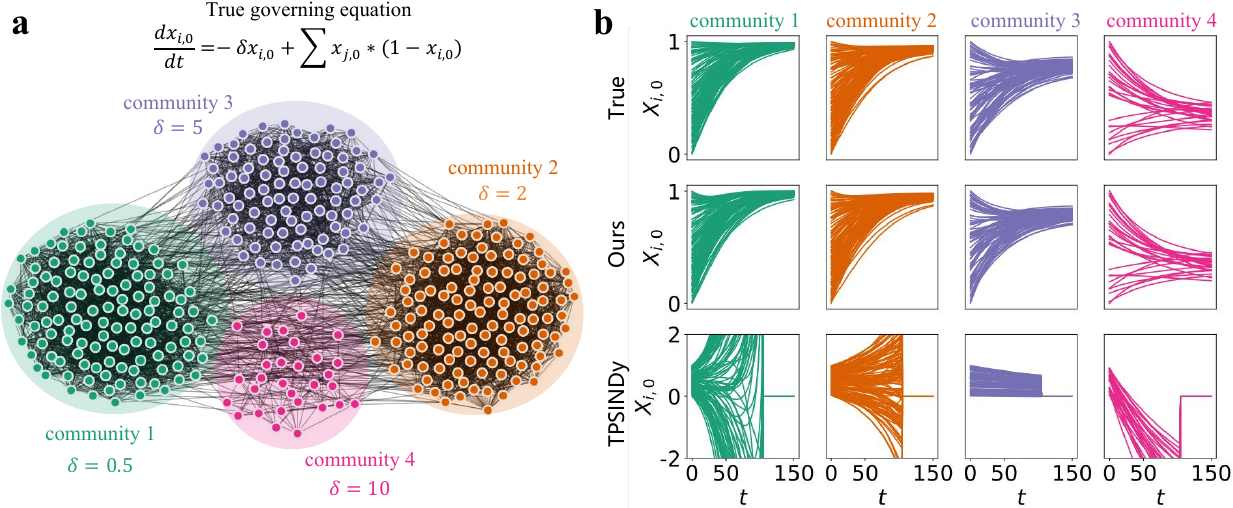}
\caption{Results on heterogeneous epidemic transmission in communities. \textbf{a.} Four heterogeneous transmission equations by assigning different recovery rates ($\delta$) in the epidemic equation, where $x_{i,0}:=I_{i}$ means the probability of an individual $i$ being susceptible.
\textbf{b.} Comparison of the state prediction curves generated by the governing equations inferred from observations at sampling nodes within each community.
Our SFR has successfully recovered the phenomena exhibited by heterogeneous transmission equations.
}\label{exp4}
\end{figure*}

Infectious diseases pose a substantial burden on global economies and public health \cite{msemburi2023estimates}. 
Understanding how transmission occurs allows us to effectively prevent and control large-scale epidemic outbreaks.
Herein, we apply our SFR to infer heterogeneous epidemic transmission patterns and discover new laws that govern global epidemic outbreaks in the real world.

\textbf{Heterogeneous epidemic transmission in communities.}
We construct four heterogeneous transmission equations by assigning different recovery rates ($\delta$) in the epidemic equation, i.e., $\frac{dI_{i}(t)}{dt}=-\delta I_{i}(t)+\beta\sum_{j=1}^{N}A_{ij}S_{i}(t)I_{j}(t)$, where $S_{i}(t)=1-I_{i}(t)$ represents the probability of an individual $i$ being susceptible, while $I_{i}(t)$ describes the probability of an individual being infected with an epidemic \cite{pastor2015epidemic}.
Specifically, we set the infection rate ($\beta$) to 1 and $\delta$ is chosen from $\{0.5, 2, 5, 10\}$, and then start simulating based on these equations in each community, as shown in Fig.~\ref{exp4}(a).
The basic reproduction numbers ($R_0=\delta/\beta$) for the four transmission equations are as follows: 2 (similar to Mpox \cite{grant2020modelling}), 0.5 (comparable to MERS \cite{kucharski2015role}), 0.2, and 0.1. 
An outbreak will die out if $R_0<1$, while if $R_0>1$, an outbreak will occur. Fig.~\ref{exp4}(b) clearly illustrates a significant increase in the number of infections in community 1. 
In contrast, communities 2 and 3 have established regional transmission, while the disease has had minimal impact on community 4.
Our SFR has successfully recovered the phenomena exhibited by heterogeneous transmission equations due to the introduction of local sampling strategies. However, TPSINDy cannot achieve that. 
The specific discovered equations can be found in Appendix \ref{secA5}.

\begin{figure*}[t]
\centering
\includegraphics[width=0.95\textwidth]{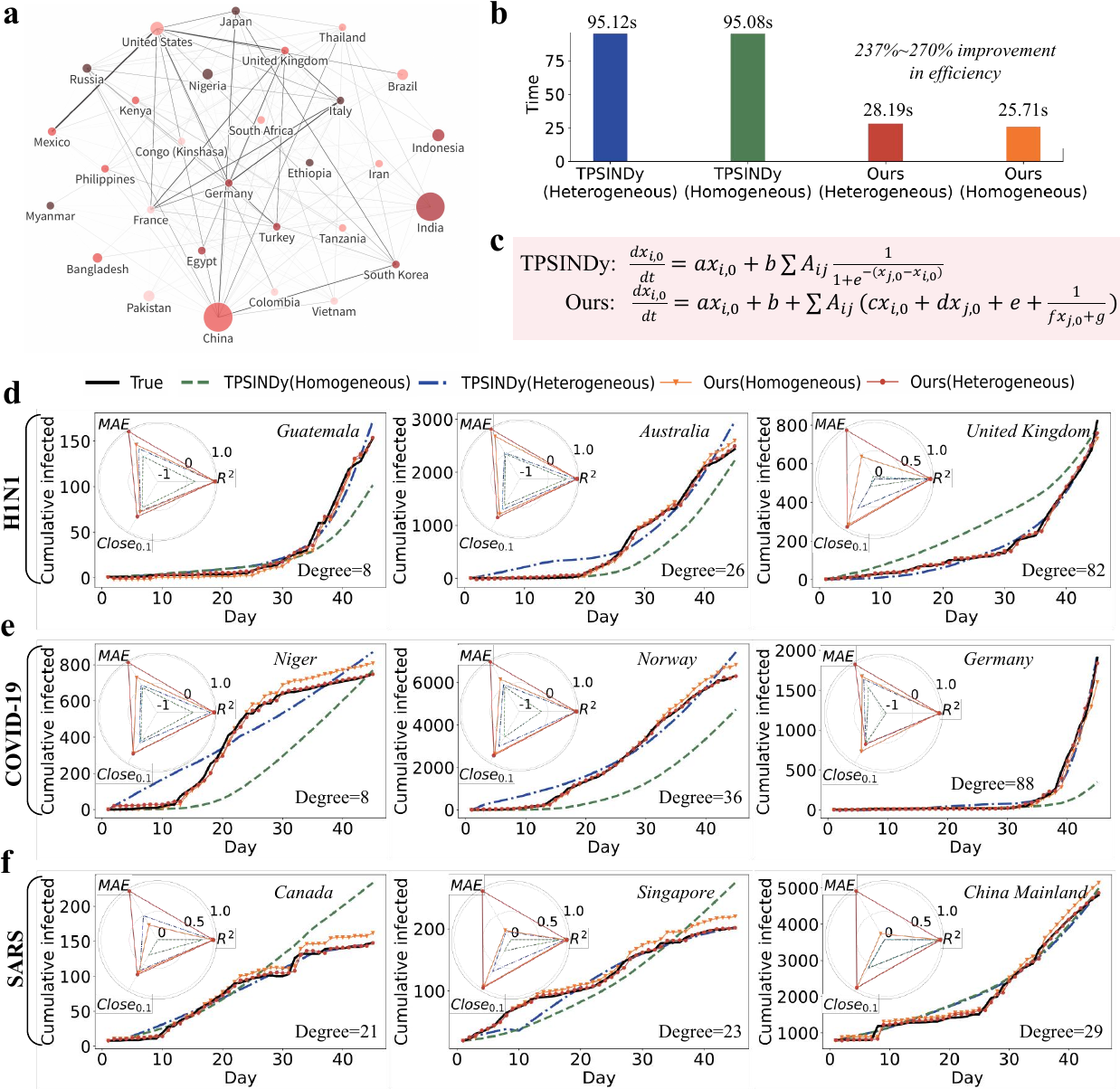}
\caption{
Results of inferring the transmission laws of real-world global epidemic outbreaks.
\textbf{a.} The worldwide airline network, only displaying countries or regions with populations over 50 million. Node size indicates population, while edge width reflects route flow.
\textbf{b.} Comparison of the time spent on discovering transmission laws.
\textbf{c.} Comparison of transmission laws discovered by TPSINDy and ours.
\textbf{d-f.} Comparison of the number of cases over time in various countries or regions generated by TPSINDy and ours on H1N1 (d), COVID-19 (e), and SARS (f). The embedding subplots show the comparison of various standardized evaluation indicators, with higher values being preferable.
}\label{exp5}
\end{figure*}

\textbf{Real-world global epidemic transmission.}

We collect daily global spreading data on H1N1 \cite{dataweb}, SARS \cite{dataweb}, and COVID-19 \cite{dong2020interactive}, and use the worldwide airline network retrieved from OpenFights \cite{openflights} as a directed, weighted empirical topology to build an empirical system of real-world global epidemic transmission, as shown in Fig.~\ref{exp5}(a).
Only the early data, prior to government interventions (e.g., the first 45 days), are taken into account to preserve the diseases' intrinsic spread dynamics.
We use the original infection data for H1N1 across all countries or regions to discover the transmission law, which are illustrated in Fig.~\ref{exp5}(c). The homogeneous equation version indicates that the transmission equations are same across all countries or regions, including their constants. In contrast, the heterogeneous equations mean that while the skeleton of the transmission equations remains consistent, the constants differ based on the specific characteristics of each region.
In terms of execution efficiency, our model has an improvement of 237 to 270 percent, enabling the discovery of the transmission law in less than half a minute (see Fig.~\ref{exp5}(b)).
In terms of equation form, the coupling term in the equation identified by TPSINDy is bounded, i.e., $\frac{b}{1+\exp{-(x_{j,0}-x_{i,0})}}$.
When the difference in the number of infections between neighbors and oneself exceeds 5, which often occurs, it suggests that either the neighbors have no impact at all, or there is a fixed impact represented by a constant $b$ value, which restricts the influence of neighbors. 
In contrast, the coupling term in our discovered equation is unbounded, i.e., $cx_{i,0}+dx_{j,0}+e+\frac{1}{fx_{j,0}+g}$, which more intuitively reflects the influence of neighbors on itself and leads to more accurate fitting results, regardless of whether the cases are homogeneous or heterogeneous (see Fig.~\ref{exp5}(d)).
We apply the transmission law identified in H1N1 to the data from SARS and COVID-19, focusing on learning the constants in the equations, which assumes that the transmission patterns of these various epidemics are similar, while acknowledging that the disease transmission characteristics (constants) differ.
As shown in Fig.~\ref{exp5}(e-f), the equation we discovered demonstrates the most effective results, indicating that we have uncovered a general law of transmission for global epidemic outbreaks, particularly one that describes the three epidemics with greater precision.
More comparison results and the discovered equations for all countries or regions can be found in Appendix \ref{secA54}.

\section*{Discussion}\label{sec3}

We presented a computational tool for accurately and efficiently discovering governing equations from observations on complex networks.
Beginning with generating a comprehensive dataset that encompasses a wide range of simulated network observation environments, we introduced a local sampling strategy to decouple global topological features and physical priors to simplify the mathematical form, and designed a set-to-sequence model with dual branches to establish a connection between observed data and the corresponding equations.
By pre-training the symbolic foundation regressor, we have extended the power of the pre-trained models to symbolic regression on complex networks. The advantage of pre-training allows our model to generalize across various new scenarios without extensive searching or learning, significantly reducing computational costs while maintaining the accuracy of the recovered equations.
By comparing our SFR with state-of-the-art techniques in different scenarios, including non-network symbolic regression, symbolic regression on networks, and varied types of network dynamics, the results demonstrated that our model reconstructs the most accurate equations with the highest efficiency. Furthermore, we uncovered new transmission laws that align more closely with data from three real global epidemic outbreaks. In summary, our work addresses the challenges posed by complex phenomena that are difficult to explain and involve numerous correlated variables, thereby facilitating the exploration and discovery of natural science laws.

Although the effectiveness and applicability of our model have been thoroughly validated, we still face more challenging scenarios in the real world, such as systems with high-order interactions, non-deterministic systems, partial differential systems, and non-autonomous systems.   Fortunately, the fundamental characteristics of our model provide significant potential for various applications, such as advanced training techniques on more data involving higher-order interactions.
An exciting development for this model is to combine with Large Language Models (LLMs) that have access to a vast amount of common knowledge. This integration could enhance the model's output by producing scientifically meaningful equations. 
To explore this further, we have conducted a preliminary attempt at LLM fusion—putting specific terms into the symbolic regression process to produce equations that better meet the desired requirements. These terms can be automatically extracted from historical equations using LLMs or generated through manual prompts, improving the interaction between the pre-trained model and users, thus increasing its controllability.
Preliminary results suggest that our proposed model can effectively leverage the knowledge contained in predefined terms during interactions with LLMs (more results can be found in Appendix \ref{secA6}). This finding opens up opportunities for future exploration on integrating our model with LLMs. Additionally, knowledge representation is multimodal, encompassing sound, images, videos, and more. Exploring how to leverage multimodal data to improve the accuracy of our model will be a fascinating direction.

\section*{{Method}}\label{sec4}
In this section, we present a detailed introduction to the key components of the proposed symbolic foundation regressor (SFR), including the creation of the corpus, construction of the model, and setup of applications.
\subsection*{Creation of the corpus}\label{sec41}
\textbf{Creating a collection of equations:}
The generation of equations in a corpus primarily relies on random expression trees. We can control the depth of the tree, the permitted operators and variables, and the probability of each operator's occurrence to produce the desired equations in a targeted manner.
The specific generation steps are as follows:
\begin{itemize}
\item[1.] Uniformly sample the number of binary operators $b$ within the range $[0,b_{max}]$ and the number of unary operators $u$ within the range $[0,u_{max}]$, where $b_{max}$ and $u_{max}$ represent the maximum allowed number of binary and unary operators, respectively.

\item[2.] Generate and sample an expression tree with $b$ non-leaf nodes, following the method outlined in \cite{lample2019deep}.

\item[3.] For each non-leaf node, sample a binary operator from the occurrence probability distribution of the binary operators, i.e.,  $P_b$.

\item[4.] For each leaf node, sample variable from $x_{i}$ or a constant $c$ within the range $[c_{min},c_{max}]$.

\item[5.] Randomly select a node whose subtree has a depth smaller than \( d_{depth} \), insert a new parent node with a unary operator sampled from the occurrence probability distribution of the unary operators, i.e., \( P_u \), and repeat this process \( u \) times.

\item[6.] Convert the produced expression tree into prefix expression for generating $f^{(self)}$.

\item[7.] Repeat steps 1 to 6 one additional time, replacing $x_{i}$ with $x_{i}$ or $x_{j}$ in step 4 to produce $f^{(inter)}$.

\item[8.] Combine $f^{(self)}$ and $f^{(inter}$to obtain the equation, i.e., $f^{(self)}+\sum{A_{ij}f^{(inter)}}$, where the topological structure $A_{ij}$ will be assigned values when generating data.

\item[9.] Verify the rationality, effectiveness and repeatability of the generated equation.
\end{itemize}

By repeating the above process, we can generate a large number of valid governing equations on complex networks.
The specific settings about generation parameters and distributions can be found in Appendix \ref{secA1}.

\textbf{Synthesizing observational data:}
We first select a governing equation from the generated corpus and then sample a topological structure from the topological space. By combining the sampled equation with the topological structure, we can derive the complete governing equation on a complex network, i.e., $y_i=f^{(self)}(x_i)+\sum{A_{ij}f^{(inter)}(x_i,x_j)}$. Next, by sampling values for \(x_i\in \mathbb{R}^d\) and \(x_j \in \mathbb{R}^d\) in the domain space, i.e., a standard normal distribution, we can calculate the corresponding \(y_i\), which allows us to obtain the observed data, i.e., $\mathcal{O}=\{o_1, o_2,...,o_i,...\}$, where $o_i=\{x_{i},\{x_{j}\}_{j \in \mathcal{N}},y_{i}\}$.
Note that, when generating the training data, the number of sampled central nodes is limited to less than one-tenth of the total number of nodes, which helps prevent the model from learning the features of the global topological structure.
Additionally, the number of sampled data points is approximately 200, simulating the sparse scenarios commonly encountered in real-world situations where collecting large amounts of observations can be challenging.

\subsection*{Construction of the model}\label{sec42}

Our SFR primarily relies on transformers with a bifurcated structure, consisting of a data representation model ($enc$), a self branch model ($dec_{self}$), and an interaction branch model ($dec_{inter}$).

\textbf{Data representation model ($enc$):}
The data representation model is primarily used to map the sampled set of observed data to the representation space, i.e. $h=enc(\mathcal{O})$, where $\mathcal{O}$ is the observations and $h$ is the data representation.
It is mainly composed of five components, including a float embedding layer ($emb^{754}$), a self-state embedding layer ($emb^{x_{i}}$), an interaction state embedding layer ($emb^{x_{j}}$), an output embedding layer ($emb^{y_{i}}$), and 
a joint embedding layer ($emb^{all}$).
$emb^{754}$ is grounded in the IEEE 754 Standard for Floating-Point Arithmetic \cite{ieee754-2019}, embedding floating-point numbers in $\mathcal{O}$ as their binary representations to mitigate gradient issues in calculations, particularly those arising from outliers in $y_{i}$, i.e., $\mathcal{O}^{754}=emb^{754}(\mathcal{O})$, where $\mathcal{O}^{754}=\{(x_{i}^{754},\{x_{j}^{754}\}_{j \in \mathcal{N}},y_{i}^{754}),...\}$ is the binary representations of observations.
Next, we feed $\mathcal{O}^{754}$ into their respective embedded layers, i.e.,  $emb^{x_{i}},emb^{x_{j}},emb^{y_{i}}$, for encoding and combine into $\mathcal{O}^{emb}=\{(emb^{x_{i}}(x_{i}^{754}),emb^{x_{j}}(\{x_{j}^{754}\}_{j \in \mathcal{N}}),emb^{y_{i}}(y_{i}^{754})),...\}$. 
Note that, unlike the direct vector-to-vector mappings of $emb^{x_{i}}$ and $emb^{y_{i}}$, $emb^{x_{j}}$ embeds a set as a vector, causing us to implement it using Deep Sets \cite{Deepsets}, which provides a neural network architecture for effectively handling the input of variable-sized sets.
Then, the data representation can be obtained by use a joint embedding layer $emb^{all}$, i.e., $h=emb^{all}(\mathcal{O}^{emb})$.
Since $\mathcal{O}^{emb}$ is still a set, we apply Set Transformer \cite{lee2019set} here to implement $emb^{all}$ for better capturing the contributions of each data point in the observed set while maintaining the characteristics of both permutation invariance and linear complexity in attention computation overhead.

\textbf{Self and interaction branch models ($dec_{self}$ and $dec_{inter}$):}
After obtaining the data representation $h=enc(\mathcal{O})$, $h$ will enter $dec_{self}$ and $dec_{inter}$ simultaneously to obtain functions \( f^{(self)} \) and \( f^{(inter)} \), the prefix expression of the equation, in an autoregressive manner, i.e.,
$p(e^{(self)}_{k+1}|h,e^{(self)}_{1},...,e^{(self)}_{k})=dec_{self}(h,e^{(self)}_{1},...,e^{(self)}_{k})$
and
$p(e^{(inter)}_{k+1}|h,e^{(inter)}_{1},...,e^{(inter)}_{k})=dec_{inter}(h,e^{(inter)}_{1},...,e^{(inter)}_{k})$,
where $e_k$ represents the expression symbol generated at step $k$.
For example, the corresponding target sequence of the formula $sin(cx+x^2)$ should be the token sequence $[sin \ add \ mul \ c \ x \ pow2 \ x]$. 
Using the extensive data-equation pairs generated earlier, we employ cross-entropy loss to pre-train the model thoroughly.
The detailed model architecture and specific hyper-parameters can be found in Appendix \ref{secA14}.

\subsection*{Setup of applications}\label{sec43}
Applying the pre-trained model to handle specific tasks can be divided into three main steps, including data pre-processing, generating equations via propagating forward, and equation post-processing.

\textbf{Pre-processing:}
For large-scale observations, according to the central limit theorem \cite{ross2014introduction}, the sampled data follows a normal distribution, so we perform a distribution scaling transformation on $x_{i}$ to match the training distribution, i.e.,
$\hat{x}_{i}=(x_{i}-\mu)/\sigma$, where $\mu$ and $\sigma$ are the statistical mean and standard deviation of the data.
For the identification of differential equations, since the observations are only $\{x_{i},\{x_{j}\}_{j \in \mathcal{N}}\}$, we need to perform difference calculations on the observations $\mathcal{O}$ to obtain $\frac{dx_{i}}{dt}$ as $y_{i}$.
Specifically, we approximate the derivatives through a five-point finite difference method \cite{gautschi2011numerical}:
$
\frac{dx_{i}}{dt}=\frac{x_{i}(t-2t_{\delta})-8x_{i}(t-t_{\delta})+8x_{i}(t+t_{\delta})-x_{i}(t+2t_{\delta})}{12t_{\delta}},
$
where $t_{\delta}$ represents the time interval.
Then, we perform clustering on sampled data $x_{i}$ and sample data from each class to remove correlations, and perform normal distribution sampling to obtain $x_{i}$, then, the distribution scaling transformation is applied.

\textbf{Generating equations via propagating forward:} The pre-processed data is fed into the pre-trained model SFR for equation regression. 
We use beam search during the decoding process, which can search for the optimal prefix expression sequence $e$ in a wider range compared to greedy algorithms. Generally, we set the beam size ($N_{beam}$) to 10.

\textbf{Post-processing:} After generating \(N_{beam}\) output equations, we employ optimization algorithms like BFGS \cite{nocedal2006numerical} to refine the constants within the equations and identify the most accurate one. This process enhances the accuracy of the equation regression. 
To finalize the equations, we apply an inverse distribution scaling transformation to ensure they are correctly formatted.
Assuming the distribution of the sampled data is $x_{i}\sim \mathbf{N}(\mu,\sigma)$ and the scaling sampled data is $x_{i}^\prime\sim \mathbf{N}(0,1)$, the process is as follows:
$
f^{(self)}(x_{i}^\prime)=f^{(self)}(\frac{x_{i}-\mu}{\sigma}),f^{(inter)}(x_{i}^\prime,x_{j}^\prime)=f^{(inter)}(\frac{x_{i}-\mu}{\sigma},\frac{x_{j}-\mu}{\sigma}).
$
In addition, we can impose specific limitations during the beam search process to arrive at the correct form of the equation. These limitations can be based on domain knowledge from experts or large language models. For example, if it's known that \textit{"the conduction equation of a certain force should include a cosine function term, such as cos(x)"}, this term can be incorporated into the decoding search process as a token.
Methods like constant and formal simplification are used to create equations that offer more scientific significance while ensuring accuracy.
More detailed pre-processing and post-processing methods can be found in the Appendix \ref{secA15}.

% \begin{itemize}
% \item Network symbolic regression: For the trained model, given a set of data from unknown equations, we use a beam search strategy during the decoding process to sample candidate equations $F^{(self)}$ and $F^{(neigh)}$ from the decoded sequence. Afterwards, we use BFGS to fit and correct the constant values in the equations by minimizing the squared loss between the true and predicted equations output values. It is worth noting that during the training process, the model only samples data from the standard normal distribution, and to adapt to different testing environments, we utilize affine transformation to adjust the distribution range of the data before inverse transforming the predicted equations to restore them to the true equations. Therefore, for data $X$ in unknown scenarios, we first need to perform decorrelation processing and normal distribution sampling from the data $X$ to obtain $X' \sim N(\mu,\sigma)$, then we utilize affine transformation to adjust the distribution range of the data $X'' \sim N(0,1)$, and then feed the transformed data into the model. Finally, we inverse transform the equation fitted by the model to obtain the correct equation, $y'=y(\sigma X''+\mu)$.

% \item Others regression: For general symbolic regression, we only fitting the $F^{self}$. For dynamics symbolic regression, we need to use differential computation to obtain the original input $Y_{i}$ and then proceed to the next steps.

% \end{itemize}

\subsection*{Performance measures}\label{sec44}
The performance measures for evaluating the methods in this work are as follows:

$R^2$ score, the coefficient of determination, is used to evaluate the fitting degree of the regression model, with a range $[-\infty,1]$, and the closer it is to $1$, the better the performance. The formula for calculating $R^2$ score is as follows:
$
R^2=1-\frac{\sum^{N}_{i=1}(x_{i}(t)-\hat{x_{i}}(t))^2}{\sum^{N}_{i=1}(x_{i}(t)-\bar{x_{i}}(t))^2},
$
where the $x_{i}(t)$ and $\hat{x}_{i}(t)$ represent the ground truth and prediction result of node $i$ at time $t$, $\bar{x_{i}}(t)$ is the average of $x_{i}(t)$ over $N$ nodes. 

$Close_{p}$ measure the accuracy of regression equations by evaluating the percentage of sampling points that satisfy the relative error precision, the formula for calculating $Close_{p}$ is as follows:
$
Close_{p}=\sum_{i=1}^{N}\frac{1}{N}I(|\frac{(x_{i}(t)-\hat{x_{i}}(t)}{x_{i}(t)}|\leq p),
$
where the $p$ are the relative error precision of $0.001,0.01,0.1$, and $I$ represents when the relative error between the ground truth and prediction result is less than $p$, $I=1$, on the contrary, it is $0$.

MAPE (Mean Absolute Percentage Error) and MAE (Mean Absolute Error), used to evaluate the error between the ground truth and prediction result, represents the absolute percentage error and absolute error between the ground truth and prediction result, with a range  $[0,\infty]$, the smaller the value, the more accurate it is. The formula for calculating MAPE and MAE is as follows:
$
MAPE=\frac{1}{N}(\sum_{i=1}^{N}|\frac{(x_{i}(t)-\hat{x_{i}}(t)}{x_{i}(t)}|)\times 100\%$,
$MAE=\frac{1}{N}(\sum_{i=1}^{N}|(x_{i}(t)-\hat{x_{i}}(t)|)
$.

\bibliography{sn-article}

\newpage

\onecolumn

\begin{appendices}

\begin{center}
{\Large{\textbf{APPENDICES for \textit{Symbolic Foundation Regressor on Complex Networks}}}}
\end{center}

\tableofcontents

\newpage

\section{More details on the method}\label{secA1}
\subsection{The specific distribution of equations in the corpus.}\label{secA11}
\begin{figure*}[h]
\centering
\includegraphics[width=1.0\textwidth]{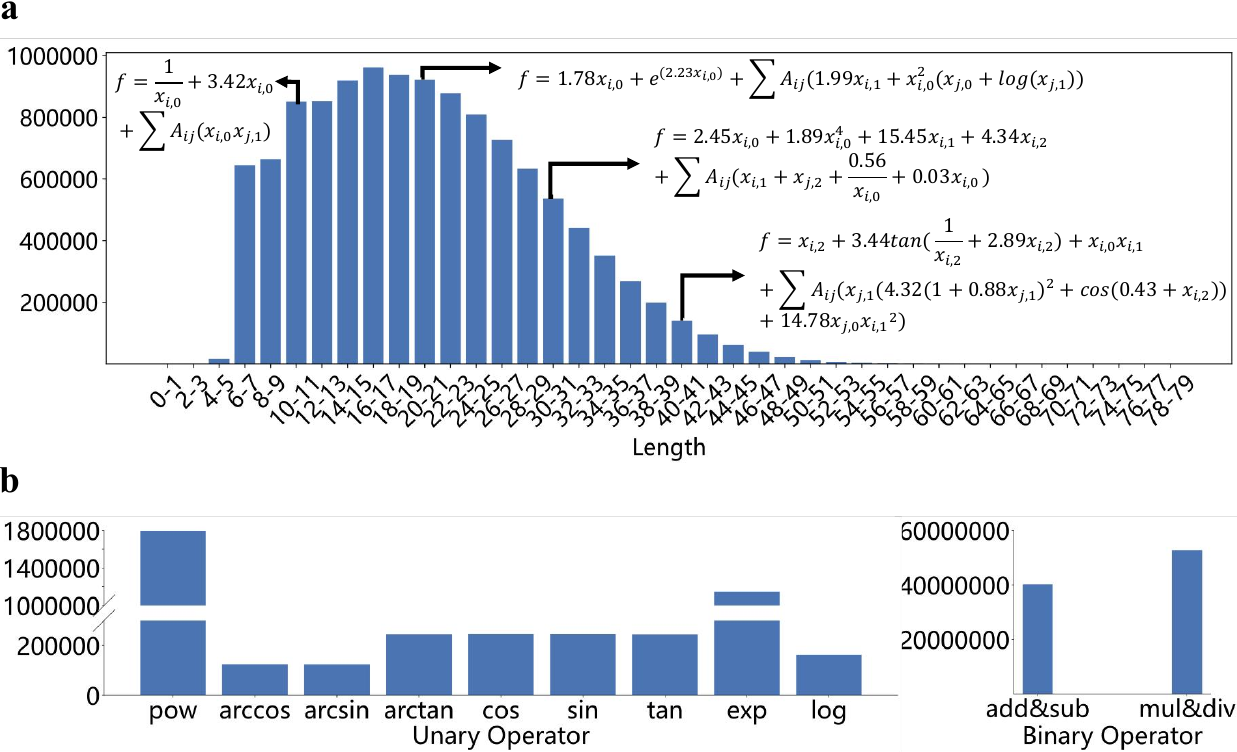}
\caption{The distribution of equation length and operators in the corpus. 
\textbf{a.} The equation lengths primarily range from 4 to 50 characters, with examples provided. The highest proportion of equations is in the length range of 16-17, totaling approximately 962,000.
\textbf{b.} The proportion of binary operators is higher, modeling the relationship between variables. For unary operators, aside from the $pow$ operator, which needs to represent multiple operations, the number of other unary operators is roughly the same.}\label{a_fig1}
\end{figure*}

We present the distribution of equation length and operators in the corpus (see Fig. \ref{a_fig1}). 
The length of the equations primarily ranges from 4 to 50 characters, which aligns with scientific intuition. Equations that are too short or too long may lack practical significance in complex networks.
Some operators can be replaced with other combinations to reduce regression categories and improve regression accuracy, for example, $/$ can be replaced with $\times$, $pow$, and the constant 1. 
Compared to other unary operators, the number of $pow$ is relatively large because $\sqrt{x}$, $x^2$, $\frac{1}{x}$ all need to be represented through $pow$. 
In terms of dimensions, the number of equations in each dimension is roughly the same.

\subsection{The generation parameters}\label{secA12}
We also list specific parameter settings for generating equations in Table \ref{A_tab1}, including the maximum dimension of each equation, denoted as $D$, the maximum number of binary operator, $b_{amx}$, and the maximum number of unary operator, $u_{amx}$. Additionally, the range of constant values is given as $[c_{min},c_{max}]$.
The occurrence probabilities of binary and unary operators, $P_b$ and $P_u$, are detailed in Table \ref{A_tab2} and \ref{A_tab3}. 
Although the occurrence probability of trigonometric function terms is relatively low, the model encounters a sufficient number of these terms during the training process due to the large number of operators in each equation and the total number of equations.

\begin{table}[h]
\caption{Parameters in corpus construction setting}\label{A_tab1}%

\begin{tabular}{cccccc}
\toprule
Parameter &  $D$        &      $b_{max}$  & $u_{max}$     &  $c_{min}$  &$c_{max}$ \\                             
\midrule
Value & 3          & 5        & 5 & -20  &20\\ 
\botrule
\end{tabular}
\end{table}

\begin{table}[h]
\caption{The occurrence probabilities of binary operators $P_b$}\label{A_tab2}%
\begin{tabular}{ccccc}
\toprule
Binary Operators & add        &      mul  & sub     &  div   \\                             
\midrule
Probability      & 0.375      & 0.375         & 0.125  & 0.125\\ 
\botrule
\end{tabular}
\end{table}

\begin{table}[h]
\caption{The occurrence probabilities of unary operators $P_u$}\label{A_tab3}%
\begin{tabular}{ccccccc}
\toprule
Unary Operators &inv        &      pow2  & exp     &  sin,cos,tan  &arcsin,arccos,arctan & log   \\                             
\midrule
Probability     &0.5       &      0.3 & 0.1     &  0.04  &0.04 & 0.02  \\ 
\botrule
\end{tabular}
\end{table}

\subsection{The check rules}\label{secA12}
After generating equations, we set some rules to ensure their applicability. 
Firstly, we assess the rationality of equation skeletons.
This includes checking whether the length, dimensions, and number of operators exceed acceptable limits, identifying any duplicates, and ensuring there are no nested trigonometric or exponential terms.
Additionally, we verify that the equation $f^{(inter)}$ includes $x_j$. 
Importantly, $f^{(self)}$ is allowed to have no equation.

Next, we evaluate the validity of equations. 
When specific coefficients are introduced into the equation skeleton, we perform domain and value checks, along with checks for invalid conditions (such as division by zero), ensuring that the sampled data can be computed correctly within the model.
It's worth noting that since random constant sampling occurs during training, some rules will be enforced throughout this process, and these two checks are designed to maximize the equation's validity.

These checks do not require extensive generation time. 
With code optimization and multi-threaded processing, we can generate millions of equations in just hours.

\subsection{The specific model setting}\label{secA14}
\begin{figure*}[t]
\centering
\includegraphics[width=1.0\textwidth]{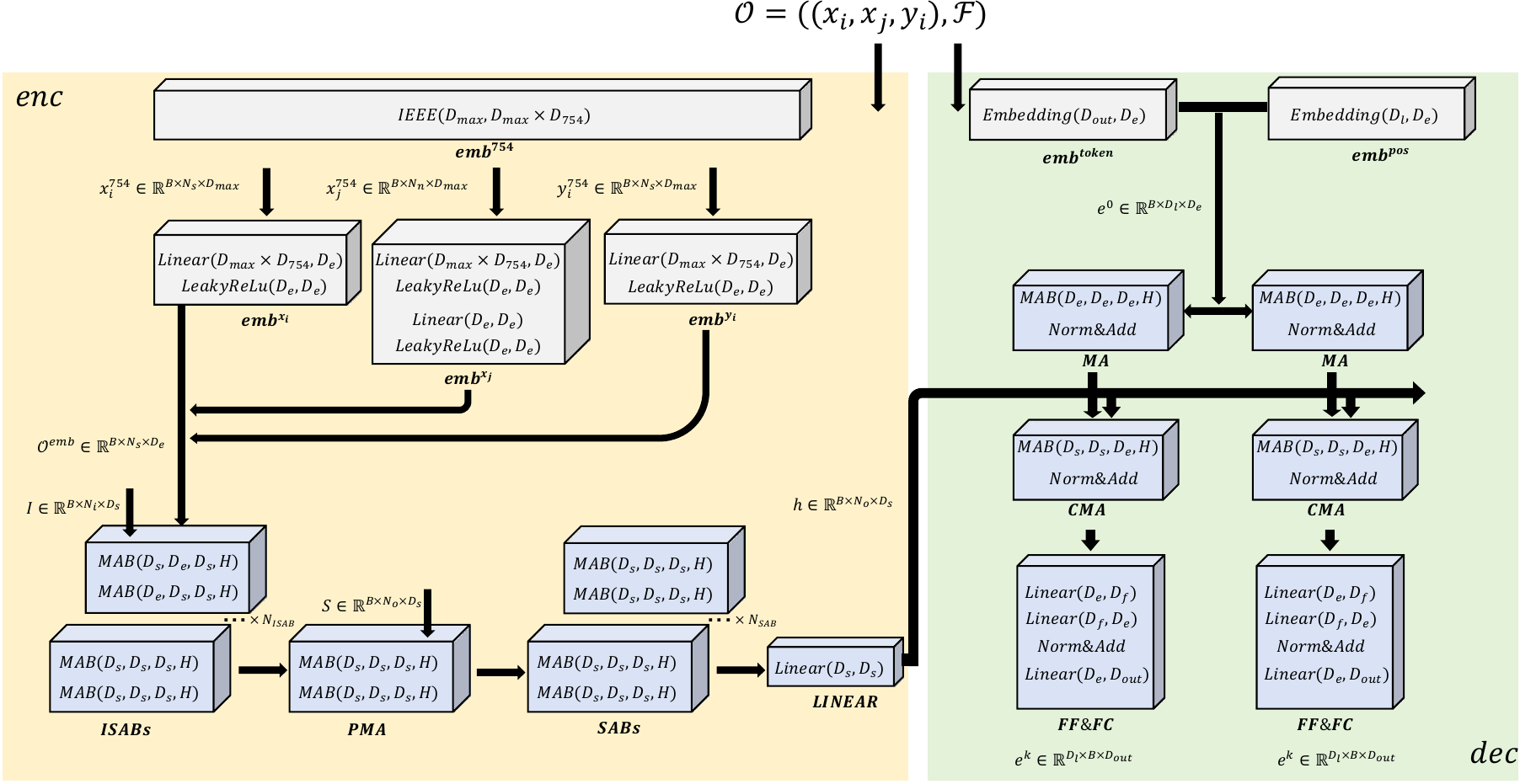}
\caption{The specific model architecture.}\label{a_fig2}
\end{figure*}

We provide a more specific model architecture and data flow (see Fig. \ref{a_fig2}), and we list the specific roles of each module:

\begin{itemize}
\item $emb^{754}$: $emb^{754}$ is an encoding layer based on the IEEE Standard for Floating-Point Arithmetic \cite{ieee754-2019}, converting data $\{x_{i},\{x_{j}\}_{j \in \mathcal{N}},y_{i}\}$ into binary floating-point encoding $\{x_{i}^{754},\{x_{j}^{754}\}_{j \in \mathcal{N}},y_{i}^{754}\}$ to avoid gradient problems during the calculation process.
\item $emb^{x_i},emb^{x_j},emb^{y_i}$,: $\{emb^{x_i},emb^{x_j},emb^{y_i}\}$ are data encoding layer consisting of Linear and LeakyRelu layers, which maps binary encoding to $D_e$-dimension encoding. Their respective results are integrated together to form $\mathcal{O}^{emb}$.
\item $ISAB$: ISAB (Induced Set Attention Block) consists of MAB layers, which are the multi-head attention blocks in the Transformer. The elements in the set data $\mathcal{O}^{emb}$ are performed attention calculation in MAB, and high-order interactions can be encoded by stacking multiple blocks. In addition, by introducing the unit vector $I$, $\mathcal{O}^{emb}$ can be projected onto a low dimensional space through $I$, reducing computational complexity ($O(N_s^2)\rightarrow O(N_sN_i)$).
\item $PMA$: PMA (Pooling Multihead Attention) is used to aggregate ISAB encoding into $N_o$ features.
\item $SAB$: SAB (Set Attention Block) further models the relationship between PMA outputs through attentions.
\item $emb^{token},emb^{pos}$: $emb^{token},emb^{pos}$ perform token encoding and positional encoding on the equation $\mathcal{F}$.
\item $MA,CMA,FF,FC$: MA (Multihead Attention), CMA (Cross Multihead Attention), FF (Feed-Forward layers ), FC (Full-Connected layers) are components in the decoder of the Transformer.
\end{itemize}

The parameter settings for each layer are outlined in Table \ref{A_tab4}. The model was trained on a hardware setup that includes {64 Intel(R) Xeon(R) silver 4314 CPUs @ 2.40GHZ} and an NVIDIA GeForce GTX 4090 GPU with 24GB of memory. The pre-training process lasted for 20 epochs, utilized a batch size of \(B\), and took approximately 16 days to complete.

\begin{table}[h]
\caption{The setting of parameters.}\label{A_tab4}%
\begin{tabular}{cccccccccccc}
\toprule
$B$ &$D_{max}$       &      $D_{754}$  & $D_{e},D_{s},D_{f}$     &  $D_{l},D_{out}$ &$N_d$ &$N_s$       &      $N_i$  & $N_o$     &  $N_{SAB}$  &  $N_{ISAB}$  \\                             
\midrule
$16$     &$3$       &      $32$ & $512$     &  $50$  & $6$     &$200$       &      $50$ & $20$     &  $2$  &  $3$\\ 
\botrule
\end{tabular}
\end{table}

\subsection{The specific application process}\label{secA15}
We provide a specific process in Algorithm \ref{algo1}.
For data from an unknown scene, we need to process it before feeding it into the model. Firstly, we sample $N$ nodes based on their in-degree, where nodes with higher in-degree may contain more information (line 2). Afterwards, $GaussianMixture(clusters=N^{cluster})$ clustering is used for the data on each node, and $T^{cluster}$ data points are sampled from each class to eliminate strong correlations between data points (line 3-9). Then, we calculate the mean $\mu$ and variance $\sigma$ of the data and perform normal distribution sampling, and for each node, we sample $T$ data points to obtain $N \times T$ data points (line 10-14). Finally, we standardize the data $X$ and fed $X^{\prime}$ into the model for fitting $f^{\prime}(X^{\prime})$. We then perform an inverse transformation on the fitted equation $f^{\prime}(X^{\prime})$ to obtain the final result $f(X)$ (line 15-18). Some post-processing methods can be applied to the output equation (line 19). 
\clearpage

\begin{algorithm}
\caption{Application process}\label{algo1}
\begin{algorithmic}[1]
\Require ${X}^{sample},N,T,N^{cluster},T^{cluster}$

\State ${X}^{sample}$ sampled from network dynamics
\State Select $N$ nodes based in-degree of nodes
\For{$i$ to $N$ do}:
    \State $clusters \Leftarrow GaussianMixture(X^{sample}_{i} \in \mathbb{R}^{T^{sample}*D})$
    \For{$j$ to $N^{clusters}$ do}:
        \State Randomly select $T^{clusters}$ sampling points $X^{sample}_{i} \in \mathbb{R}^{T^{clusters}*D}$ 
    \EndFor
\EndFor
\State Integrate clustered data to obtain $X^{cluster} \in \mathbb{R}^{N*(T^{clusters}\times N^{clusters})*D}$
\State Calculate mean and variance, $\mu,\sigma \Leftarrow X^{cluster}$
\State Perform normal distribution $\mathcal{N}(\mu,\sigma)$ sampling on data $X^{cluster}$ to obtain $X$
\For{$i$ to $N$ do}:
    \State Select $T$ sampling points $X_{i} \in \mathbb{R}^{T*D}$
\EndFor
\State Distribution transformation: $X^{\prime} \Leftarrow \frac{X-\mu}{\sigma},X^{\prime} \sim \mathcal{N}(0,1) \Leftarrow X \sim \mathcal{N}(\mu,\sigma)$
\State Fitting equations by beam search with $N_{beam}$ beam size: $f^{\prime}(X^{\prime}) \Leftarrow Model(X^{\prime})$
\State BFGS constants optimization
\State Inverse transformation: $f(X)=f^{\prime}(\frac{X-\mu}{\sigma})$
\State Optional post-processing operations
\State Obtain final result $f(X)$
\end{algorithmic}
\end{algorithm}

We list some post-processing methods for equation optimization.
\begin{itemize}
\item Simplification coefficient: For coefficients that are less than the threshold $C_s$ and have almost no impact on the accuracy of the equation, we simplify the coefficients and terms to 0, such as $y=0.0000134x_{i,0}+3.84x_{i,1} \rightarrow y=3.84x_{i,1}$, and generally set $C_s$ to $10e-4$.
\item Simplified form: For equations with complex form, we simplify the overall form of the equation to make it more concise.
\item Combining with genetic algorithm: For precise but complex equations, we use this equation as the starting point of genetic algorithm or directly optimize the equation form through genetic algorithm to make it more researchable.
\item Combining with LLM: Optimizing equations through domain knowledge, as detailed in Appendix \ref{secA6}.
\end{itemize}

\clearpage

\section{More details on classical non-network symbolic regression}\label{secA2}

\subsection{AI-Feynman}\label{secA21}
AI-Feynman \cite{udrescu2020ai} contains physical model that come from common laws in classical physics \cite{feynman2018feynman}, such as Newton's laws, thermodynamic equations, electromagnetic equations, etc., which is commonly used to evaluate the performance of symbolic regression methods. Each physical model in AI-Feynman corresponds to a equation and we select 100 equations as the dataset. These equations have 1-3 independent variables, for variables exceeding 3, we replace some of their independent variables with random constants, for example, $F=\frac{Gm_{1}m_{2}}{(x_{2}-x{1})^2+(y_{2}-y{1})^2+(z_{2}-z{1})^2}$ is replaced to $y=\frac{x_{i,0}x_{i,1}x_{i,2}}{(c_{5}-c{4})^2+(c_{3}-c{2})^2+(c_{1}-c{0})^2}$, where $c_i$ is the random constant. Note that all methods are evaluated on replaced equations to ensure fairness.

\subsection{USE-F}\label{secA22}
The USE-F is a set of unseen synthetic equations with only self parts $f^{(self)}$, describing more diverse and complex scenarios with independent variables. It should be noted that the USE-F is constructed in the same way as the corpus but has no intersection with the corpus. The dataset contains a total of 30,000 $f^{(self)}$ equations with 0-60 length, 1-3 dimension and 0-40 operators. Compared to AI-Feynman, these equations are on average significantly longer and more complex.

\subsection{Details on experimental setting}\label{secA23}
For experiments on AI-Feynman (see Fig. \ref{exp1}(a)), the test set consists of all 100 equations in the AI-Feynman. The number of input data (IN-Domain) and prediction of unknown data (OUT-Domain) are 200 and 1000, sampled from a normal distribution of random parameters $\mathbf{N}(a,b)$.

For performance experiments on USE-F (see Fig. \ref{exp1}(a)), 1500 equations are sampled randomly for evaluation each time, and for experiments on equations with various lengths, dimensions, operators and test points in USE-F (see Fig. \ref{exp1}(b,c)), the number of sampled equations is 200 (such as $200$ equations with dimension 1, 200 equations with length between 5-10, etc.) each time.
For the experiment on test points on USE-F, the number of test points varies, and for other experiments, the number of input data (IN-Domain) and prediction of unknown data (OUT-Domain) are 200 and 1000, sampled from the distribution of $\mathbf{N}(0,1)$ and $\mathbf{N}(0,10)$.

The parameters for PySR, the library for SINDy, and the models for E2E and NeSymReS follow the configurations specified in their respective published papers or latest version in Github. 
All methods are evaluated using the same test data, ensuring the fairness.

\begin{figure*}[t]
\centering
\includegraphics[width=1.0\textwidth]{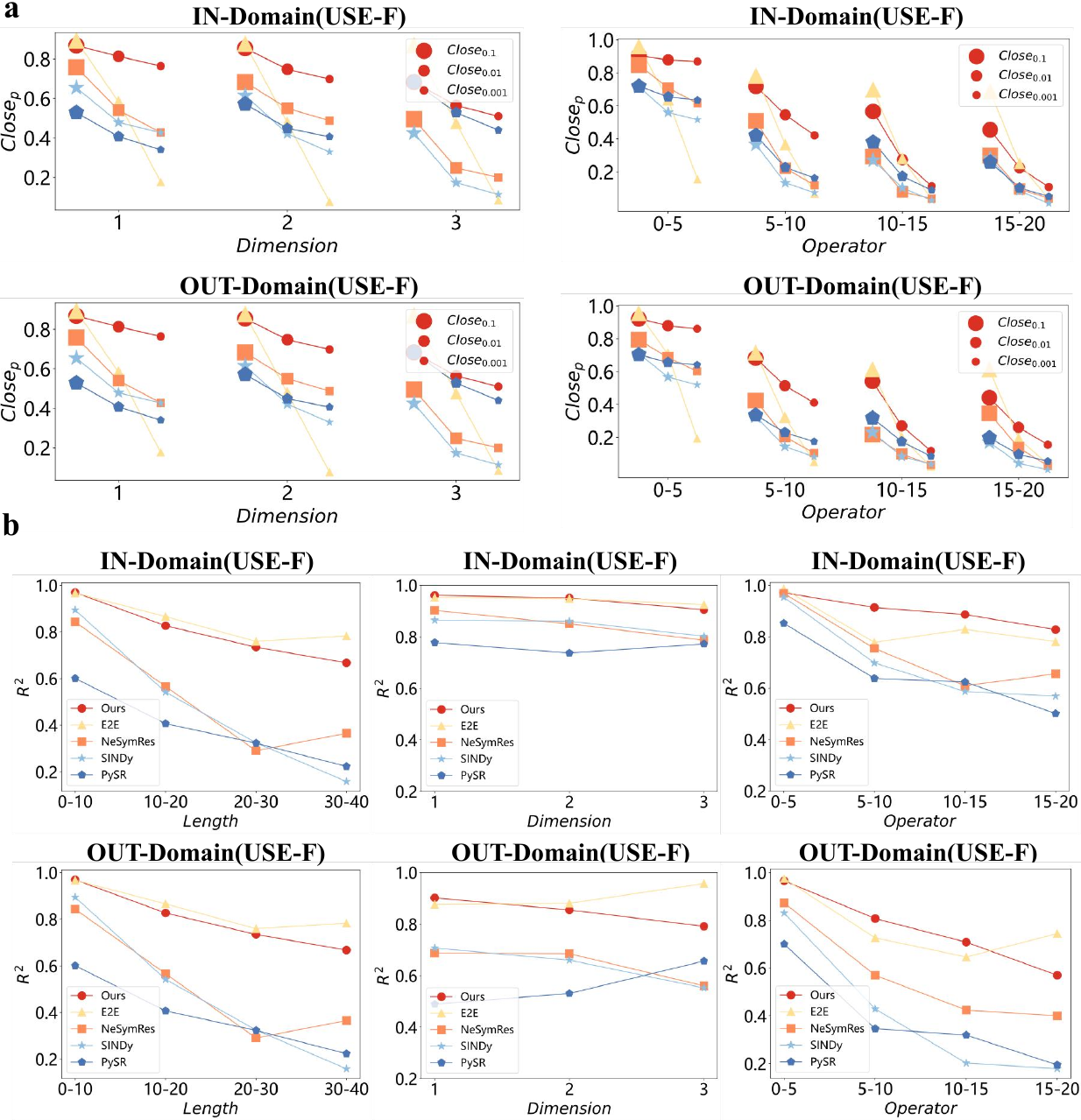}
\caption{\textbf{a.}The effect of the dimension and the number of operator on classical non-network symbolic regression, the precision gradually increases until $Close_{0.001}$. \textbf{b.}The effect of the length, dimension and the number of operator on classical non-network symbolic regression ($R^2$).}\label{a_exp1_1}
\end{figure*}

\subsection{More experimental results and regression analyses}\label{secA24}
The number of operator in equations and dimension of equations also affect the performance of model.
As the number increases, the performance of these methods tends to decline (see Fig.~\ref{a_exp1_1}(a)).
Our SFR outperforms most baselines, especially in more stringent measurement. 
Although E2E achieves optimal results at some low-precision, its lack of a constant optimization module, resulting in a significant decrease in performance at high-precision.
The diversity of the datasets leads to performance limitations of the SINDy and PySR, and fine-tuning parameters for each equation is time-consuming.
The insufficient construction of the corpus and simple constant optimization might be the main reasons for the poor performance of NeSymRes.

We also show the $R^{2}$ performance of all methods on the equations with different length, dimension and the number of operator (see Fig.~\ref{a_exp1_1}(b))
For the $R^{2}$ index, it places more emphasis on the degree of fit and does not require high accuracy, therefore our SFR achieve the best results on most indicators, and E2E is better than ours in some indicators.

\begin{figure*}[h]
\centering
\includegraphics[width=1.0\textwidth]{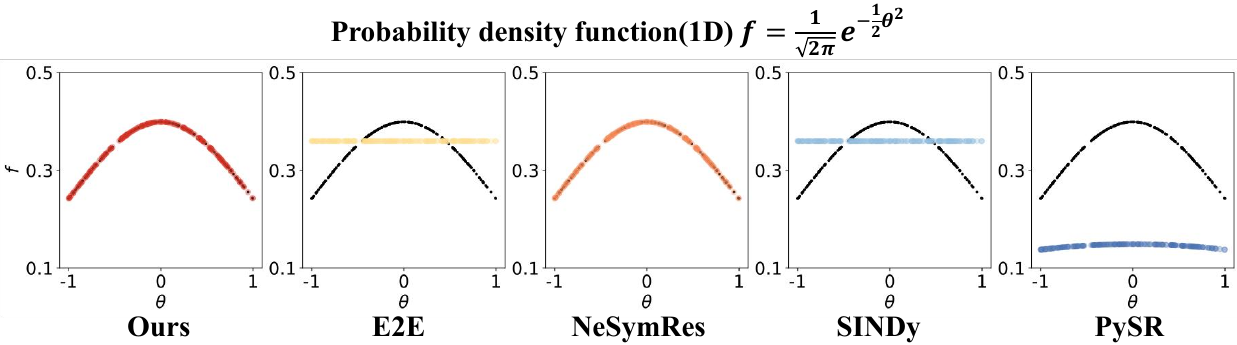}
\caption{Results of symbolic regression on equations in AI-Feynman with 1 dimension (equation curve), the black line represents the true equation, while the horizontal line indicates the failure of the regression equation.}\label{a_exp1_2}
\end{figure*}

\begin{figure*}[t]
\centering
\includegraphics[width=1.0\textwidth]{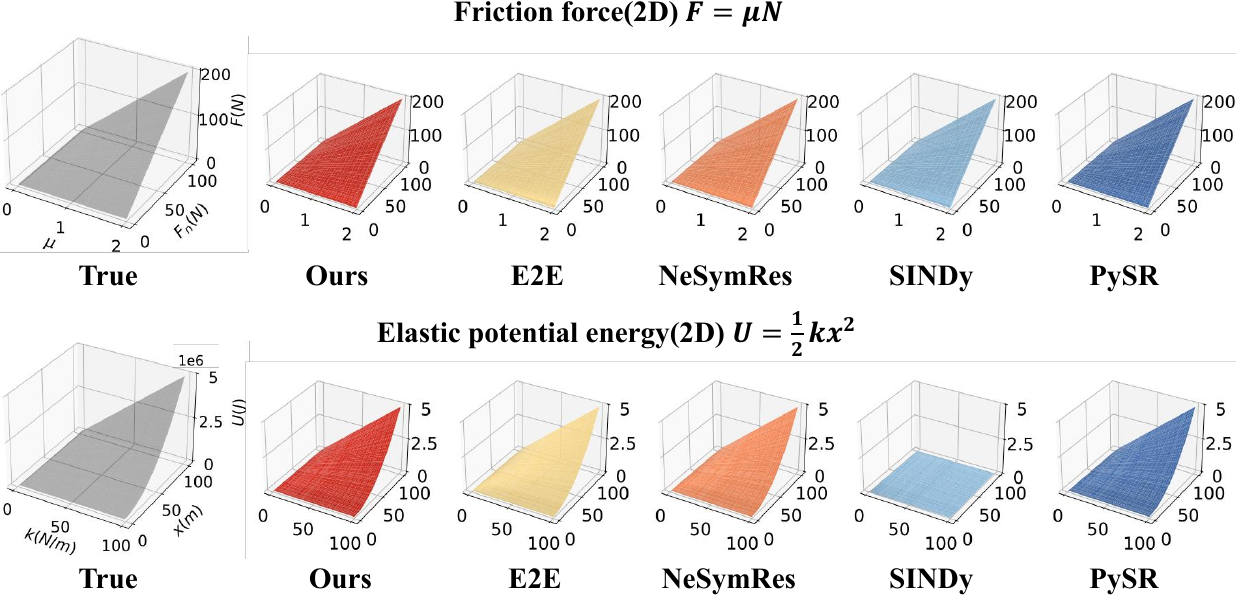}
\caption{Results of symbolic regression on equations in AI-Feynman with 2 dimensions (equation surface), the gray surface on the left is the true equation}\label{a_exp1_3}
\end{figure*}

We provide more visualized equation regression results.
In Fig.~\ref{a_exp1_2}-\ref{a_exp1_3} and Table.~\ref{A_tab5}-\ref{A_tab7}, we present the regression results of the probability density function (one dimension (1D)), friction force (2D), elastic potential energy equations for a normal distribution (2D), angular distribution in proton decay (3D) and magnetic moment (3D) equations. 

For the equations with 3 dimensions, considering the $y$, it is difficult to represent 4 variables with figures, so we only present the form of equations.
Most methods can regress to the correct equation, and we use the $-$ to represent failed regression equations (equations where the model regression fails or are meaningless).

\begin{table}[h]
\caption{Results of symbolic regression on equations in AI-Feynman with 1 dimension (equation form).}\label{A_tab5}%
\begin{tabular}{cc}
\toprule
\diagbox{Method}{Physical Law}& Probability density function \\                             
\midrule
True     &$f=\frac{1}{\sqrt{2\pi}}e^{-\frac{1}{2}\theta^{2}}$       \\ 
Ours     &$f=0.399e^{-\frac{1}{2}\theta^{2}}$       \\ 
E2E     &$-$       \\ 
NeSymRes     &$f=0.398e^{-\frac{1}{2}\theta^{2}}$       \\ 
SINDy     &$-$       \\ 
PySR    &$f=0.054e^{cos(0.4\theta)}$       \\ 
\botrule
\end{tabular}
\end{table}

\clearpage

% \vspace{-30pt}

\begin{table}[t]
\caption{Results of symbolic regression on equations in AI-Feynman with 2 dimensions (equation form).}\label{A_tab6}%
\begin{tabular}{ccc}
\toprule
\diagbox{Method}{Physical Law}& Friction force & Elastic potential energy equations\\                             
\midrule
True     &$F=\mu N$ & $U=\frac{1}{2}kx^{2}$     \\ 
Ours     &$F=\mu N$    &  $U=\frac{1}{2}kx^{2}$ \\ 
E2E     &$F=\mu N-0.008N$  & $U=\frac{1}{2}kx^{2}$    \\ 
NeSymRes     &$F=\mu N$  &   $U=\frac{1}{2}kx^{2}$  \\ 
SINDy     &$F=\mu N$  & \makecell{$U=0.22k^{2}+0.25kx+12.17x$\\$+0.175x^{2}-0.54k-6.36$}    \\ 
PySR    &$F=\mu N$    & $U=\frac{1}{2}kx^{2}$  \\ 
\botrule
\end{tabular}
\end{table}

\begin{table}[h]
\caption{Results of symbolic regression on equations in AI-Feynman with 3 dimensions (equation form).}\label{A_tab7}%
\begin{tabular}{ccc}
\toprule
\diagbox{Method}{Physical Law}& Angular distribution in proton decay & Magnetic moment\\                             
\midrule
True     &$f=\beta (1+\alpha cos \theta)$ & $\mu=\frac{qh}{4\pi m}$     \\ 
Ours     &$f=\beta (1+\alpha cos \theta)$    &  $\mu=0.079\frac{qh}{m}$ \\ 
E2E     &\makecell{$f=0.24\beta((4.4\alpha$\\$+0.19)cos(1.1\theta+0.65)+4.02)$}  & $\mu=0.39\frac{qh}{m}$    \\ 
NeSymRes     &$f=\beta(1.57-0.24\alpha)cos(0.3\theta)$  &   $\mu=0.079\frac{qh}{m}$  \\ 
SINDy     &$-$  & $-$    \\ 
PySR    &$f=\frac{\alpha}{\theta}$    & $\mu=\frac{1}{sin(0.27q)}$  \\ 
\botrule
\end{tabular}
\end{table}

\clearpage

\begin{figure*}[t]
\centering
\includegraphics[width=1.0\textwidth]{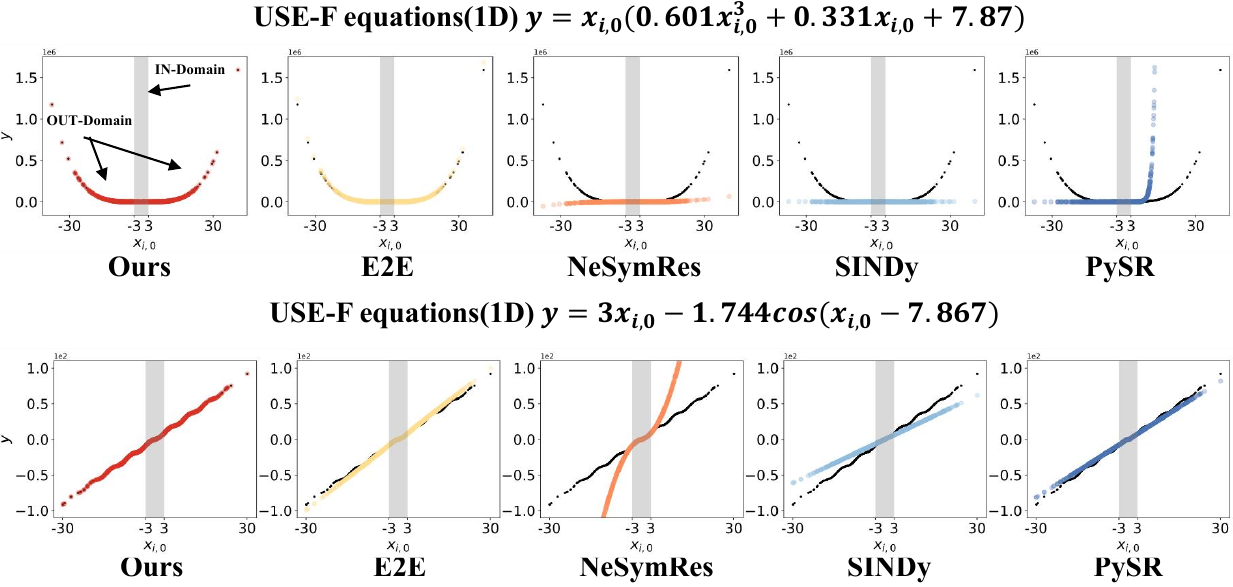}
\caption{Results of symbolic regression on equations in USE-F with 1 dimensions (equation curve), the gray area represents the IN-Domain task, and the white area represents the OUT-Domain task.}\label{a_exp1_4}
\end{figure*}

\begin{table}[t]
\caption{Results of symbolic regression on equations in USE-F with 1 dimensions (equation form).}\label{A_tab8}%
\begin{tabular}{ccc}
\toprule
Method & \multicolumn{2}{c}{Equation}\\                             
\midrule
Ours       &$y=x_{i,0}(0.601x_{i,0}^3+0.331x_{i,0}+7.87)$ & $y=3x_{i,0}-1.744sin(x_{i,0}-0.013)$\\ 
E2E        &$y=x_{i,0}(0.637x_{i,0}^3+7.996)$             & \makecell{$y=(1.303x_{i,0}+0.026)(1-$\\$arctan(0.02x_{i,0}-0.215x_{i,0}^2))$}     \\ 
NeSymRes   &$y=x_{i,0}^3+6.896x_{i,0}$                    & $y=0.643x_{i,0}|x_{i,0}|+0.908x_{i,0}+0.015$  \\ 
SINDy      &$y=3.657x_{i,0}^2+7.613x_{i,0}-1.638$         & $y=2.048x_{i,0}$    \\ 
PySR       &$y=e^{x_{i,0}+1.441}-5.016$                   & $y=x_{i,0}e^{cos(\frac{1}{x_{i,0}})}$  \\ 
\botrule
\end{tabular}
\end{table}

\begin{figure*}[t]
\centering
\includegraphics[width=1.0\textwidth]{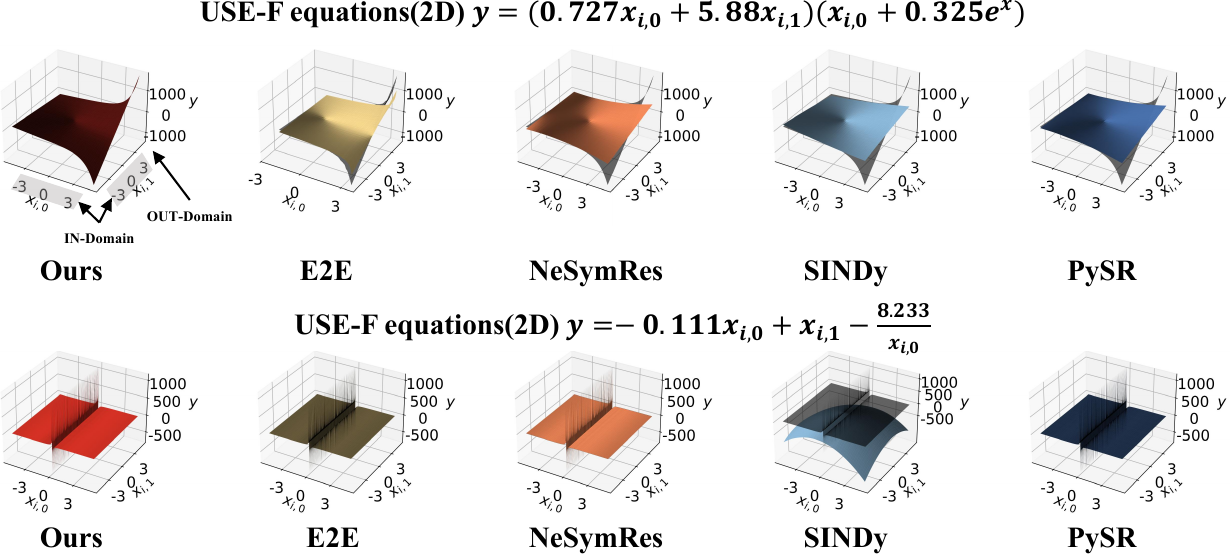}
\caption{Results of symbolic regression on equations in USE-F with 2 dimensions (equation surface).}\label{a_exp1_5}
\end{figure*}

In Fig. \ref{a_exp1_4}-\ref{a_exp1_5} and Table. \ref{A_tab8}-\ref{A_tab10}, we present the results of some representative equations in USE-F, include complex polynomials, fractions, trigonometric functions, exponential functions, etc. 
Each subfigure shows a comparison between the regression equation and the true equation (black), it can be seen that the regression equation from our method not only approaches the ground truth on curve, but also has a more consistent form with the true equations.
The gray area represents IN-Domain tasks, while other areas represent OUT-Domain tasks.
As shown in the figures, all methods achieved good results on IN-Domain tasks, but when faced with unknown data on OUT-Domain, the equation regressed from our method is more in line with the real equation, indicating that our equation results have stronger ability to predict unknown data and have the potential to discover unknown physical laws.

\clearpage

\begin{table}[t]
\caption{Results of symbolic regression on equations in USE-F with 2 dimensions (equation form).}\label{A_tab9}%
\begin{tabular}{ccc}
\toprule
Method & Equation\\                              
\midrule
Ours     &$y=(0.727x_{i,0}+0.236e^x)(x_{i,0}+8.088x_{i,1})$ \\ 
E2E     &\makecell{$y=\frac{5.897-0.214x_{i,0}}{33.378-1.589x_{i,0}}(0.654-38\sqrt{1-\frac{0.746}{0.052x_{i,0}-0.186}})(-0.07x_{i,0}$\\$-0.018)(0.944x_{i,0}+7.365x_{i,1}+0.09)$ (Wrong Domain)}  \\ 
NeSymRes     &$(y=0.686x_{i,0}+0.132)(x_{i,0}+2x_{i,0}x_{i,1}+12.13x_{i,1})$   \\ 
SINDy     &$y=2.137x_{i,0}^2+10.329x_{i,0}x_{i,1}+1.285x_{i,0}-1.071x_{i,1}^2+3.8x_{i,1}-0.112$     \\ 
PySR    &$y=x_{i,0}x_{i,1}(x_{i,0}+8.921)$    \\ 
\botrule
\end{tabular}
\end{table}

\begin{table}[t]
\caption{Results of symbolic regression on equations in USE-F with 2 dimensions (equation form).}\label{A_tab10}%
\begin{tabular}{ccc}
\toprule
Method & Equation\\                              
\midrule

Ours     &$y=x_{i,1}-0.111x_{i,0}-\frac{8.233}{x_{i,0}}$ \\ 
E2E     &$y=0.98x_{i,1}-0.144x_{i,0}-\frac{8.268}{x_{i,0}}$ \\ 
NeSymRes     &$y=-0.269x_{i,0}-\frac{8.243}{x_{i,0}}$  \\ 
SINDy     &$y=-31.203x_{i,0}^2+10.751x_{i,0}x_{i,1}-4.128x_{i,0}-21.932x_{i,1}^2-21.67x_{i,1}+89.118$\\ 
PySR    &$y=x_{i,1}-\frac{8.229}{x_{i,0}}$\\ 
\botrule
\end{tabular}
\end{table}

\clearpage

\renewcommand{\arraystretch}{2}
\begin{table}[h]
\caption{Results of symbolic regression on equations in USE-F with 3 dimensions (equation form).}\label{A_tab11}%
\begin{tabular}{ccc}
\toprule
Method & Equation & $R^2$\\                              
\midrule
True         &$y=0.53x_{i,0}+x_{i,1}+2.32x_{i,2}+10.34tan(x_{i,2}+1.28)$   & $/$ \\ 
Ours         &$y=0.558x_{i,0}+x_{i,1}+x_{i,2}+10.34tan(x_{i,2}+1.28)$      & $0.996$ \\ 
E2E          &$y=0.558x_{i,0}+0.87x_{i,1}+11tan(1.105x_{i,2}+1.288)+0.117$ & $0.994$ \\ 
NeSymRes     &$y=\frac{0.003tan(x_{i,0}+x_{i,1}-0.262)}{x_{i,0}-0.116}$  & $>-1$ \\ 
SINDy        &$-$                                                        & $>-1$ \\ 
PySR         &$y=\frac{1}{x_{i,2}-0.297}$                                  & $>-1$\\ 
\botrule
\end{tabular}
\end{table}

\renewcommand{\arraystretch}{2}
\begin{table}[h]
\caption{Results of symbolic regression on equations in USE-F with 3 dimensions (equation form).}\label{A_tab12}%
\fontsize{6}{8}\selectfont{
\begin{tabular}{ccc}
\toprule
Method & Equation & $R^2$\\                              
\midrule
True     &$y=x_{i,0}(2.097x_{i,0}(x_{i,2}-8x_{i,1})-8.539)-(1.744x_{i,0}+7.867x_{i,2})\frac{8x_{i,1}-x_{i,2}}{8x_{i,1}-x_{i,2}}$ & $/$\\ 
Ours     &$y=x_{i,0}(1.867x_{i,0}(0.1227x_{i,2}-x_{i,1})-1)-7.979x_{i,2})\frac{x_{i,1}-0.123x_{i,2}}{x_{i,1}-0.123x_{i,2}}$ & $0.971$\\ 
E2E     &\makecell{$y=-7.808x_{i,2}+(0.16-1.082x_{i,0})(1.047x_{i,0}+(0.005x_{i,2}$\\$+0.985)(1.104x_{i,0}+0.744-\frac{6.09}{10.318x_{i,1}-1.649x_{i,2}+0.34})+0.947)-0.302$} & $0.364$ \\ 
NeSymRes     &$y=-0.449x_{i,0}tan(1.179x_{i,0}-x_{i,1}+0.547)-x_{i,2}^3$ & $0.379$ \\ 
SINDy     &$y=-2.085x_{i,0}^2+1.332x_{i,0}x_{i,1}-1.547x_{i,0}-0.116x_{i,2}^2-7.993x_{i,2}+0.207$ & $0.604$\\ 
PySR    &$y=\frac{1}{sin(\frac{1}{x_{i,1}})}$ & $>-1$\\ 
\botrule
\end{tabular}
}
\end{table}

We also provide some ultra long equations with 3 dimensions, (see Table. \ref{A_tab11} and \ref{A_tab12}), in which the performance of each method has decreased to varying degrees. Our method cannot guarantee a completely consistent equation form with the true equation, but the quantitative result $R^2$ is still satisfactory compared to other methods.

\clearpage

\section{More details on symbolic regression on complex networks}\label{secA3}

\subsection{USE}\label{secA31}
The USE is an expanded dataset of USE-F, adding 30000 interaction equations $f^{(inter)}$ of the same quantity as self equations $f^{(self)}$ in USE-F. 
These self and interaction equation can be freely combined to $\{f^{(self)},f^{(inter)}\}$, with different five types of topologies to generate equations in the following form $y_{i}=f^{(self)}(x_{i})+\sum_{j=1}^{N}A_{ij}f^{(inter)}(x_{i},x_{j})$, modeling various complex networks.

\subsection{Topological structures of complex networks}\label{secA32}
\begin{itemize}
    \item \textbf{Grid}: 
    \begin{itemize}
    \item 1.Construct a grid structure with $n \times n=N$.
    \item 2.Each cell represents a network node and connect with its surrounding cells.
    \end{itemize}
    
    \item \textbf{Power Law} is Barab\'asi-Albeat network and has the characteristic of power-law degree distribution. The construction steps are as follows:
    \begin{itemize}
    \item 1.Initialize a connected graph with $n+1$ nodes (each node has at least one edge)
    \item 2.Each time a new node is added, it has $n$ edges connected to existing nodes until the number of nodes reaches $N$. The connection condition of a new node depends on the degree of the existing node.
    \end{itemize}
    
    \item \textbf{Small World} is composed of a large number of nearby nodes and randomly distributed weak connections. The construction steps are as follows:
    \begin{itemize}
    \item 1.Initialize a connected graph with $N$ nodes and $k$ neighbors for each node.
    \item 2.Randomly reconnect nodes with a probability of $p$.
    \end{itemize}

    \item \textbf{Community} is a network structure with characteristic both within and outside the community. The construction steps are as follows:
    \begin{itemize}
    \item 1.Initialize $N$ nodes and divide them into $[n_1, n_2,...,n_k]$ communities.
    \item 2.Nodes in the same community are connected with a probability of $p_{in}$, while nodes in different communities are connected with a probability of $p_{out}$
    \end{itemize}
    
    \item \textbf{Random} is Erd\H{o}s-R\'enyi network and has the characteristic of randomness. The construction steps are as follows:
    \begin{itemize}
    \item 1.Initialize $N$ nodes.
    \item 2.Connect each node with a probability of $p$.
    \end{itemize}
\end{itemize}

We use the NetworkX package \cite{hagberg2008exploring} to generate the above topological structure and also present the generation settings and a connection example for each topology in Table \ref{A_tab13}. 

\setlength{\tabcolsep}{1pt}
\begin{table}[t]
\caption{Topology setting}\label{A_tab13}%
\begin{tabular}{cccccc}
\toprule
Topology & grid& power law& small world& community& random\\                             
\midrule
$A_{ij}$&\begin{minipage}[b]{0.15\columnwidth}
		\centering
		\raisebox{-.5\height}{\includegraphics[width=0.8\linewidth]{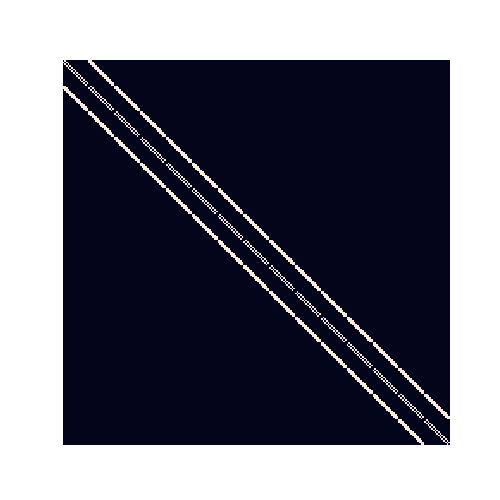}}
	\end{minipage}
    & \begin{minipage}[b]{0.15\columnwidth}
		\centering
		\raisebox{-.5\height}{\includegraphics[width=0.8\linewidth]{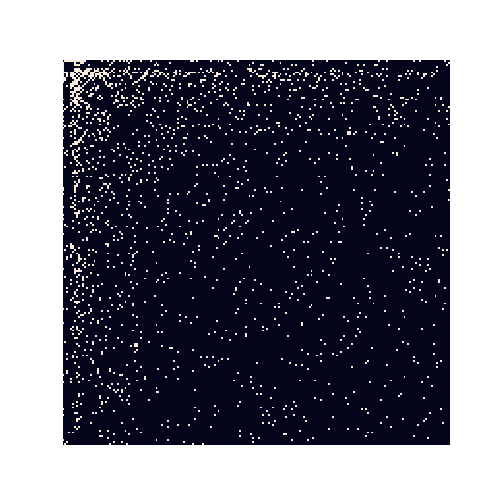}}
	\end{minipage}
    & \begin{minipage}[b]{0.15\columnwidth}
		\centering
		\raisebox{-.5\height}{\includegraphics[width=0.8\linewidth]{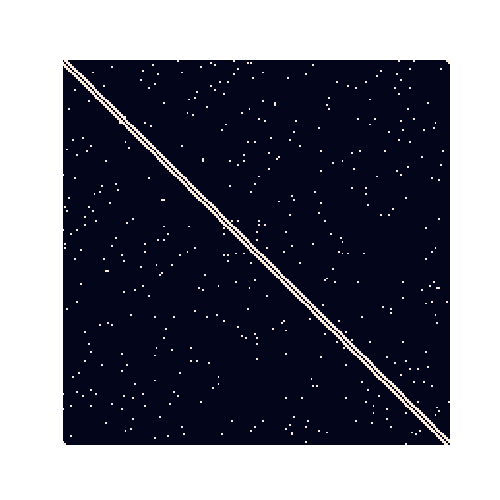}}
	\end{minipage}
    & \begin{minipage}[b]{0.15\columnwidth}
		\centering
		\raisebox{-.5\height}{\includegraphics[width=0.8\linewidth]{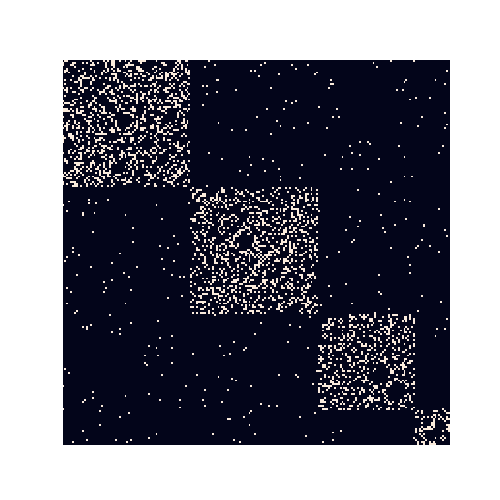}}
	\end{minipage}
    & \begin{minipage}[b]{0.15\columnwidth}
		\centering
		\raisebox{-.5\height}{\includegraphics[width=0.8\linewidth]{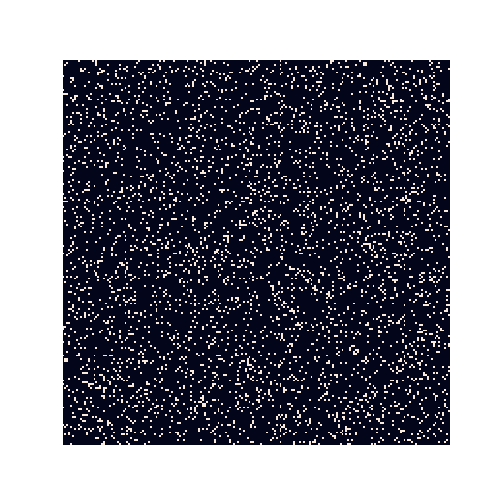}}
	\end{minipage}
    \\
Parameters & \makecell{$N \in [10,200]$}
& \makecell{$n=5$ \\ $N \in [10,200]$}
& \makecell{$k=5$ \\ $p=0.5$ \\ $N \in [10,200]$}
& \makecell{$k=4$ \\ $p_{in}=0.25$ \\$p_{out}=0.01$ \\ $N \in [10,200]$}
& \makecell{$p=0.1$ \\ $N \in [10,200]$}\\
\botrule
\end{tabular}
\end{table}

\subsection{Details on experimental setting}\label{secA33}

For performance experiments on USE (see Fig. \ref{exp2}(a)), 5000 equations are sampled randomly for evaluation each time, and for experiments on equations with various lengths, dimensions, operators and test points in USE (see Fig. \ref{exp2}(b)), the number of sampled equations is 100 (such as 200 equations with dimension 1, 200 equations with length between 5-10, etc.) each time.
Each equation will be paired with 5 randomly generated topologies (grid, power law, small world, community and random), and the topology generation rules are as described in Section \ref{secA32}.

For the experiment on test points on USE, the number of test data points varies, and for other experiments, the raw data is generated based on the distribution of $\mathbf{N}(0,1)$. 
The number of input data sampled from raw data (IN-Domain) are 200, which is the product of the number of sampled time slices and the number of sampled nodes, and the number of sampled nodes does not exceed 20.
Unknown prediction data (OUT-Domain) comes from raw data with the distribution of $\mathbf{N}(0,10)$, ranging from 2000 to 40000 in quantity (Each node on the topology randomly samples 200 time slices of data as unknown data).

\subsection{More experimental results and regression analyses}\label{secA34}

\begin{figure*}[t]
\centering
\includegraphics[width=0.9\textwidth]{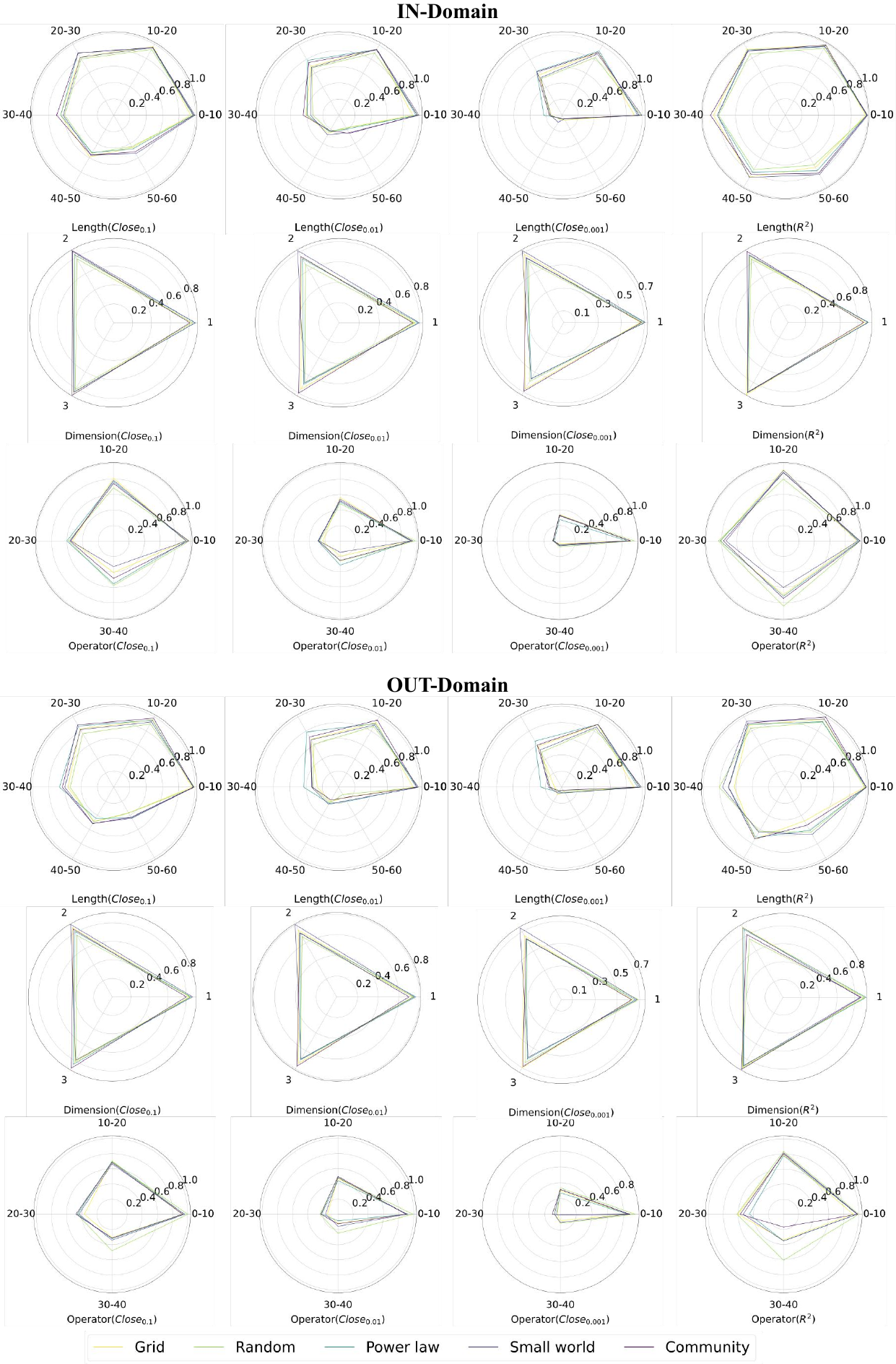}
\caption{The effect of length, dimension and the number of operator on symbolic regression on complex networks ($R^2,Close_{0.1},Close_{0.01},Close_{0.001}$).}\label{a_exp2_1}
\end{figure*}

We fully demonstrate the impact of length, dimension and the number of operator on IN and OUT-Domain tasks in Fig. \ref{a_exp2_1}. 
From the figures, it can be observed that the change in dimension has almost no impact on the performance of our model, while as the length of the equation and the number of operator increase, the performance gradually decreases.
The satisfactory results can be obtained on equations with 0-30 length or 0-20 operator, The model achieved satisfactory results, benefits from the distribution of equations in the corpus, where the number of equations with lengths below 30 far exceeds those with lengths above 30.
Although high-precision symbolic regression on ultra long equations with complex networks is still a challenge, our method can ensure the $Close_{0.1}$ and $R^2$ performance, considering that excessively long equations may not have practical significance on complex networks, such results are acceptable.
Moreover, these results demonstrate that the performance of the model is topology independent again.

\clearpage

\begin{figure*}[t]
\centering
\includegraphics[width=0.8\textwidth]{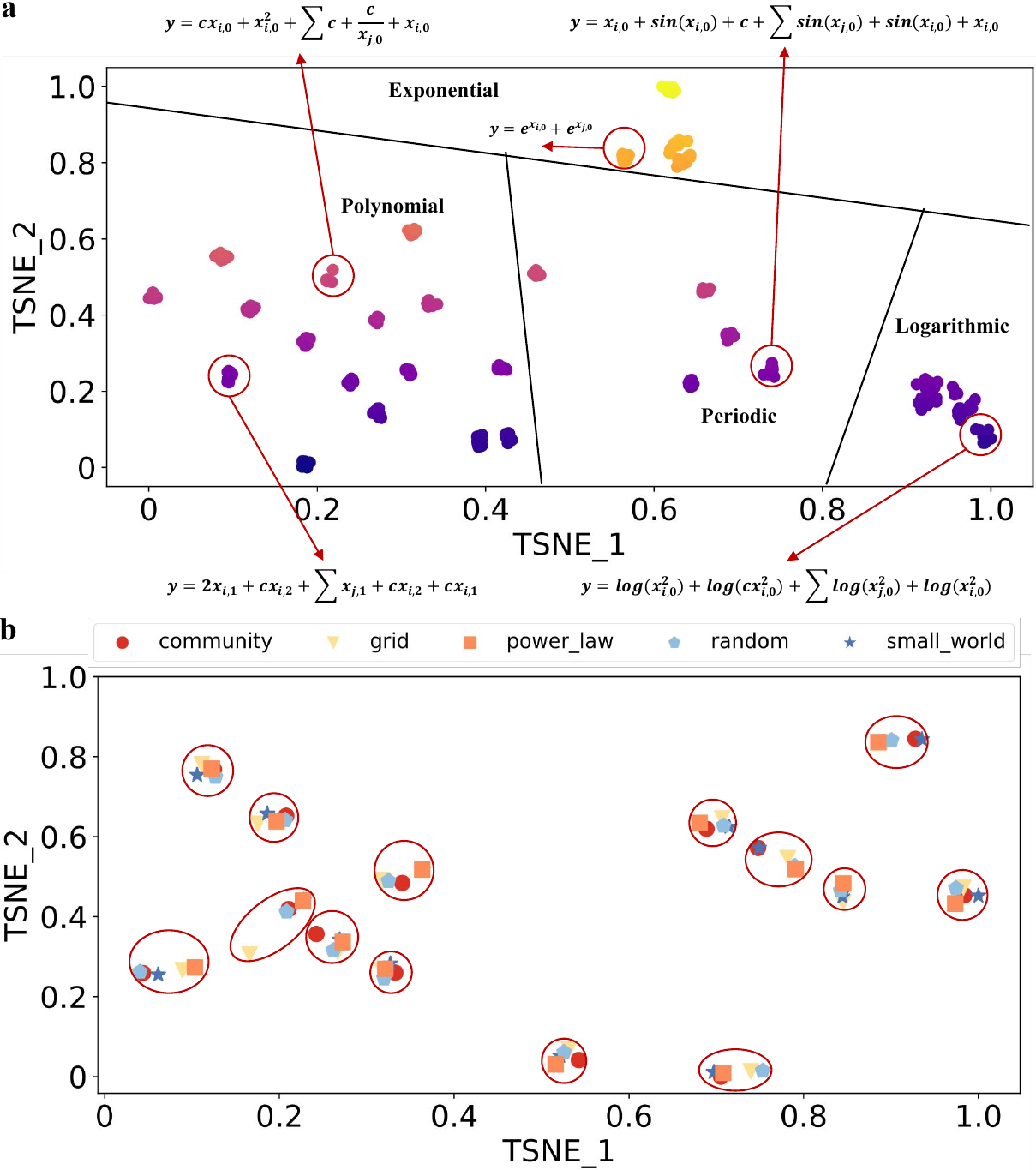}
\caption{The distribution of representation on the equations with different topologies and constants. \textbf{a.} Equations with the same skeleton are clustered together, and skeletons with similar structures are located in the same region. \textbf{b.} For the same equation, it is clustered together regardless of topology.}\label{a_exp2_2}
\end{figure*}

We also provide a more detailed representation projection as shown in Fig. \ref{a_exp2_2}.
Firstly, we randomly select network equations with different structures and the same topology. Each equation is sampled the constant for 10 times randomly. Then, we feed the data from the equations into the model representation layer and project the output into representation with 2 dimensions through t-SNE. For clarity, we select 320 equation data representations from 32 different structures for display, including periodic, exponential, logarithmic and polynomial types. As shown in Fig. \ref{a_exp2_2}(a), even with different constants, equations with the same structure tend to cluster together, such as $y=log(x_{i,0}^2)+log(3.76x_{i,0}^2)+\sum log(x_{i,0}^2)+log(x_{j,0}^2)$ and $y=log(x_{i,0}^2)+log(0.97x_{i,0}^2)+\sum log(x_{i,0}^2)+log(x_{j,0}^2)$ , and those with similar structural features are more likely to occur in the same region, such as $y=2x_{i,1}+c+x_{i,2}+\sum x_{j,1}+c+x_{i,2}+x_{i,1}$ and $y=cx_{i,0}+x_{i,0}^2+\sum c+\frac{c}{x_{j,0}}+x_{i,0}$, proving that our model has interpretable and effective representations.
In addition, we also select network equations and combine 5 types of topology for each equation to perform representation projection. It is interesting that for an equation, no matter how its topology changes, its representation gathers together (see Fig. \ref{a_exp2_2}(b)), indicating that our local sampling strategy and decoupling interaction term are effective.

We also conduct experiments on large-scale scenarios to demonstrate the superiority of our local topology sampling as shown in Fig. \ref{a_exp2_5}. Even in a scenario with 5000 nodes (LV dynamic in Appendix \ref{secA4}), our model can still ensure performance.

\begin{figure*}[t]
\centering
\includegraphics[width=0.9\textwidth]{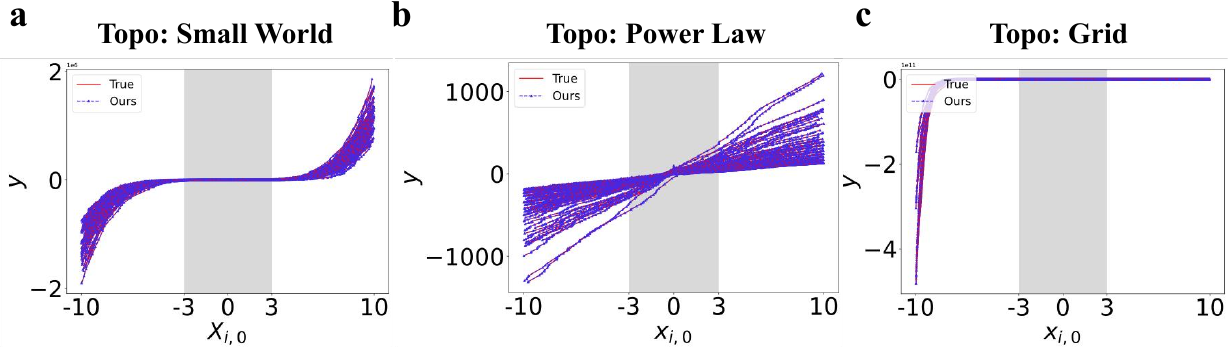}
\caption{Results of symbolic regression on equations in USE with 1 dimension (equation curve).}\label{a_exp2_3}
\end{figure*}

\begin{figure*}[t]
\centering
\includegraphics[width=0.8\textwidth]{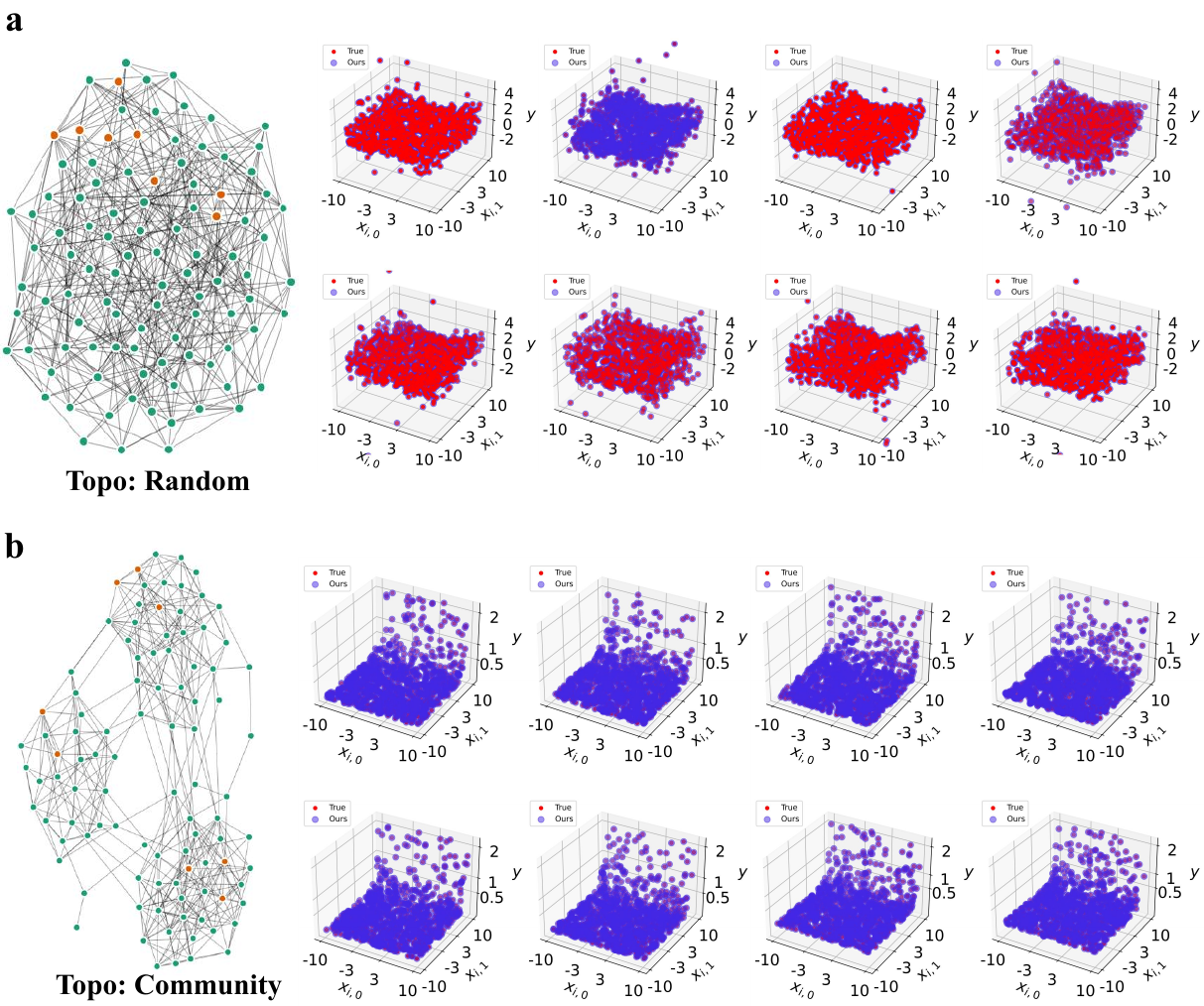}
\caption{Results of symbolic regression on equations in USE with 2 dimensions. The comparison between the true distribution and predicted distribution of the state of the selected nodes (orange) is shown on the right side.}\label{a_exp2_4}
\end{figure*}

Some specific visual examples of symbolic regression on complex networks are shown in Fig. \ref{a_exp2_3}-\ref{a_exp2_4} and Table.  \ref{A_tab14}-\ref{A_tab16}.
We select some representative equations with 1 dimension, including polynomials, fractions, exponential, trigonometric functions, etc., combined with a specific topology (100 nodes). Our method can accurately regress the equations, both on the data in IN-Domain and OUT-Domain and the form of equations.
For equations with 2 dimensions (see Fig. \ref{a_exp2_4} and Table. \ref{A_tab15}), we sample some nodes from the global topology of each equation to display the results, and we also provide specific global topologies. 
Our method can also perform accurate symbolic regression.

\begin{figure*}[h]
\centering
\includegraphics[width=1.0\textwidth]{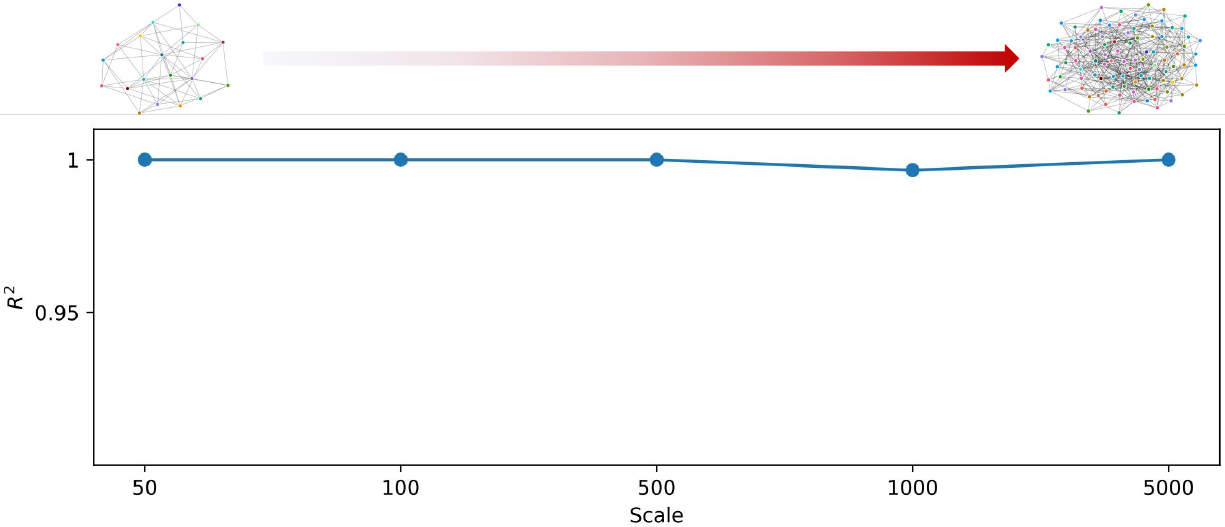}
\caption{The effect of the scale of topology on symbolic regression performance.}\label{a_exp2_5}
\end{figure*}

\clearpage

\begin{table}[h]
\caption{Results of symbolic regression on equations in USE with 1 dimension (equation form).}\label{A_tab14}%
\begin{tabular}{ccc}
\toprule
Equation ID & \multicolumn{2}{c}{Comparison}  \\                              
\midrule
\multirow{2}{*}{Fig. \ref{a_exp2_3}(a)} &True: &$y=-11.133x_{i,0}-\frac{1.285}{x_{i,0}}+\sum A_{ij}2x_{i,0}^2x_{j,0}^2$\\ 
                                        &Ours: &$y=-11.133x_{i,0}-\frac{1.285}{x_{i,0}}+\sum A_{ij}2x_{i,0}^2x_{j,0}^2$\\ 
\multirow{2}{*}{Fig. \ref{a_exp2_3}(b)} &True: &$y=-2.521x_{i,0}-4.707logx_{i,0}^2+\sum A_{ij}x_{j,0}+2.771x_{i,0}+cosx_{j,0}$\\ 
                                        &Ours: &$y=-2.521x_{i,0}-4.707logx_{i,0}^2+\sum A_{ij}x_{j,0}+2.771x_{i,0}+cosx_{j,0}$\\ 
\multirow{2}{*}{Fig. \ref{a_exp2_3}(c)} &True: &$y=0.015x_{i,0}^2+0.972x_{i,0}-\sum A_{ij}0.016x_{i,0}-5.712x_{j,0}-13.645e^{-2.22x_{i,0}}$\\ 
                                        &Ours: &$y=0.015x_{i,0}^2+0.972x_{i,0}-\sum A_{ij}0.016x_{i,0}-5.712x_{j,0}-13.645e^{-2.22x_{i,0}}$\\ 
\botrule
\end{tabular}
\end{table}

\begin{table}[t]
\caption{Results of symbolic regression on equations in USE with 2 dimension (equation form).}\label{A_tab15}%
\begin{tabular}{ccc}
\toprule
Equation ID & \multicolumn{2}{c}{Comparison}  \\                              
\midrule
\multirow{2}{*}{Fig. \ref{a_exp2_4}(a)} &True: &$x_{i,1}+\frac{1}{x_{i,1}}+4.48x_{i,0}+5.68x_{i,0}^2+\sum A_{ij}4.28x_{i,1}+\frac{x_{i,1}}{0.243+0.08x_{i,1}}+7.10x_{i,0}x_{j,0}$\\ 
                                        &Ours: &$x_{i,1}+\frac{1}{x_{i,1}}+4.48x_{i,0}+5.68x_{i,0}^2+\sum A_{ij}4.28x_{i,1}+\frac{x_{i,1}}{0.243+0.08x_{i,1}}+7.10x_{i,0}x_{j,0}$\\ 
\multirow{2}{*}{Fig. \ref{a_exp2_4}(b)} &True: &$y=x_{i,0}+0.075x_{i,1}+0.982e^{x_{i,1}}+\sum A_{ij}x_{j,0}+0.575x_{i,0}+\frac{1.458}{x_{j,1}}$\\ 
                                        &Ours: &$y=x_{i,0}+0.075x_{i,1}+0.982e^{x_{i,1}}+\sum A_{ij}x_{j,0}+0.575x_{i,0}+\frac{1.458}{x_{j,1}}$\\ 

\botrule
\end{tabular}
\end{table}

We also present some challenging network symbolic regression tasks, which involve equations with long structures and 3 independent variables, combined with a complex network topology.
From the Table. \ref{A_tab16}, although it is not possible to directly and accurately regress highly complex equations, our method can still obtain an approximate structure and a high $R^2$ metric, which means that it is possible to obtain a more accurate equation form by further constant optimization or by incorporating the result as the initial population into a genetic algorithm.

\begin{sidewaystable}
\large
% \begin{table}[h]
\caption{Results of symbolic regression on equations in USE with 3 dimension (equation form).}\label{A_tab16}%
\begin{tabular}{ccc}
\toprule
\multicolumn{2}{c}{Comparison} & $R^2$  \\                              
\midrule
True: &$y=x_{i,0}x_{i,1}+x_{i,1}+2x_{i,2}+4.44x_{i,0}+\sum A_{ij}0.091+x_{j,1}+0.083x_{j,2}+0.93x_{j,0}+\frac{0.826}{x_{j,0}}$& $/$\\
Ours: &$y=x_{i,0}x_{i,1}+x_{i,1}+1.987x_{i,2}+4.43x_{i,0}+\sum A_{ij}0.091+x_{j,1}+0.09x_{j,2}+x_{j,0}+\frac{0.826}{x_{j,0}}$& $>0.9$\\ 
True: &$y=x_{i,0}+\frac{0.329}{5.069-x_{i,1}}(x_{i,2}+0.598x_{i,0}x_{i,2}+0.04x_{i,0})+\sum A_{ij}0.231x_{i,0}+x_{i,1}+2.78x_{i,2}+x_{j,2}$& $/$\\
Ours: &$y=x_{i,0}+0.038(x_{i,2}+0.54x_{i,0}x_{i,2})+\sum A_{ij}0.231x_{i,0}+x_{i,1}+2.724x_{i,2}+x_{j,2}$& $>0.9$\\ 
True: &$y=x_{i,1}+\frac{1}{x_{i,2}}+(1.793+x_{i,1})^2+2_{i,2}+3.65e^{x_{i,2}}+\sum A_{ij}(x_{j,1}-\frac{4.926}{x_{i,1}(x_{j,2}-14.061)^2})(x_{i,1}+x_{j,0}+\frac{1}{x_{i,1}})$& $/$\\
Ours: &$y=x_{i,1}^2+\frac{1}{x_{i,2}}+11.8x_{i,1}+x_{i,2}-0.132e^{3.711x_{i,2}}+\sum A_{ij}x_{j,1}(x_{i,1}+x_{i,2}+x_{j,0}+\frac{26.634}{x_{i,1}})$& $>0.9$\\ 
\botrule
\end{tabular}
% \end{table}
\end{sidewaystable}

\clearpage

\section{More details on inferring interpretable network dynamics}\label{secA4}

\subsection{Network dynamics}\label{secA41}
\begin{itemize}
\item Biochemical Dynamics (Bio)\cite{voit2000computational}: It is used to describe the dynamic biochemical process of PPI (Protein-Protein Interactions), and its equation form is as follows: $\frac{dx_{i}}{dt}=F_{i}-B_{i}x_{i}(t)+\sum_{j=1}^{N}A_{ij}x_{i}(t)x_{j}(t)$, where $x_{i}(t)$ represents the protein $i$ concentration at time $t$, $F_{i}$ represents the average influx rate of proteins and $B_{i}$ represents the average degradation rate of proteins. 

\item Epidemic Dynamics (Epi)\cite{pastor2015epidemic}: It is used to describe the dynamic spread of infectious diseases, and its equation form is as follows: $\frac{dx_{i}}{dt}=-\delta_{i} x_{i}(t)+\sum_{j=1}^{N}A_{ij}(1-x_{i}(t))x_{j}(t)$, where $x_{i}(t)$ represents the infection probability of node $i$ and the $\delta_{i}$ represent the rate of recovery.

\item Gene Regulatory Dynamics (Gene)\cite{mazur2009reconstructing}: It is used to describe the dynamic regulation of genes,and its equation form is as follows: $\frac{dx_{i}}{dt}=-B_{i}x_{i}(t)^f+\sum_{j=1}^{N}A_{ij}\frac{x_{j}(t)^h}{{x_{j}(t)^h}+1}$, where $x_{i}(t)$ represents the expression of gene $i$, $B_i$ represent the decay rate and $h$ represents the Hill coefficient.

\item Heat Diffusion Dynamics (Heat)\cite{zang2020neural}: It is used to describe the dynamic heat diffusion, and its equation form is as follows: $\frac{dx_{i}}{dt}=\sum_{j=1}^{N}A_{ij}k_{i}(x_{j}(t)-x_{i}(t))$, where $x_{i}(t)$ represents the heat change of node $i$ and the $k_{i}$ represent the rate of heat change.

\item Mutualistic Interaction Dynamics (Mutu)\cite{gao2016universal}: It is used to describe the dynamic ecology of interactions between population, and its equation is as follows: $\frac{dx_{i}}{dt}=b_i+x_{i}(t)(1-\frac{x_{i}}{k_{i}})(\frac{x_{i}(t)}{c_{i}}-1)+\sum_{j=1}^{N}A_{ij}\frac{x_{i}(t)x_{j}(t)}{d_{j}+e_{i}x_{i}(t)+h_{i}x_{i}(t)}$, where $x_{i}(t)$ represents the population richness of population $i$, $b_i$ represents the migration probability, $k_i$ represents population growth rate and $c_i$ represents population decline threshold.

\item Lotka-Volterra Model (LV)\cite{macarthur1970species}: It is used to describe the dynamic population of species in competition, and its equation is as follows: $\frac{dx_{i}}{dt}=x_{i}(t)(\alpha_{i}-\theta_{i}x_{i}(t))-\sum_{j=1}^{N}A_{ij}x_{i}(t)x_{j}(t)$, where $x_{i}(t)$ represents the population size of species $i$, $\alpha_{i}$ and $\theta_{i}$ represent the growth parameters of species.
\end{itemize}

\subsection{Details on experimental setting}\label{secA42}

To ensure the reproducibility of our findings, we set the parameters of each network dynamics and display them in Table \ref{A_tab17}. 
The topology setting is still as shown in Section \ref{secA32}. The number of nodes in each dynamic scenario under various topologies is $100-500$. 
Each node is set with an appropriate initial value $x_{i}(0)$ to simulate its dynamic changes during the time period from $t=0$ to $T_R$ (IN-Domain) or $T_P$ (OUT-Domain) with a step-size $t_{\delta}$.
The data is needs to be preprocessed before being fed into the model, we perform distribution sampling and distribution scaling on the data (see Method section).
We sample 200 data points in $[0,T_R]$ as the observations from dynamics.
\begin{table}[t]
\caption{Network dynamics setting}\label{A_tab17}%
\begin{tabular}{ccccccc}
\toprule
Dynamic Scene   & Parameter setting     & $x_{i}(0)$    & $t_{\delta}$  &$T_R$  &$T_P$\\                             
\midrule
Bio          & $F_{i}=1,B_{i}=-1$        & $x_{i}(0) \sim \mathbf{U}(0,2)$    & $0.0001$  &$0.1$  &$0.5$\\ 
Epi          & $\delta_{i}=1.0$       & $x_{i}(0) \sim \mathbf{U}(0,1)$    & $0.001$   &$1$    &$5$\\ 
Gene         & $B_{i}=1,f=1,h=2$        & $x_{i}(0) \sim \mathbf{U}(0,2)$    & $0.01$    &$5$    &$10$\\ 
Heat         & $k=1$        & $x_{i}(0) \sim \mathbf{U}(0,1)$    & $0.01$    &$1$    &$5$\\ 
Mutu   & $b_i=1,k_i=5,c_i=1,d_i=5,e_i=0.9,h_i=0.1$        & $x_{i}(0) \sim \mathbf{U}(0,2)$    & $0.001$   &$1$    &$5$\\ 
LV           & $\alpha_{i}=0.5,\theta_{i}=1.0$        & $x_{i}(0) \sim \mathbf{U}(0,5)$    & $0.0001$  &$0.1$  &$0.5$\\ 

\botrule
\end{tabular}
\end{table}

\begin{figure*}[t]
\centering
\includegraphics[width=1.0\textwidth]{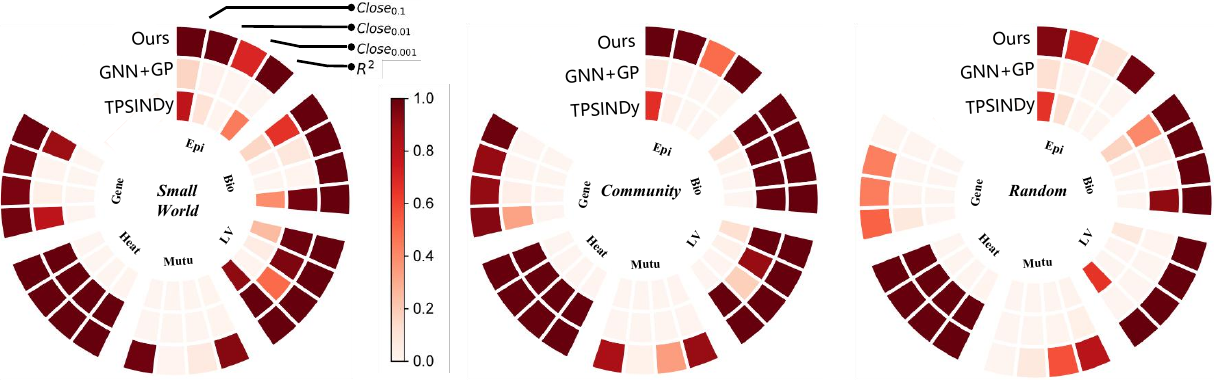}
\caption{Result of symbolic regression on network dynamics under small world, community and random topologies.}\label{a_exp3_1}
\end{figure*}

\subsection{More experimental results and regression analyses}\label{secA43}
\begin{figure*}[t]
\centering
\includegraphics[width=1.0\textwidth]{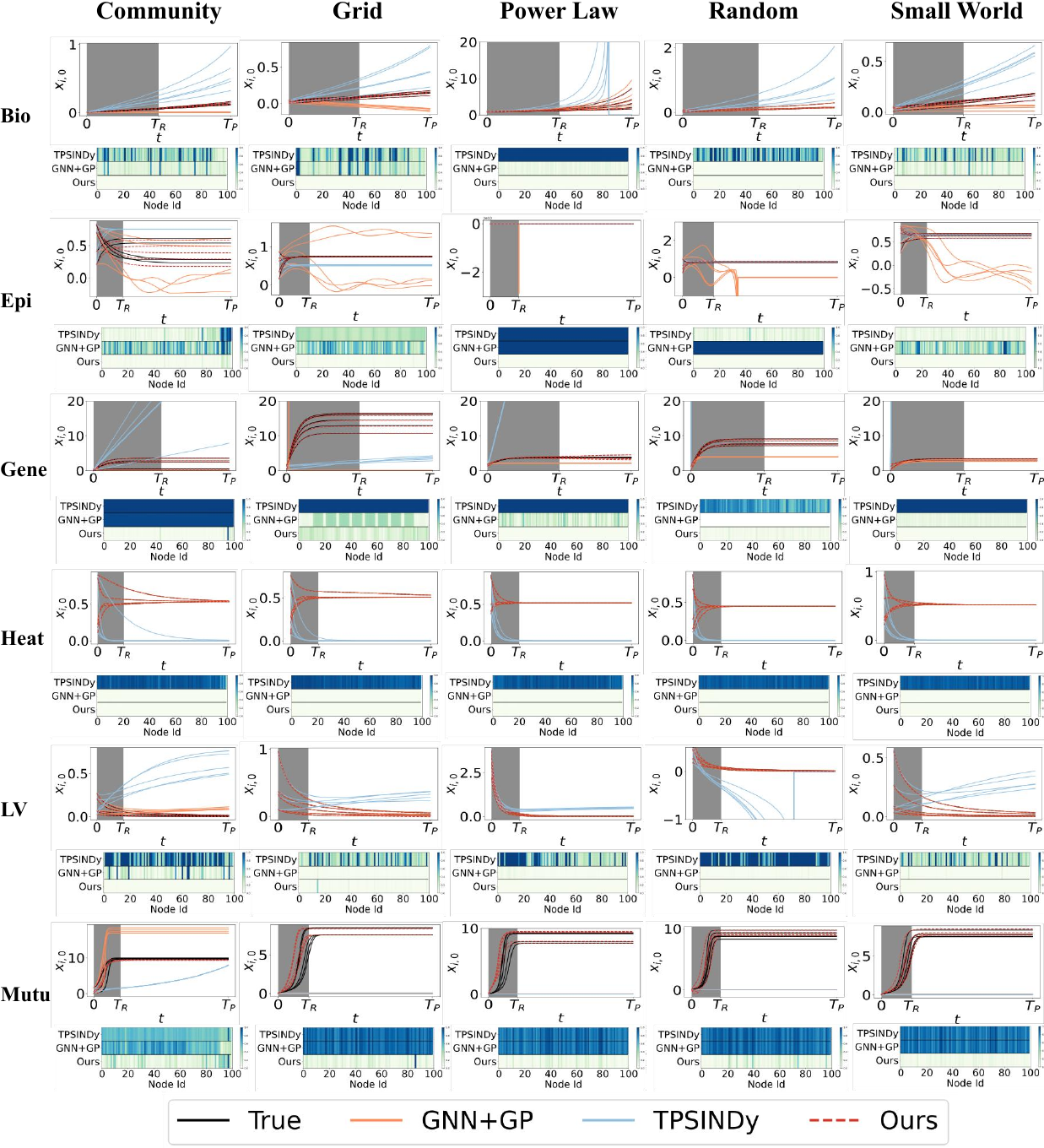}
\caption{Result of symbolic regression on network dynamics under small world, community and random topologies.}\label{a_exp3_2}
\end{figure*}

We present the performance of our model in dynamic scenarios under other topologies (Small World, Community and Random), and our method has the best performance (see Fig .\ref{a_exp3_1}). 
A billion level corpus is key, compared to TPSINDy and GNN+GP, a larger learning space can effectively regress and find suitable dynamic equations. 
It is worth noting that in order to compare the optimal performance of the SOTA method, its input data far exceeds 200, while we still only need 200.
To compare the performance of various methods in more detail, we also randomly sample some nodes for each dynamics, set initial values for each node, and calculate its trajectory based on the equations regressed from the model. 
Our method has the highest trajectory fitting degree in all dynamics (see Fig .\ref{a_exp3_2}). 
However, due to the fixed base library setting, the equations regressed from TPSINDy cannot even be integrated in some scenarios to obtain the correct trajectory. 
We also compare the trajectories of all nodes through the RMSE metric, which is present at the bottom of each trajectory figures. As shown, the color band of our method is the shallowest and purest, indicating the high accuracy of our method. 
We normalize the MAPE in order to make the color bands easier to distinguish.

The specific regression equations are listed in Table \ref{A_tab18}. 
Note that some equations may not be consistent with the ground truth, but from the results, they appear to be equivalent forms of the ground truth.

\clearpage

\begin{sidewaystable}[t]

% \begin{table}[h]
\renewcommand\arraystretch{4}

\caption{Results of symbolic regression on on interpretable network dynamics (Bio).}\label{A_tab18}%
\fontsize{8}{10}\selectfont{
\begin{tabular}{lccc}
\toprule
\multirow{2}{*}{Topology}   & \multicolumn{3}{c}{Comparison\#Bio:  $\frac{dx_{i,0}}{dt}=1 - x_{i,0} + \sum A_{ij}( x_{j,0}  x_{i,0})$}\\
                            &Ours&TPSINDy&GNN+GP\\
\midrule
Community                  
                            &\makecell[l]{$\frac{dx_{i,0}}{dt}=1 - x_{i,0}$\\\hspace{2em}$+ \sum A_{ij}x_{j,0}x_{i,0}$}
                            &\makecell[l]{$\frac{dx_{i,0}}{dt}=0$\\\hspace{2em}$+\sum A_{ij}(0.840 x_{j,0}  x_{i,0} + 0.651)$}
                            &\makecell[l]{$\frac{dx_{i,0}}{dt}=1.913x_{i,0} - 14.995\sin(0.225x_{i,0})$\\\hspace{2em}$ + \sum A_{ij}x_{j,0}x_{i,0}$}\\
Grid                        
                            &\makecell[l]{$\frac{dx_{i,0}}{dt}=1 - x_{i,0}$\\\hspace{2em}$+ \sum A_{ij}x_{j,0}x_{i,0}$}
                            &\makecell[l]{$\frac{dx_{i,0}}{dt}=0$\\\hspace{2em}$+\sum A_{ij}(0.874 x_{j,0}  x_{i,0} + 0.651)$}
                            &\makecell[l]{$\frac{dx_{i,0}}{dt}=-1.130 - x_{i,0}$\\\hspace{2em}$ +\sum A_{ij} x_{j,0}x_{i,0}$}\\
Power Law                   
                            &\makecell[l]{$\frac{dx_{i,0}}{dt}=1 - x_{i,0}$\\\hspace{2em}$+ \sum A_{ij}x_{j,0}x_{i,0}$}
                            &\makecell[l]{$\frac{dx_{i,0}}{dt}=0$\\\hspace{2em}$+\sum A_{ij}(0.956 x_{j,0} x_{i,0} + 0.513)$}
                            &\makecell[l]{$\frac{dx_{i,0}}{dt}=1.599x_{i,0} + e^{(-6.909x_{i,0})}$\\\hspace{2em}$+\sum A_{ij}x_{i,0}(-\frac{0.076\,x_{i,0}x_{j,0}}{x_{i,0} + x_{j,0}} + x_{j,0})$}\\
Random                      
                            &\makecell[l]{$\frac{dx_{i,0}}{dt}=1 - x_{i,0}$\\\hspace{2em}$+ \sum A_{ij}x_{j,0}x_{i,0}$}
                            &\makecell[l]{$\frac{dx_{i,0}}{dt}=0$\\\hspace{2em}$+\sum A_{ij}(0.914 x_{j,0}  x_{i,0} + 0.630)$}
                            &\makecell[l]{$\frac{dx_{i,0}}{dt}=\log((x_{i,0}e^{(3.848x_{i,0} - 2.848|x_{i,0} - 18.175|)} + 2.716)e^{-x_{i,0}})$\\\hspace{2em}$+ \sum A_{ij}x_{j,0}x_{i,0}$}\\
Small World                 
                            &\makecell[l]{$\frac{dx_{i,0}}{dt}=1 - x_{i,0}$\\\hspace{2em}$+ \sum A_{ij}x_{j,0}x_{i,0}$}
                            &\makecell[l]{$\frac{dx_{i,0}}{dt}=0$\\\hspace{2em}$+\sum A_{ij}((0.940 x_{j,0} x_{i,0} + 0.651)$}
                            &\makecell[l]{$\frac{dx_{i,0}}{dt}=x_{i,0}(7.531\sin(x_{i,0}^{0.25}) - \sin(1.692\sin(\sqrt{x_{i,0}})) - 1.0)$\\\hspace{2em}$+ \sum A_{ij}x_{i,0}|0.132x_{i,0} - x_{j,0}|$}\\

\botrule
\end{tabular}
}
% \end{table}
\end{sidewaystable}

\clearpage

\begin{sidewaystable}[t]

\renewcommand\arraystretch{4}
\caption{Results of symbolic regression on on interpretable network dynamics (Epi).}\label{A_tab18}%
\fontsize{8}{10}\selectfont{
\begin{tabular}{lccc}
\toprule
\multirow{2}{*}{Topology}   & \multicolumn{3}{c}{Comparison\#Epi:  $\frac{dx_{i,0}}{dt}=-2x_{i,0} + \sum A_{ij}(x_{j,0} - x_{i,0}x_{j,0})$}\\
                            &Ours&TPSINDy&GNN+GP\\
\midrule
Community                  
                            &\makecell[l]{$\frac{dx_{i,0}}{dt}=\frac{-19.702x_{i,0} - 1.234}{6.515x_{i,0} + 7.72}$\\$+ \sum A_{ij}(-0.647x_{i,0} + 0.311x_{j,0} - 0.157\arctan(16.43x_{j,0} - 15.826) + 0.21)$}
                            &\makecell[l]{$\frac{dx_{i,0}}{dt}=-4.29x_{i,0}^2 + 2.45$\\$+\sum A_{ij}0$}
                            &\makecell[l]{$\frac{dx_{i,0}}{dt}=-x_{i,0}$\\$ - \sum A_{ij}(x_{j,0} - x_{j,0}^2)$}\\
Grid                        
                            &\makecell[l]{$\frac{dx_{i,0}}{dt}=-1.267 x_{i,0}^2 + 0.279 x_{i,0}-0.946$\\$-\sum A_{ij}(0.442x_{i,0}-0.213x_{j,0}^2-0.334x_{j,0}+0.437x_{i,0}x_{j,0}-0.39)$}
                            &\makecell[l]{$\frac{dx_{i,0}}{dt}=-4.93x_{i,0}^2$\\$+ \sum A_{ij}0.337x_{j,0}$}
                            &\makecell[l]{$\frac{dx_{i,0}}{dt}=-x_{i,0}$\\$ + \sum A_{ij}(x_{j,0} - x_{j,0}^2)$}\\
Power Law                   
                            &\makecell[l]{$\frac{dx_{i,0}}{dt}=-1.914x_{i,0}-0.045$\\$+\sum A_{ij}(0.797x_{i,0}^2x_{j,0}-0.559x_{i,0}^2-2.213x_{i,0}x_{j,0}+0.843x_{i,0}+1.446x_{j,0}-0.311)$}
                            &\makecell[l]{$\frac{dx_{i,0}}{dt}=0.68x_{i,0}$\\$+ \sum A_{ij}0.1x_{j,0}$}
                            &\makecell[l]{$\frac{dx_{i,0}}{dt}=-x_{i,0}$\\$ + \sum A_{ij}(x_{j,0} - x_{j,0}^2)$}\\
Random                      
                            &\makecell[l]{$\frac{dx_{i,0}}{dt}=0.047-2.048x_{i,0}$\\$+\sum A_{ij}(0.22x_{i,0}+1.226xx_{j,0}-1.32x_{i,0}x_{j,0}-0.162)$}
                            &\makecell[l]{$\frac{dx_{i,0}}{dt}=-5.05x_{i,0}^2 + 3.29$\\$+ \sum A_{ij}0.04x_{j,0}$}
                            &\makecell[l]{$\frac{dx_{i,0}}{dt}=-x_{i,0}$\\$ + \sum A_{ij}(x_{j,0} - x_{j,0}^2)$}\\
Small World                 
                            &\makecell[l]{$\frac{dx_{i,0}}{dt}=\frac{-43.027x_{i,0}^4 + 54.464x_{i,0}^3- 8.113x_{i,0}^2 + 15.701x_{i,0} + 3.53}{24.59x_{i,0}^2 - 29.75x_{i,0} - 4.659
}$\\$-\sum A_{ij}(0.027x_{i,0}-(0.704x_{i,0}-1)^2(1.164x_{j,0}+0.722))$}
                            &\makecell[l]{$\frac{dx_{i,0}}{dt}=-3.65x_{i,0}^2 + 1.56$\\$+\sum A_{ij}0$}
                            &\makecell[l]{$\frac{dx_{i,0}}{dt}=-x_{i,0}$\\$ + \sum A_{ij}(x_{j,0} - x_{j,0}^2)$}\\

\botrule
\end{tabular}
}
% \end{table}
\end{sidewaystable}

\clearpage

\begin{sidewaystable}[t]

\renewcommand\arraystretch{4}
\caption{Results of symbolic regression on on interpretable network dynamics (Gene).}\label{A_tab18}%
\fontsize{6}{8}\selectfont{
\begin{tabular}{lccc}
\toprule
\multirow{2}{*}{Topology}   & \multicolumn{3}{c}{Comparison\#Gene:  $\frac{dx_{i,0}}{dt}=-2x_{i,0}+ \sum A_{ij}\frac{x_{j,0}^2}{1 + x_{j,0}^2}$}\\
                            &Ours&TPSINDy&GNN+GP\\
\midrule
Community                  
                            &\makecell[l]{$\frac{dx_{i,0}}{dt}=-0.027x_{i,0}^2 - 1.858x_{i,0} - 0.23$\\$+ \sum A_{ij}\frac{0.019x_{j,0}^3 - 0.188x_{j,0}^2 + 1.062x_{j,0} + 2.80}{0.225x_{j,0}^2 - 1.467x_{j,0} + 7.676}$}
                            &\makecell[l]{$\frac{dx_{i,0}}{dt}=0$\\$+\sum A_{ij}(0.78)$}
                            &\makecell[l]{$-1.798x_{i,0}$\\$ + \sum A_{ij}(0.455x_{j,0}cos(\frac{0.169x_{i,0}}{x_{j,0}-2.403}))$}\\
Grid                        
                            &\makecell[l]{$\frac{dx_{i,0}}{dt}=0.362x_{i,0}-2.158$\\$+\sum A_{ij}(0.579 - 0.204x_{i,0}+\frac{1}{9.349(0.353x_{j,0}-1)^2+2.027})$}
                            &\makecell[l]{$\frac{dx_{i,0}}{dt}=0$\\$+\sum A_{ij}(\frac{2.02x_{j,0}^2}{1+x_{j,0}^2})$}
                            &\makecell[l]{$-$}\\
%\frac{dx_{i,0}}{dt}=-x_{i,0}sin(sin(sin(sin(2x_{i,0})+0.602))+0.796)-x_{i,0}$\\$ - \sum A_{ij}(\frac{x_{j,0}}{x_{j,0}-\frac{0.046}{x_{i,0}-2.108}+\frac{0.975}{x_{j,0}}})
Power Law                   
                            &\makecell[l]{$\frac{dx_{i,0}}{dt}=-2.253x_{i,0}-0.404$\\$+\sum A_{ij}(0.084x_{j,0}-0.396+\frac{1}{-0.336x_{j,0}+1.405})$}
                            &\makecell[l]{$\frac{dx_{i,0}}{dt}=-0.17x_{i,0}$\\$+\sum A_{ij}(\frac{0.04x_{j,0}^2}{1+x_{j,0}^2})$}
                            &\makecell[l]{$\frac{dx_{i,0}}{dt}=x_{i,0}(0.149x_{i,0}^2(x_{i,0} - 1.307)(x_{i,0} - 1) - 2)$\\$+ \sum A_{ij}\frac{x_{j,0}^2}{x_{j,0}(0.031x_{i,0}^3 + x_{j,0}) + 0.979}$}\\
Random                      
                            &\makecell[l]{$\frac{dx_{i,0}}{dt}=-1.987x_{i,0}+0.004$\\$+\sum A_{ij}(0.522x_{j,0}-0.101)$}
                            &\makecell[l]{$\frac{dx_{i,0}}{dt}=0$\\$+\sum A_{ij}(-8.29\frac{x_{j,0}^2}{1 + x_{j,0}^2} + 64.39)$}
                            &\makecell[l]{$\frac{dx_{i,0}}{dt}=x_{i,0}(0.119x_{i,0}^2 - 2.278)$\\$+ \sum A_{ij}\frac{-x_{i,0}(0.017x_{i,0}^2 - 0.057)(x_{j,0}^2 + 1.151) + x_{j,0}^2}{x_{j,0}^2 + 1.151}$}\\
Small World                 
                            &\makecell[l]{$\frac{dx_{i,0}}{dt}=-2.001x_{i,0}+0.002$\\$-\sum A_{ij}(0.22x_{j,0}^2-0.958x_{j,0}+0.245)$}
                            &\makecell[l]{$\frac{dx_{i,0}}{dt}=0$\\$+\sum A_{ij}(-16.51\tanh(x_{i,0}x_{j,0}) + 106.54)$}
                            &\makecell[l]{$\frac{dx_{i,0}}{dt}=x_{i,0}(-\sin(\cos(0.506x_{i,0}) + 0.746) - 1)$\\$+\sum A_{ij}\sin \sqrt \frac{x_{i,0}x_{j,0}^3}{x_{i,0}x_{j,0}^2\sqrt{e^{x_{i,0}}} + 2.086}$}\\

\botrule
\end{tabular}
}
% \end{table}
\end{sidewaystable}

\clearpage

\begin{sidewaystable}[t]

% \begin{table}[h]
\renewcommand\arraystretch{4}
\caption{Results of symbolic regression on on interpretable network dynamics (Heat).}\label{A_tab18}%
\fontsize{8}{10}\selectfont{
\begin{tabular}{lccc}
\toprule
\multirow{2}{*}{Topology}   & \multicolumn{3}{c}{Comparison\#Heat:  $\frac{dx_{i,0}}{dt}=0+ \sum A_{ij}(x_{j,0} - x_{i,0})$}\\
                            &Ours&TPSINDy&GNN+GP\\
\midrule
Community                  
                            &\makecell[l]{$\frac{dx_{i,0}}{dt}=0+ \sum A_{ij}(x_{j,0} - x_{i,0})$}
                            &\makecell[l]{$\frac{dx_{i,0}}{dt}=0+ \sum A_{ij}(0.1x_{j,0} - x_{i,0})$}
                            &\makecell[l]{$\frac{dx_{i,0}}{dt}=0+ \sum A_{ij}(x_{j,0} - x_{i,0})$}\\
Grid                        
                            &\makecell[l]{$\frac{dx_{i,0}}{dt}=0+ \sum A_{ij}(x_{j,0} - x_{i,0})$}
                            &\makecell[l]{$\frac{dx_{i,0}}{dt}=0+ \sum A_{ij}(0.1x_{j,0} - x_{i,0})$}
                            &\makecell[l]{$\frac{dx_{i,0}}{dt}=0+ \sum A_{ij}(x_{j,0} - x_{i,0})$}\\
Power Law                   
                            &\makecell[l]{$\frac{dx_{i,0}}{dt}=0+ \sum A_{ij}(x_{j,0} - x_{i,0})$}
                            &\makecell[l]{$\frac{dx_{i,0}}{dt}=0+ \sum A_{ij}(0.1x_{j,0} - x_{i,0})$}
                            &\makecell[l]{$\frac{dx_{i,0}}{dt}=0+ \sum A_{ij}(x_{j,0} - x_{i,0})$}\\
Random                      
                            &\makecell[l]{$\frac{dx_{i,0}}{dt}=0+ \sum A_{ij}(x_{j,0} - x_{i,0})$}
                            &\makecell[l]{$\frac{dx_{i,0}}{dt}=0+ \sum A_{ij}(0.1x_{j,0} - x_{i,0})$}
                            &\makecell[l]{$\frac{dx_{i,0}}{dt}=0+ \sum A_{ij}(x_{j,0} - x_{i,0})$}\\
Small World                 
                            &\makecell[l]{$\frac{dx_{i,0}}{dt}=0+ \sum A_{ij}(x_{j,0} - x_{i,0})$}
                            &\makecell[l]{$\frac{dx_{i,0}}{dt}=0+ \sum A_{ij}(0.1x_{j,0} - x_{i,0})$}
                            &\makecell[l]{$\frac{dx_{i,0}}{dt}=0+ \sum A_{ij}(x_{j,0} - x_{i,0})$}\\

\botrule
\end{tabular}
}
% \end{table}
\end{sidewaystable}

\clearpage

\begin{sidewaystable}[t]

% \begin{table}[h]
\renewcommand\arraystretch{4}
\caption{Results of symbolic regression on on interpretable network dynamics (LV).}\label{A_tab18}%
\fontsize{8}{10}\selectfont{
\begin{tabular}{lccc}
\toprule
\multirow{2}{*}{Topology}   & \multicolumn{3}{c}{Comparison\#LV:  $\frac{dx_{i,0}}{dt}=0.5 x_{i,0} - x_{i,0}^2 - \sum A_{ij}( x_{j,0}  x_{i,0})$}\\
                            &Ours&TPSINDy&GNN+GP\\
\midrule
Community                  
                            &\makecell[l]{$\frac{dx_{i,0}}{dt}=-0.999 x_{i,0}^2 + 0.50 x_{i,0}$\\\hspace{2em}$- \sum A_{ij}(-0.999x_{i,0} x_{j,0})$}
                            &\makecell[l]{$\frac{dx_{i,0}}{dt}=0$\\\hspace{2em}$+\sum A_{ij}(-1.06x_{j,0} x_{i,0}+0.551)$}
                            &\makecell[l]{$\frac{dx_{i,0}}{dt}=(x_{i,0} - 0.79) (-x_{i,0} + e^{(0.012 x_{i,0})} - 1.451)$\\\hspace{2em}$ - \sum A_{ij}(0.999x_{i,0} x_{j,0})$}\\
Grid                        
                            &\makecell[l]{$\frac{dx_{i,0}}{dt}=-1.00 x_{i,0}^2 + 0.499 x_{i,0}$\\\hspace{2em}$-\sum A_{ij}(0.999x_{i,0}  x_{j,0})$}
                            &\makecell[l]{$\frac{dx_{i,0}}{dt}=0$\\\hspace{2em}$+\sum A_{ij}(-1.15x_{j,0} x_{i,0}+0.266)$}
                            &\makecell[l]{$\frac{dx_{i,0}}{dt}=x_{i,0} ( -1.026x_{i,0} + 0.017e^{(0.349x_{i,0})} + 0.509 )$\\\hspace{2em}$ - \sum A_{ij} x_{j,0}  x_{i,0} - 0.0006 | \sqrt{ e^{(x_{i,0})} }|)$}\\
Power Law                   
                            &\makecell[l]{$\frac{dx_{i,0}}{dt}=-1.00 x_{i,0}^2 + 0.50 x_{i,0}$\\\hspace{2em}$-\sum A_{ij}(1.00x_{i,0}  x_{j,0})$}
                            &\makecell[l]{$\frac{dx_{i,0}}{dt}=0$\\\hspace{2em}$+\sum A_{ij}(-1.039x_{j,0}  x_{i,0}+0.328)$}
                            &\makecell[l]{$\frac{dx_{i,0}}{dt}=- x_{i,0} ( x_{i,0} - 0.431 )$\\\hspace{2em}$+\sum A_{ij}(- x_{i,0} ( x_{j,0} - 0.011 ))$}\\
Random                      
                            &\makecell[l]{$\frac{dx_{i,0}}{dt}=-0.999 x_{i,0}^2 + 0.50 x_{i,0}$\\\hspace{2em}$-\sum A_{ij}(0.999x_{i,0} x_{j,0})$}
                            &\makecell[l]{$\frac{dx_{i,0}}{dt}=0$\\\hspace{2em}$+\sum A_{ij}(-1.08x_{j,0}  x_{i,0}-0.5)$}
                            &\makecell[l]{$\frac{dx_{i,0}}{dt}=x_{i,0} (0.486 - 0.998x_{i,0})$\\\hspace{2em}$+\sum A_{ij}(-x_{j,0} (x_{i,0}))$}\\
Small World                 
                            &\makecell[l]{$\frac{dx_{i,0}}{dt}=-0.99 x_{i,0}^2 + 0.49 x_{i,0} + 2.4e^{-5}$\\\hspace{2em}$-\sum A_{ij}1.00x_{i,0}  x_{j,0}$}
                            &\makecell[l]{$\frac{dx_{i,0}}{dt}=0$\\\hspace{2em}$+\sum A_{ij}(-1.16x_{j,0}  x_{i,0}+0.275)$}
                            &\makecell[l]{$\frac{dx_{i,0}}{dt}=- x_{i,0} ( x_{i,0} - 0.51 )$\\\hspace{2em}$+\sum A_{ij}(- x_{j,0} ( x_{i,0}))$}\\

\botrule
\end{tabular}}

% \end{table}
\end{sidewaystable}

\clearpage

\begin{sidewaystable}[t]
\scriptsize
% \begin{table}[h]
\renewcommand\arraystretch{4}
\caption{Results of symbolic regression on on interpretable network dynamics (Mutu).}\label{A_tab18}%
\begin{tabular}{llll}
\toprule
\multirow{2}{*}{Topology}   & \multicolumn{3}{c}{Comparison\#Mutu:  $-0.2x_{i,0}^3 + 1.2x_{i,0}^2 - x_{i,0} + 1 + \sum A_{ij}\frac{x_{i,0}x_{j,0}}{5 + 0.9x_{i,0} + 0.1x_{j,0}}$}\\
                            &Ours&TPSINDy&GNN+GP\\
\midrule
Community                  
                            &\makecell[l]{$\frac{-1.458x_{i,0}^3 + 19.397x_{i,0}^2 - 81.254x_{i,0} + 121.834}{1.120x_{i,0}^2 - 8.278x_{i,0} + 16.618}$\\$+ \sum A_{ij}\frac{1}{(10.691(0.169x_{i,0} - 1)^2 + 1.023)(1.338(0.169x_{j,0} - 1)^2 + 0.303)}$}
                            &\makecell[l]{$\frac{dx_{i,0}}{dt}=0.406x_{i,0}$\\$ + \sum A_{ij}0.28(x_{j,0} - x_{i,0})$}
                            &\makecell[l]{$\frac{dx_{i,0}}{dt}=0.722x_{i,0}^2 - 0.170x_{i,0}^3 + 1$\\$ +\sum A_{ij}(0.192x_{i,0}x_{j,0} + 0.004x_{j,0})$}\\
Grid                        
                            &\makecell[l]{$\frac{dx_{i,0}}{dt}=-0.266x_{i,0} - 0.195e^{0.653x_{i,0}} + 1.55$\\$+\sum A_{ij}(0.265x_{i,0}+0.415x_{j,0}-0.297)$}
                            &\makecell[l]{$-$}
                            &\makecell[l]{$-$}\\
Power Law                   
                            &\makecell[l]{$\frac{dx_{i,0}}{dt}=11.044 - 3.977x_{i,0}$\\$+\sum A_{ij}\frac{0.002x_{i,0}^2 + 0.178x_{i,0} + 0.567x_{j,0} - 0.722}{0.092x_{i,0}^2 - 0.955x_{i,0} + 3.333}$}
                            &\makecell[l]{$-$}
                            &\makecell[l]{$-$}\\
Random                      
                            &\makecell[l]{$\frac{dx_{i,0}}{dt}=\frac{30.671x_{i,0}^2-323.132x_{i,0}+729.502}{0.366x_{i,0}-15.8}$\\$+\sum A_{ij}0.281x_{j,0}-2.041-\frac{3.596}{(0.138x_{i,0}-1)^2(0.037x_{j,0}+1)^2-1.63}$}
                            &\makecell[l]{$-$}
                            &\makecell[l]{$-$}\\
Small World                 
                            &\makecell[l]{$\frac{dx_{i,0}}{dt}=-2.353x_{i,0}-0.117x_{i,0}^2+12.396$\\$-\sum A_{ij}\frac{3.392(0.19x_{j,0}+1)^2}{8.775(0.207x_{i,0}-1)^2+4.927}$}
                            &\makecell[l]{$-$}
                            &\makecell[l]{$-$}\\

\botrule
\end{tabular}
% \end{table}
\end{sidewaystable}

\clearpage

\section{More details on on inferring the transmission laws of epidemics}\label{secA5}
% \subsection{Heterogeneous epidemic transmission}\label{secA51}

\subsection{Real-world global epidemic transmission}\label{secA51}
We collect daily cumulative infected numbers of three classic epidemics, SARS \cite{dataweb}, H1N1 \cite{dataweb}, and COVID-19 \cite{dong2020interactive}, and map a directed weighted empirical topology using the global aviation network in OpenFights \cite{openflights} to construct an empirical real-world global epidemic transmission system. Specifically, SARS collects 117 days of data from 37 countries and regions (for the convenience of subsequent expression, it is referred to as nodes), H1N1 collects 74 days of data from 130 nodes, and COVID-19 collects 158 days of data from 174 nodes. To maintain the transmission characteristics of the epidemic itself, we only consider early data before government intervention, which we define as the first 45 days. For example, if a node collects the first infected on May 1st, the data from May 1st to June 14th will be used.

\subsection{Details on experimental setting}\label{secA53}
For experiments on heterogeneous epidemic transmission, we set the number of topology nodes to 360, and the number of nodes in the four communities is ${120,120,90,30}$, respectively.
Each community independently samples 200 data points (in IN-Domain), which will be fed into the model for equation regression.

For experiments on real-world global epidemic transmission, considering the number of air passengers and the population of each node, the adjacency matrix of the topology result is modified to $\tilde{A_{ij}}=\frac{N_p}{\sum_{i=1}^n}A_{ij}$, where the $N_p$ represents the total passengers daily and $P_i$ is the population of node $i$.
We only randomly sample 30 days of data within 45 days to simulate scenarios where it is difficult to collect information during the early stages of epidemics outbreak.
Note that when calculating the interaction dynamics $f^{(inter)}(x_{i},x_{j})$ of a node $i$ at time $t$, the data of all interaction nodes $x_{j}$ comes from the the calendar date corresponding to that node at time $t$, to avoid the influence of collection offset caused by different propagation start times of each node on the interaction information.
All test data still needs to be preprocessed (see Method section) before being fed into the model, the distribution of data is scaled into a distribution that is as close as possible to the training data.

\subsection{More experimental results and regression analyses}\label{secA54}
\begin{figure*}[t]
\centering
\includegraphics[width=1.0\textwidth]{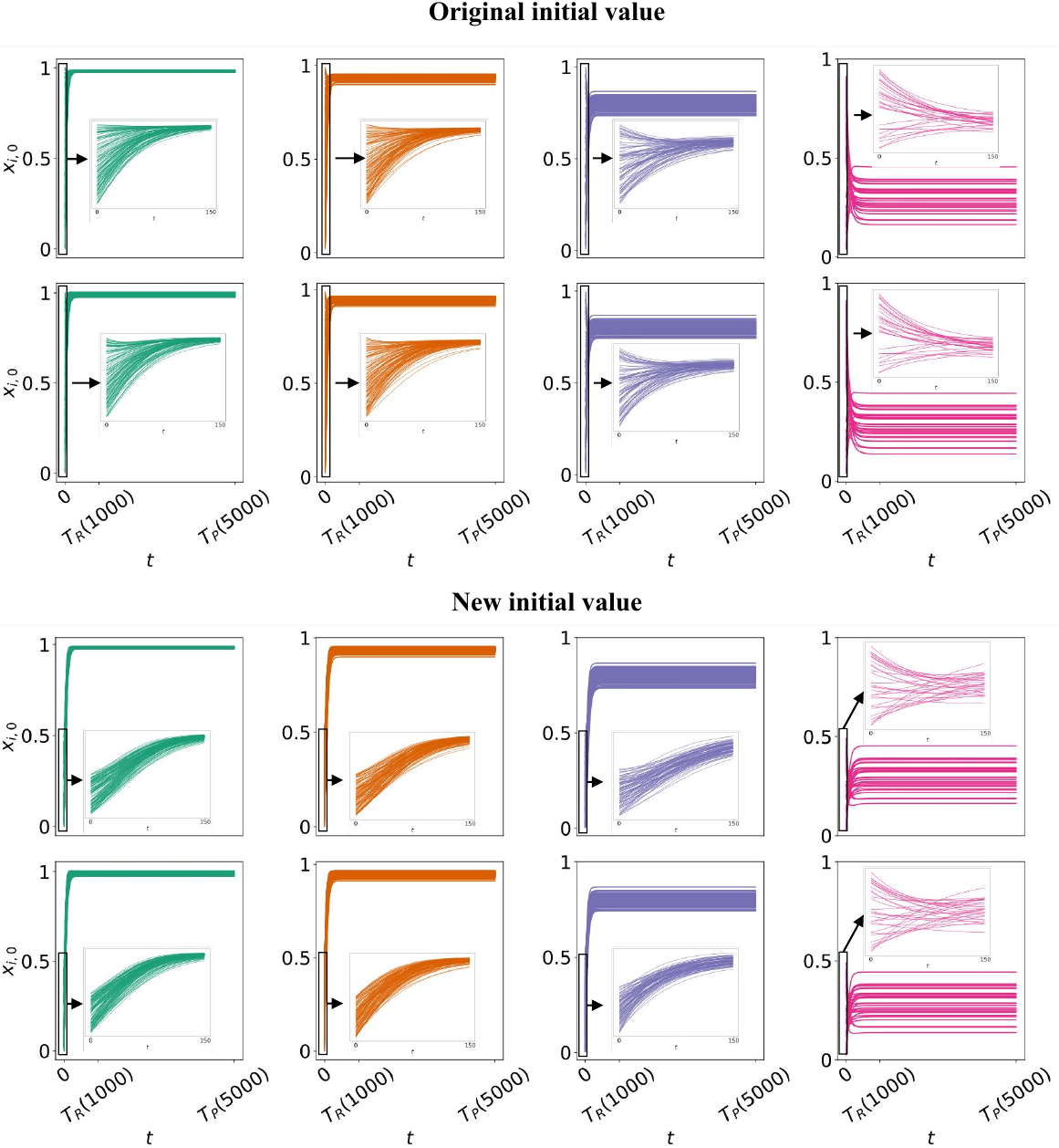}
\caption{\textbf{a.} Heterogeneous epidemic transmission prediction on longer time. \textbf{b.} Heterogeneous epidemic transmission prediction on new initial value, the distribution of 
 the initial value is $\mathbf{U}(0,0.5)$.}\label{a_exp4_1}
\end{figure*}

We first provide the specific regression equation in the experiments on heterogeneous epidemic transmission (see Table. \ref {A_tab18}). The equation regressed by our method is equivalent to the original equation after experimental verification. Longer length predictions and new initial value predictions can also demonstrate the effectiveness of equivalent equations (see Fig .\ref{a_exp4_1}).

% \clearpage

\begin{sidewaystable}[t]
\setlength\tabcolsep{20pt}
\renewcommand\arraystretch{3}
\caption{Results of symbolic regression on heterogeneous epidemic transmission.}\label{A_tab18}%
\begin{tabular}{cccc}
\toprule
\multirow{2}{*}{Community ID}  & \multicolumn{3}{c}{Comparison}\\
                                &True&Ours&TPSINDy\\
\midrule
1                       &\makecell{$\frac{dx_{i,0}}{dt} = -0.5 x_{i,0} +$\\$ \sum x_{j,0} (1 - x_{i,0})$}
                        &\makecell{$\frac{dx_{i,0}}{dt} = -11.66x_{i,0}^2 + 17.81x_{i,0} - 7.135 -$\\$ \sum A_{ij}(0.796x_{i,0}+0.762x_{j,0}^3+0.97x_{j,0}^2+0.554)$}
                        &\makecell{$\frac{dx_{i,0}}{dt} = -0.683 x_{i,0}^2 -$\\$ \sum A_{ij} 1.199 (x_{j,0} - x_{i,0})$} \\
2                       &\makecell{$\frac{dx_{i,0}}{dt} = -2 x_{i,0} + $\\$ \sum x_{j,0} (1 - x_{i,0})$}
                        &\makecell{$\frac{dx_{i,0}}{dt} = -11.789x_{i,0}^2 + 12.725x_{i,0} - 3.312 + $\\$ \sum A_{ij}(-0.583x_{i,0} - 0.114x_{j,0}^2+0.292x_{j,0}+0.44)$}
                        &\makecell{$\frac{dx_{i,0}}{dt} = -0.466 x_{i,0}^2 - $\\$ \sum A_{ij} 0.391 (x_{j,0} - x_{i,0})$}\\
3                       &\makecell{$\frac{dx_{i,0}}{dt} = -5 x_{i,0} +$\\$ \sum x_{j,0} (1 - x_{i,0})$}
                        &\makecell{$\frac{dx_{i,0}}{dt} = -7.047x_{i,0}^2 + 2.142x_{i,0} +$\\$ \sum A_{ij}(-0.647x_{i,0} -0.192x_{j,0}^2-0.406x_{j,0} + 0.403)$}
                        &\makecell{$\frac{dx_{i,0}}{dt} = -4.192 x_{i,0}^2 - $\\$\sum A_{ij} 0.007 (x_{j,0} - x_{i,0})$}\\
4                       &\makecell{$\frac{dx_{i,0}}{dt} = -10 x_{i,0} +$\\$ \sum x_{j,0}  (1 - x_{i,0})$}
                        &\makecell{$\frac{dx_{i,0}}{dt} = -3.294x_{i,0}^2 + 0.851x_{i,0} - 6.932 + $\\$\sum A_{ij}(0.685x_{i,0}-1) ^2(1.13x_{j,0}+0.03$}
                        &\makecell{$\frac{dx_{i,0}}{dt} = -0.683 x_{i,0}^2 - $\\$ \sum A_{ij} 1.199 (x_{j,0} - x_{i,0})$}\\
\botrule
\end{tabular}
\end{sidewaystable}

\begin{figure*}[t]
\centering
\includegraphics[width=0.95\textwidth]{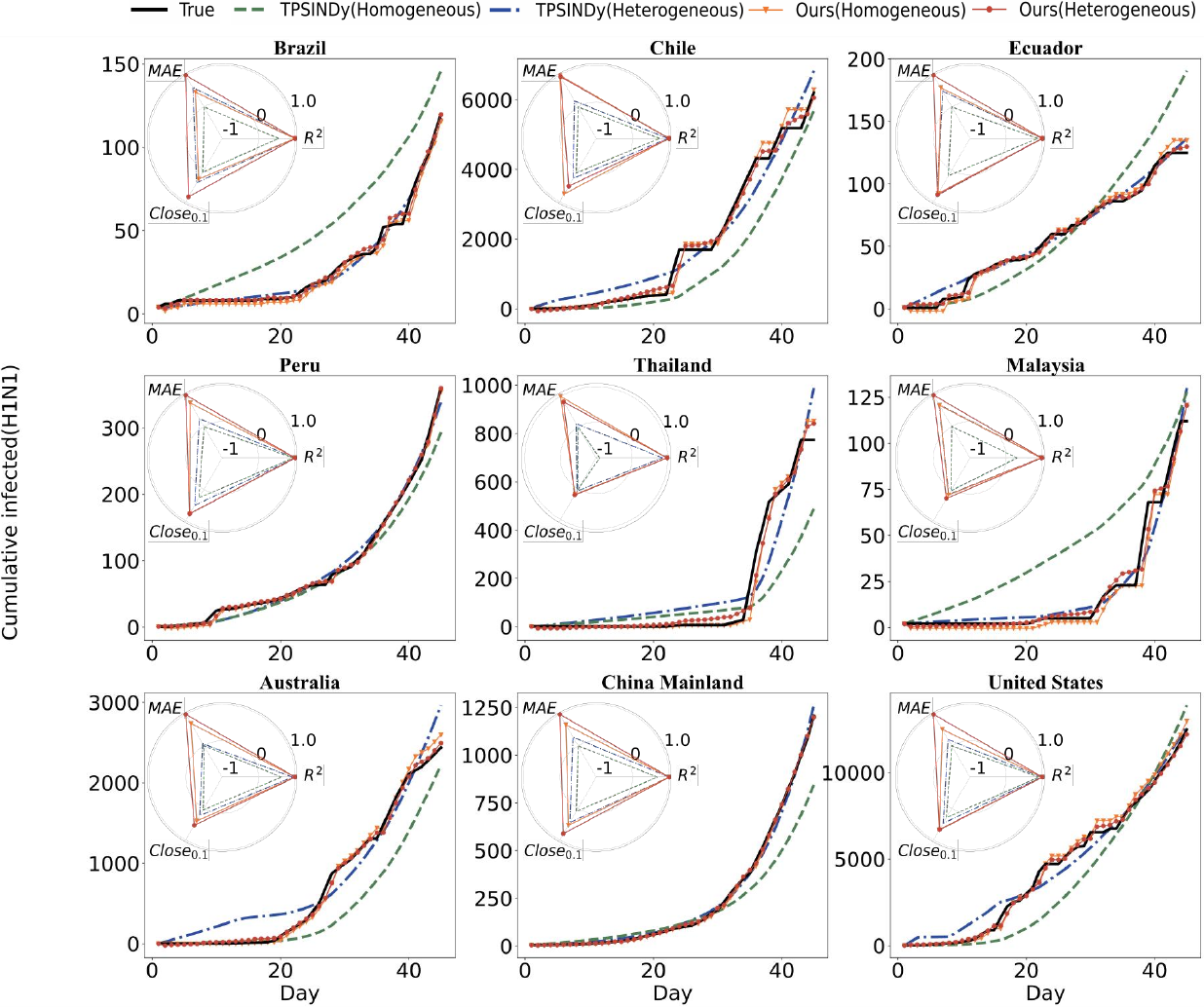}
\caption{Comparison of epidemic transmission equations in all nodes on H1N1. Each subgraph represents the cumulative infected curve calculated by each regressed equation, as well as the error with the ground truth ($R^2,Close_{0.1},MAE$).}\label{a_exp4_3}
\end{figure*}

\begin{figure*}[t]
\centering
\includegraphics[width=0.95\textwidth]{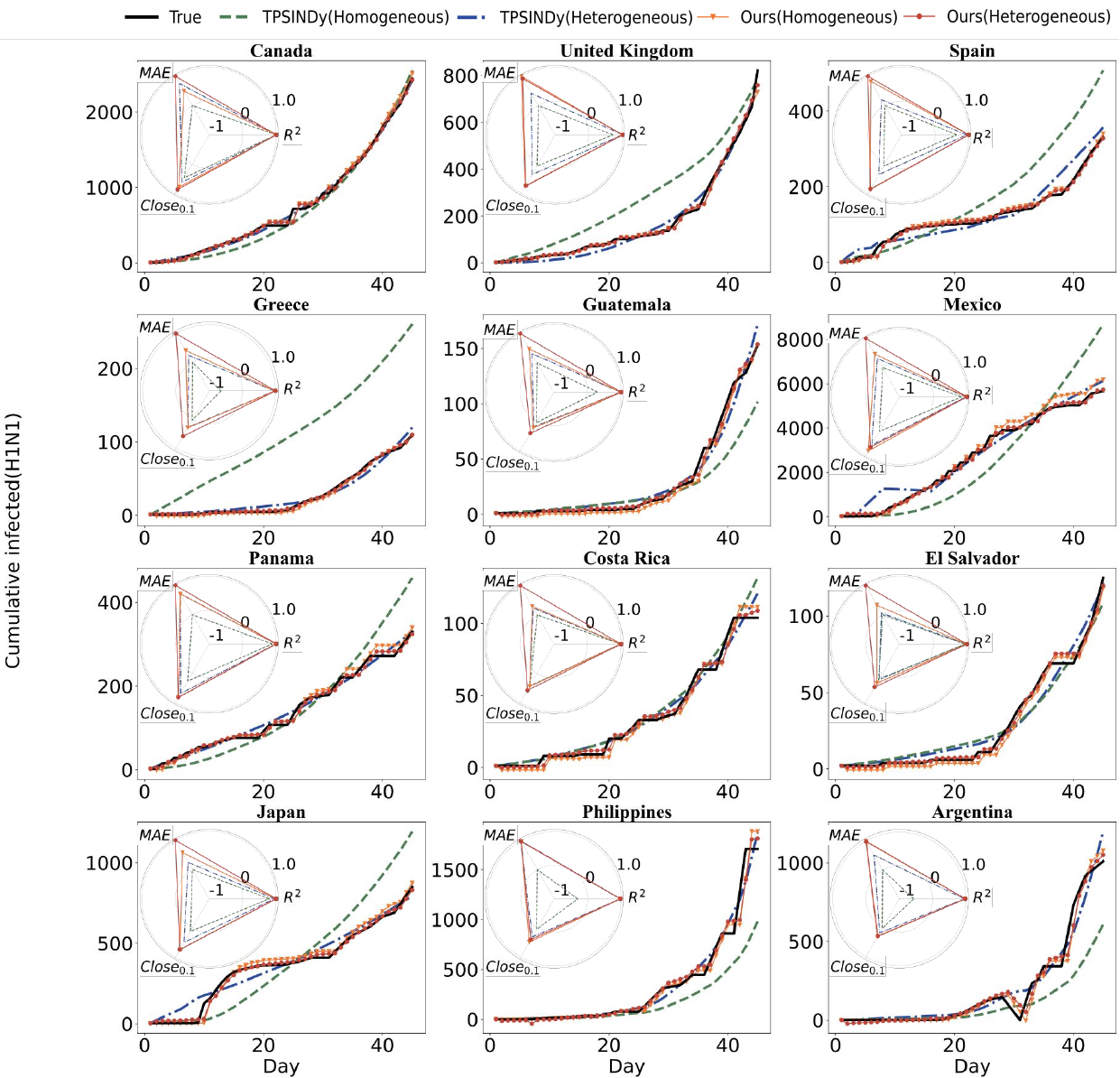}
\caption{Comparison of epidemic transmission equations in all nodes on H1N1.}\label{a_exp4_2}
\end{figure*}

\begin{figure*}[t]
\centering
\includegraphics[width=1.0\textwidth]{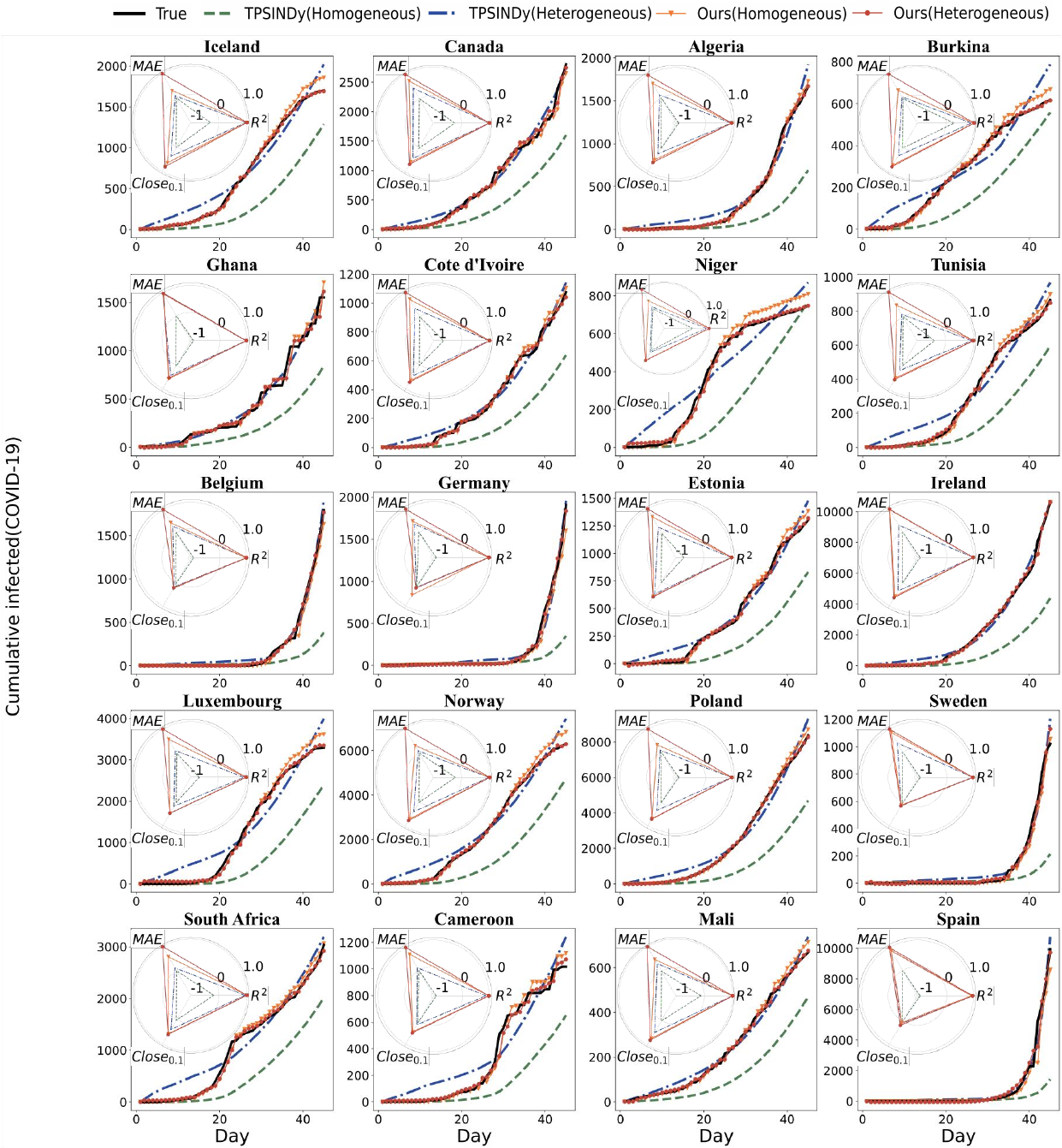}
\caption{Comparison of epidemic transmission equations in all nodes on COVID-19.}\label{a_exp4_4}
\end{figure*}

\begin{figure*}[t]
\centering
\includegraphics[width=0.7\textwidth]{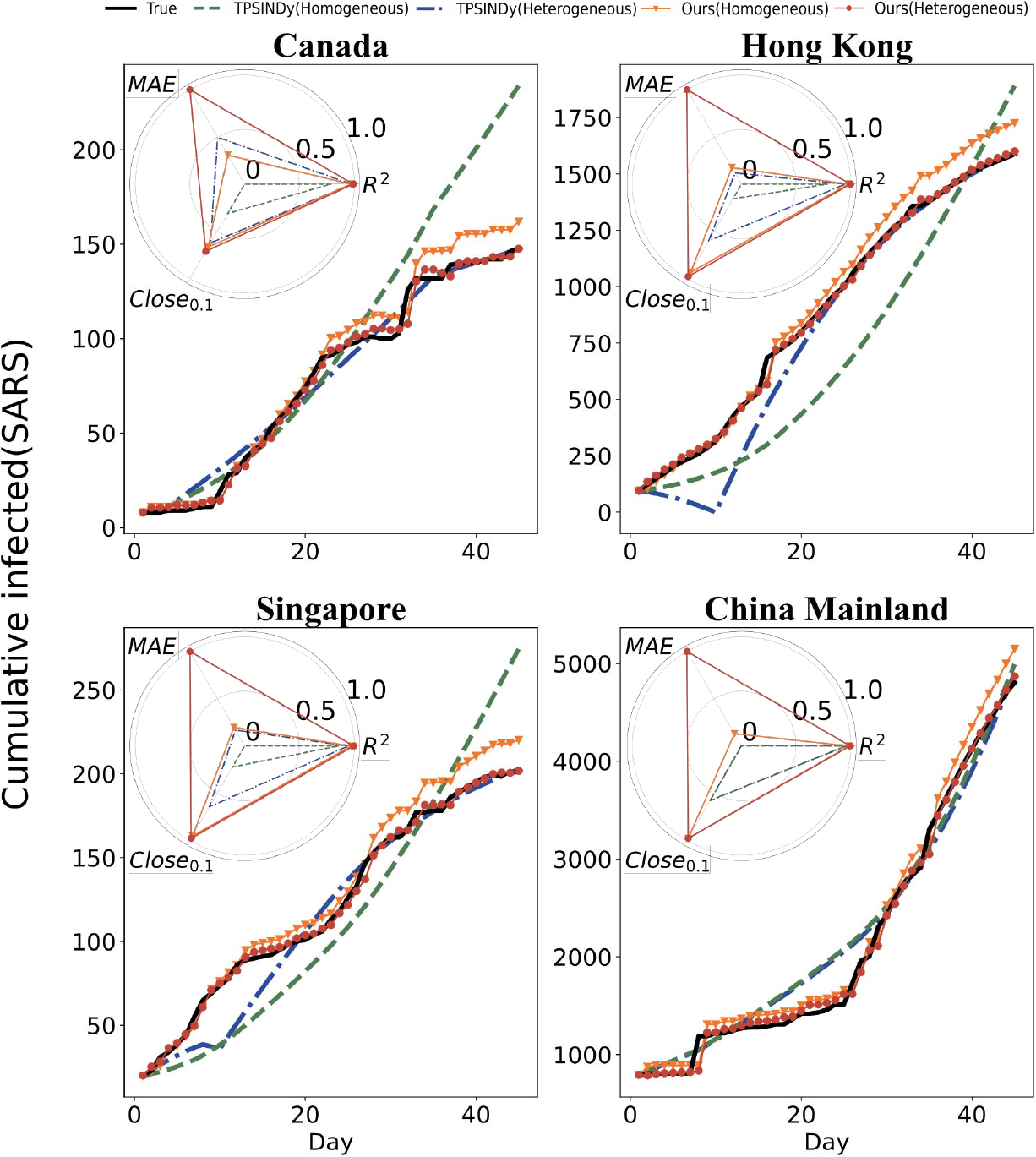}
\caption{Comparison of epidemic transmission equations in all nodes on SARS.}\label{a_exp4_5}
\end{figure*}

\clearpage
We provide the results of the regression equation at other nodes on H1N1, SARS and COVID-19.
For H1N1, the homogeneous equation is regressed from the all test data, and the heterogeneous equation of each node is obtained by optimizing the constants of the homogeneous equation based on the data of each node.
As shown in Fig \ref{a_exp4_2}-\ref{a_exp4_3}, our isomorphic equation is more capable of reflecting the initial transmission characteristics of epidemics at each node, and the optimized heterogeneous equation is more in line with the true transmission curve of each node, compared with TPSINDy. To test the generality of the obtained equations, We directly apply the equation regressed only from H1N1 data to SARS and COVID-19 and and obtain heterogeneous equations through data from each node.
The cumulative infected generated based on our equation is closer to the ground truth (see Fig \ref{a_exp4_4}-\ref{a_exp4_5}), indicating that our method has stronger generalization ability.
Noted that due to the number of infected ranging from 10 to 100000, we use $Close_{0.1}$ and MAE.

\clearpage

\begin{table}[h]

\renewcommand\arraystretch{1.8}
\caption{Specific parameters in equations on H1N1.}\label{A_tab19}%
\begin{tabular}{lll}
\toprule
Node & Ours & TPSINDy\\
                
\midrule
All & \makecell[l]{$a=1.103 , b=2.855$     \\$c=-0.0004 , d=-0.0005 , e=0.538 , f=-0.049 , g=-159.111$} &\makecell[l]{$a=0.040$\\$b=105.160$}\\
Canada & \makecell[l]{$a=16.398 , b=4.906$     \\$c=18.688 , d=0.002 , e=6.611 , f=9.788 , g=11.216$} &\makecell[l]{$a=0.058$\\$b=22.732$}\\
United Kingdom & \makecell[l]{$a=3.573 , b=0.196$    \\$c=-0.749 , d=-0.008 , e=0.610 , f=29.378 , g=-71.376$} &\makecell[l]{$a=0.108 $\\$ b=-0.525$}\\
Spain & \makecell[l]{$a=633.117 , b=344.593$    \\$c=-234.318 , d=0.013 , e=-126.620 , f=1680.625 , g=-992.798$} &\makecell[l]{$a=0.012 $\\$ b=17.050$}\\
Greece & \makecell[l]{$a=-1.140 , b=-0.215$    \\$c=2.755 , d=0.009 , e=1.006 , f=- 360.972 , g=1443.838$} &\makecell[l]{$a=0.093 $\\$ b=0.521$}\\
Guatemala & \makecell[l]{$a=-1.840 , b=-2.757$    \\$c=57.232 , d=0.018 , e=20.727 , f=1477.795 , g=533.003$} &\makecell[l]{$a=0.139 $\\$ b=2.740$}\\
Mexico & \makecell[l]{$=0.822 , b=80.409$    \\$c=0.289 , d=-0.015 , e=53.556 , f=19.122 , g=1102.118$} &\makecell[l]{$a=-0.022 $\\$ b=488.844 $}\\
Panama  & \makecell[l]{$a=0.512 , b=-2.516$    \\$c=2.165 , d=0.07 , e=0.153 , f=156.925 , g=96.461$} &\makecell[l]{$a=0.023 $\\$ b=74.139$}\\
Costa Rica & \makecell[l]{$a=-6.657 , b=-1.111$    \\$c=80.620 , d=0.018 , e=0.119 , f=-156.157 , g=-91887.221$} &\makecell[l]{$a=0.065 $\\$ b=7.881$}\\
El Salvador & \makecell[l]{$a=0.619 , b=- 1.020$    \\$c=5.803 , d=0.008 , e=0.056 , f=-4.589 , g=-375.397$} &\makecell[l]{$a=0.091 $\\$ b=4.335 $}\\
Japan & \makecell[l]{$a=-0.863 , b=4.172$    \\$c=2.286 , d=0.032 , e=3.613 , f=-672.750 , g=417.308$} &\makecell[l]{$a=0.016 $\\$ b=75.001$}\\
Philippines & \makecell[l]{$a=-8.320 , b=-120.541$    \\$c=47.371 , d=1.095 , e=354.446 , f=26.562 , g= -26.563$} &\makecell[l]{$a=0.142 $\\$ b=10.367$}\\
Argentina & \makecell[l]{$a=-17.142 , b=-37.776$    \\$c=146.885 , d=0.408 , e=26.043 , f=22.729 , g=-227.262$} &\makecell[l]{$a=0.145 $\\$ b=16.403 $}\\
Brazil & \makecell[l]{$a=-2.927 , b=-1.308$    \\$c=14.665 , d=0.001 , e=1.053 , f=7.184 , g= -93.375$} &\makecell[l]{$a=0.119 $\\$ b=-3.423$}\\
Chile & \makecell[l]{$a=0.712 , b=-299.531$    \\$c=2.032 , d=2.747 , e=-24.770 , f=2.230 , g=-2.228$} &\makecell[l]{$a=0.071 $\\$ b=1342.554$}\\
Ecuador & \makecell[l]{$a=0.924 , b=-0.420$    \\$c=-0.271 , d=0.052 , e=1.110 , f=317.152 , g=265.366$} &\makecell[l]{$a=0.025 $\\$ b=46.512$}\\
Peru & \makecell[l]{$a=-0.272 , b=3.335$    \\$c=12.058 , d=-0.017 , e=0.326 , f=-20.606 , g=-1958.933$} &\makecell[l]{$a=0.088 $\\$ b=5.479$}\\
Thailand & \makecell[l]{$a=-4.659 , b=-6.702$    \\$c=10.451 , d=0.595 , e=-3.398 , f=211.541 , g=-349.450$} &\makecell[l]{$a=0.159 $\\$ b=9.869$}\\
Malaysia & \makecell[l]{$a=-67.280 , b=-3.111$    \\$c=167.315 , d=0.089 , e=4.312 , f=-743.10 , g=-4662.720$} &\makecell[l]{$a=0.178 $\\$ b=-0.928$}\\
Australia & \makecell[l]{$a=-3.199 , b=-39.275$    \\$c=9.161 , d=0.380 , e=- 17.971 , f=-14.079 , g=56.303$} &\makecell[l]{$a=0.085 $\\$ b=89.574 $}\\
China & \makecell[l]{$a=-0.068 , b=-11.693$    \\$c=0.886 , d=0.024 , e=-12.426 , f=8.312 , g=0.028$} &\makecell[l]{$a=0.122 $\\$ b=3.296$}\\
United States & \makecell[l]{$a=-2.333 , b=3.957$    \\$c=0.943 , d=0.452 , e=11.136 , f=-264.038 , g=1177.158$} &\makecell[l]{$a=0.056 $\\$ b=438.925$}\\
\hline
Ours:&\multicolumn{2}{l}{$\frac{dx_{i,0}}{dt}=ax_{i,0}+b+\sum A_{ij}(cx_{i,0}+dx_{j,0}+e+\frac{1}{fx_{j,0}+g})$}\\
TPSINDy:&\multicolumn{2}{l}{$\frac{dx_{i,0}}{dt}=ax_{i,0}+b\sum A_{ij}\frac{1}{1+e^{-(x_{j,0}-x_{i,0})}}$}\\
\botrule
\end{tabular}
\end{table}

\clearpage

\begin{table}[h]

\renewcommand\arraystretch{1.8}
\caption{Specific parameters in equations on COVID-19.}\label{A_tab19}%
\begin{tabular}{lll}
\toprule
Node & Ours & TPSINDy\\
                
\midrule
All & \makecell[l]{$a=1.103 , b=2.855$     \\$c=-0.0004 , d=-0.0005 , e=0.538 , f=-0.049 , g=-159.111$} &\makecell[l]{$a=0.074$\\$b=7.13$}\\
Iceland & \makecell[l]{$a=1.108, b=8.012$ \\$c=0.608, d=-0.044, e=2.062, f=-305.231, g=-293.528$} &\makecell[l]{$a=0.042$ \\ $b=706.343 $}\\
Canada & \makecell[l]{$a=46.924 , b=158.631$ \\ $c=-57.657 , d=0.0001 , e=-182.065 , f=9369.583 , g=2323.033$} &\makecell[l]{$a=0.060$\\$ b=26.534$}\\
Algeria & \makecell[l]{$a=-23.233 , b=-2.851$ \\ $c=154.532 , d=0.075 , e=-2.640 , f=-38.008 , g=-259.349$} &\makecell[l]{$a=0.012$\\$ b=123.971$}\\
Burkina Faso & \makecell[l]{$ a=1.039 , b=6.626$ \\ $c=1.831 , d=-0.195 , e=1.524 , f=-32.223 , g=64.446$} &\makecell[l]{$a=0.011$\\$ b=922.518$}\\
Ghana & \makecell[l]{$a=0.995 , b=-10.572$ \\ $c=-0.805 , d=0.072 , e=-0.684 , f=32.287 , g=33.454$} &\makecell[l]{$a=0.076$\\$ b=262.872$}\\
Cote d'Ivoire & \makecell[l]{$a=0.647 , b=-1.709$ \\ $c=1.563 , d=0.816 , e=-2.957 , f=166.895 , g=-132.336$} &\makecell[l]{$a=0.067$\\$ b=189.751$}\\
Niger  & \makecell[l]{$ a=-0.993 , b=20.614$ \\ $c=0.934 , d=-0.845 , e=-0.911 , f=18.258 , g= 9.215$} &\makecell[l]{$a=0.028 $\\$ b=970.443$}\\
Tunisia & \makecell[l]{$a=0.136 , b=-4.475$ \\ $c=8.718 , d=0.194 , e=-2.805 , f=8175.448 , g=-18914.28$} &\makecell[l]{$a=0.049 $\\$ b=77.134$}\\
Belgium & \makecell[l]{$a=-36.852 , b=-22.534$ \\ $c=76.289 , d=0.399 , e=-55.262 , f=-153.424 , g=0.006$} &\makecell[l]{$a=0.237 $\\$ b=8.970 $}\\
Germany & \makecell[l]{$ a=-2.739 , b=-1.537$ \\ $c=1.387 , d=0.195 , e=-1.086 , f=-13.210 , g=277.411$} &\makecell[l]{$a=0.291 $\\$ b=-6.556$}\\
Estonia & \makecell[l]{$ a=1.747 , b=47.225$ \\ $c=-19.799 , d=0.256 , e=-52.340 , f=20.919 , g=-0.0002$} &\makecell[l]{$a=0.056 $\\$ b=358.731$}\\
Ireland & \makecell[l]{$a=16.564 , b=-46.331$ \\ $c=138.148 , d=0.003 , e=38.312 , f=-18.845 , g=101.045$} &\makecell[l]{$a=0.099 $\\$ b=163.027$}\\
Luxembourg & \makecell[l]{$a=-44.031 , b=29.240$ \\ $c=546.251 , d=-0.113 , e=418.154 , f=83.217 , g=77.527$} &\makecell[l]{$a=0.047 $\\$ b=623.419$}\\
Norway & \makecell[l]{$ a=0.855 , b=26.478$ \\ $c=0.775 , d=-0.050 , e=9.815 , f=665.812 , g=632.194$} &\makecell[l]{$a=0.042 $\\$ b=515.854$}\\
Poland & \makecell[l]{$a=-0.038 , b=-12.039$ \\ $c=2.417 , d=0.041 , e=-5.419 , f=-29.190 , g=15.205$} &\makecell[l]{$a=0.072 $\\$ b=201.713$}\\
Sweden & \makecell[l]{$ a=-4.991 , b=18.458$ \\ $c=16.816 , d=0.089 , e=0.237 , f=-44.419 , g=44.417$} &\makecell[l]{$a=0.276 $\\$ b=7.410$}\\
South Africa & \makecell[l]{$a=25.737 , b= 25.417$ \\ $c=-141.837 , d=-0.009 , e=16.654 , f=-34.385 , g=18.0317$} &\makecell[l]{$a=0.045 $\\$ b=652.840$}\\
Cameroon & \makecell[l]{$a=0.688 , b=10.144$ \\ $c=1.992 , d=0.464 , e=-60.827 , f=151.537 , g=135.525$} &\makecell[l]{$a=0.064 $\\$ b=417.863$}\\
Mali & \makecell[l]{$ a=0.932 , b=-7.206$ \\ $c=0.499 , d=0.894 , e=-0.439 , f=28.823 , g=6.681$} &\makecell[l]{$a=0.052 $\\$ b=201.635 $}\\
Spain & \makecell[l]{$a=587.169 , b=-34.708$ \\ $c=-252.665 , d=0.554 , e=-12.002 , f=4.395 , g=-35.623$} &\makecell[l]{$a=0.361 $\\$ b=5.163$}\\
\hline
Ours:&\multicolumn{2}{l}{$\frac{dx_{i,0}}{dt}=ax_{i,0}+b+\sum A_{ij}(cx_{i,0}+dx_{j,0}+e+\frac{1}{fx_{j,0}+g})$}\\
TPSINDy:&\multicolumn{2}{l}{$\frac{dx_{i,0}}{dt}=ax_{i,0}+b\sum A_{ij}\frac{1}{1+e^{-(x_{j,0}-x_{i,0})}}$}\\
\botrule
\end{tabular}
\end{table}

\clearpage

\clearpage

\begin{table}[h]

\renewcommand\arraystretch{1.8}
\caption{Specific parameters in equations on SARS.}\label{A_tab19}%
\begin{tabular}{lll}
\toprule
Node & Ours & TPSINDy\\
                
\midrule
All & \makecell[l]{$a=1.103 , b=2.855$     \\$c=-0.0004 , d=-0.0005 , e=0.538 , f=-0.049 , g=-159.111$} &\makecell[l]{$a=0.074$\\$b=7.13$}\\
Canada & \makecell[l]{$a=1151.383 , b=2.058$ \\ $c=-123.860 , d=0.003 , e=0.078 , f=300.515 , g=-1451.585$} &\makecell[l]{$a=0.009$ \\ $b=4.941$}\\
Hong Kong & \makecell[l]{$a=2.184 , b=1.144$ \\ $c=-0.205 , d=0.0002 , e=6.909 , f=-53.296 , g=-86.233$} &\makecell[l]{$a=-0.055$\\$ b=495.126$}\\
Singapore & \makecell[l]{$a=-928.491 , b=1589.761$ \\ $c=147.121 , d=-0.003 , e=-250.814 , f=-41.359 , g=-126.697$} &\makecell[l]{$a=-0.044$\\$ b=116.412$}\\
China Mainland & \makecell[l]{$ a=4382.935 , b=3.857$ \\ $c=-231.848 , d=0.042 , e=-2.813 , f=-1849.291 , g=22.359$} &\makecell[l]{$a=0.045$\\$ b=5.003e-6$}\\
\hline
Ours:&\multicolumn{2}{l}{$\frac{dx_{i,0}}{dt}=ax_{i,0}+b+\sum A_{ij}(cx_{i,0}+dx_{j,0}+e+\frac{1}{fx_{j,0}+g})$}\\
TPSINDy:&\multicolumn{2}{l}{$\frac{dx_{i,0}}{dt}=ax_{i,0}+b\sum A_{ij}\frac{1}{1+e^{-(x_{j,0}-x_{i,0})}}$}\\
\botrule
\end{tabular}
\end{table}

\section{Preliminary attempts with LLM}\label{secA6}
\begin{figure*}[h]
\centering
\includegraphics[width=1.0\textwidth]{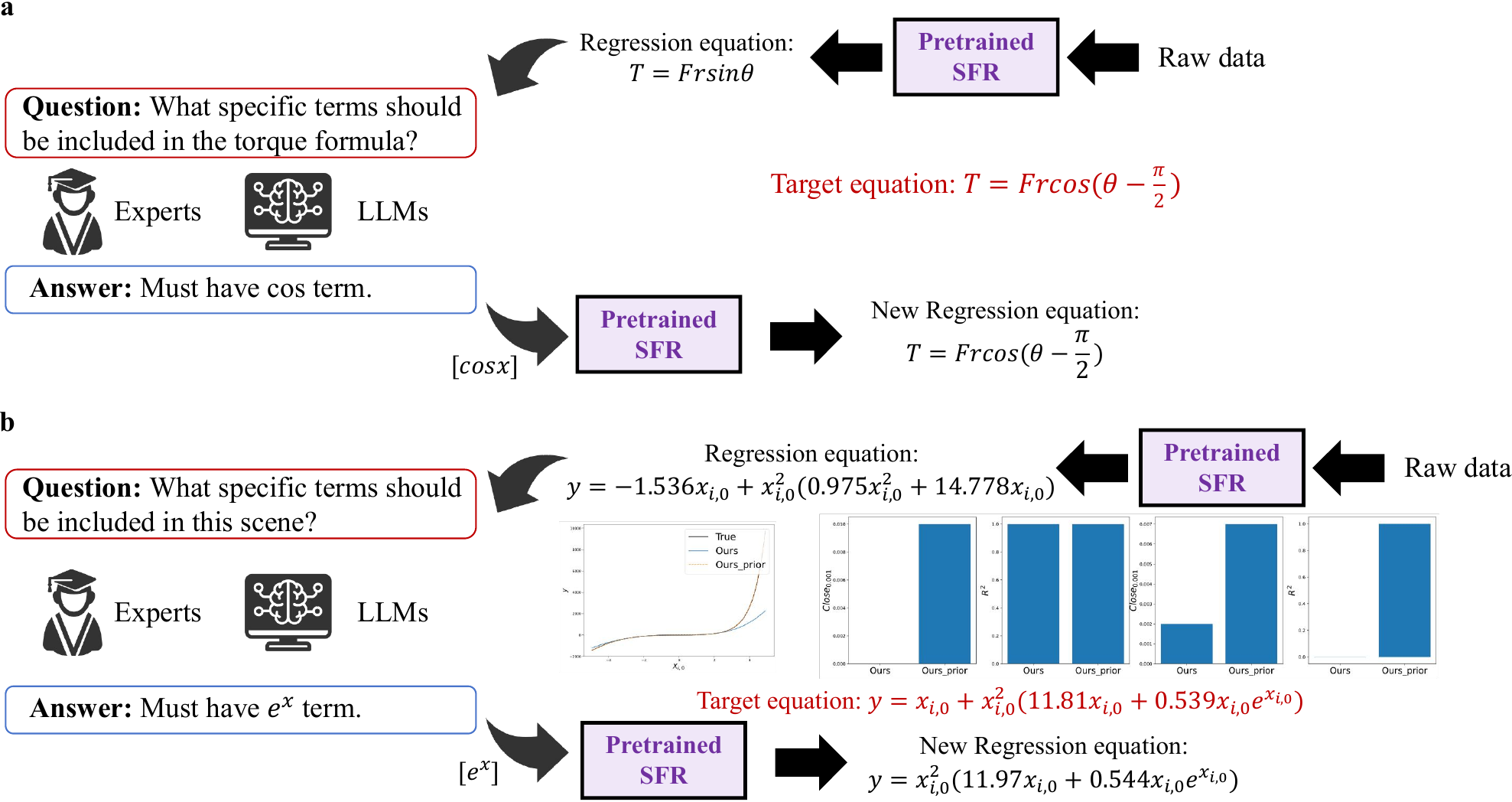}
\caption{Symbolic regression combined with domain knowledge}\label{a_exp6_1}
\end{figure*}

We use a decoding strategy called constrained beam search to embed domain knowledge into the regression process of the model. Firstly, the representation $h$ generated from the data enters the decoder. When $h$ enters the decoder $dec^{self}$ and $dec^{inter}$, the model will call the domain knowledge module to obtain the symbolic representation $e_{know}$ of domain knowledge from experts or LLMs, such as $[+, x_i, x_j]$. The representation $e_{know}$ will be forcibly added to each round of beam search $e_k$ to ensure that the final symbolic regression result contains $e_{know}$. By embedding domain knowledge into the model, we can generate specific forms of equations (see Fig .\ref{a_exp6_1}(a)) and further improve the accuracy of the model for high complexity equations (see Fig .\ref{a_exp6_1}(b)).

\end{appendices}

\end{document}